\documentclass{bionetta}

\usepackage[toc,page]{appendix}

\usepackage{pgfplots}
\usepackage{graphicx}
\usepackage{listings}
\usepackage{makecell}
\pgfplotsset{compat=1.18} 
\usepackage{float}
\usepackage{dblfloatfix}
\usepackage{ulem}
\usepackage{caption}
\usepackage{dramatist}
\usepackage{xspace}
\usepackage{pifont}
\usepackage{multirow}
\usepackage{tcolorbox}
\usepackage{xltabular}
\usepackage{longtable}
\usepackage{hyperref}
\usepackage[nameinlink]{cleveref}
\usepackage{tablefootnote}
\usepackage{amsfonts}
\usepackage{amsmath}
\usepackage{amssymb}
\usepackage{lineno}
\usepackage{multirow}
\usepackage{adjustbox}
\usepackage[bottom]{footmisc}
\usepackage{CJKutf8}
\usepackage{subfigure}
\usepackage{setspace}
\usepackage{dsfont}
\usepackage{array} 
\usepackage{tabularx} 
\usepackage{subfigure}
\usepackage{csquotes}
\usepackage[table]{xcolor}
\usepackage{lipsum}  
\usepackage{multicol} 
\usepackage{empheq}
\usepackage{pgfplotstable}
\usepackage{caption}
\usepackage{subcaption}
\usepackage{bbm}
\usepackage{arydshln}

\usepackage{siunitx}  % For the 'S' column type (number alignment)
\usepackage{nicematrix}
\usepackage{tabularx} % For tables that fit a specific width
\usepackage{booktabs} % For professional-looking tables (\toprule, \midrule, \bottomrule)
\usepackage{multirow} % To span rows in a table
\newcolumntype{C}{>{\centering\arraybackslash}X}

\usepackage[linesnumbered,ruled,vlined]{algorithm2e}

\SetKwComment{Comment}{/* }{ */}

\SetCommentSty{mycommentstyle}

\SetKwInOut{Input}{Input}
\SetKwInOut{Output}{Output}
\SetKwInOut{Return}{Return}

\usepackage[
  backend=bibtex,   % Ensure Biber is used for processing
  style=alphabetic,   % Use the abbrvnat citation style (similar to ACM)
  url=true,
  doi=true,
  isbn=true
]{biblatex}
\usepackage{listings}
\usepackage{xcolor}

\definecolor{codegreen}{rgb}{0,0.6,0}
\definecolor{codegray}{rgb}{0.5,0.5,0.5}
\definecolor{codepurple}{rgb}{0.58,0,0.82}
\definecolor{backcolour}{rgb}{0.95,0.95,0.92}

\lstdefinestyle{mystyle}{
    backgroundcolor=\color{white},   
    commentstyle=\color{codegray},
    keywordstyle=\color{blue},
    numberstyle=\tiny\color{codegray},
    stringstyle=\color{codepurple},
    basicstyle=\ttfamily\small,
    breakatwhitespace=false,         
    breaklines=true,                 
    captionpos=b,                    
    keepspaces=true,                 
    numbers=none,                    
    numbersep=5pt,                  
    showspaces=false,                
    showstringspaces=false,
    showtabs=false,                  
    tabsize=2,
    escapeinside=||, % Allows LaTeX commands inside code
    language=Python
}

\newcommand{\tick}{\textcolor{green!50!black}{\ding{51}}}
\newcommand{\cross}{\textcolor{red!70!black}{\ding{55}}}

\makeatletter
\newcommand{\smalldollar}{\mathrel{\mathpalette\small@dollar\relax}}
\newcommand{\small@dollar}[2]{%
  \vcenter{\hbox{%
    $#1\textnormal{\fontsize{0.7\dimexpr\f@size pt}{0}\selectfont\$}$%
  }}%
}
\makeatother

\begin{document}

% --- Title ---
\title{\textbf{Bionetta: Efficient Client-Side Zero-Knowledge Machine Learning Proving}
\\
\vspace{7px}
\large \textcolor{gray}{Technical Report}}

% --- Authors and contacts ---
\author{
    Rarimo \\
    \texttt{Info@rarimo.com}
    \and
    Distributed Lab \\
    \texttt{contact@distributedlab.com}
}

\date{} % No date needed

\maketitle

\begin{abstract}
  In this report, we compare the performance of our \textit{UltraGroth}-based zero-knowledge
  machine learning framework \texttt{Bionetta} to other tools of similar purpose
  such as \texttt{ezkl}, Lagrange's \texttt{deep-prove}, \texttt{zkml},
  \texttt{keras2circom}, or \texttt{zkCNN}. The
  results show a significant boost in the proving time for custom-crafted neural
  networks: they can be proven even on mobile devices, enabling numerous
  client-side proving applications. While our scheme increases the cost of
  one-time preprocessing steps, such as circuit compilation and generating
  trusted setup, our approach is, to the best of our knowledge, the only one
  that is deployable on the native EVM smart contracts without overwhelming
  proof size and verification overheads.
\end{abstract}

% --- Reusable break symbols defined with TikZ ---
\tikzset{
    outlierbreak/.pic={
        \node[inner sep=1pt, rotate=90, font=\small] at (0,3pt) {//};
        \node[font=\footnotesize, fill=white, inner sep=1pt] at (0,14pt) {($\sim$3400x)};
    }
}

% --- Data Table with safe, underscore-free identifiers ---
\pgfplotstableread{
model           bionetta_ug bionetta_g16 keras2circom zkml      ezkl      deep-prove  zkcnn
ProofSize       1.00        0.68         0.68         4.7      144.32     3388.33  26.4
VKeySize        1.00        0.97         0.97         694       1084    0.00 0.00
ProvingTime     1.00        2.90         8.20         360       430    1.75 1.14
VerifyingTime   1.00        0.95         0.95         0.99      173.39    20.89 110
}\relativedata

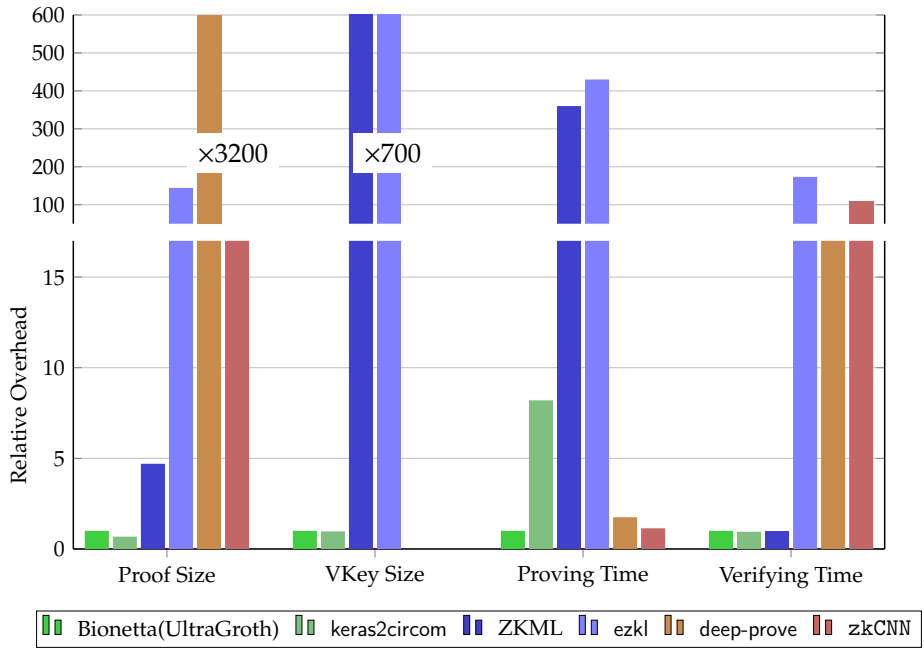
\begin{figure}[h]
    \centering
    \begin{tikzpicture}[scale=0.75]

    % -------- LOWER AXIS (0–17) --------
    \begin{axis}[
        width=450pt,
        height=200pt,
        ybar,
        bar width=12pt,
        symbolic x coords={ProofSize, VKeySize, ProvingTime, VerifyingTime},
        xtick=data,
        xticklabels={Proof Size, VKey Size, Proving Time, Verifying Time},
        ymin=0, ymax=17,
        ymajorgrids=true,
        ylabel={Relative Overhead},
        enlarge x limits=0.15,
        ytick={0, 5, 10, 15},
        axis line style={line width=0.75pt},
        ticklabel style={font=\small},
        label style={font=\small},
        legend style={font=\small, at={(0.5,-0.2)}, anchor=north, legend columns=6, column sep=0.5em},
        axis x line*=bottom,
        y axis line style={line width=0.75pt},
    ]

    % --- GROUP 1: GREENS ---
    % Bionetta (UltraGroth)
    \addplot[fill=green!75!black!75, draw=none]
        table[x=model, y expr={\thisrow{bionetta_ug} > 17 ? 17 : \thisrow{bionetta_ug}}]{\relativedata};

    % keras2circom  
    \addplot[fill=green!50!black!50, draw=none]
        table[x=model, y expr={\thisrow{keras2circom} > 17 ? 17 : \thisrow{keras2circom}}]{\relativedata};

    % --- GROUP 2: BLUES ---
    % ZKML
    \addplot[fill=blue!75!black!75, draw=none]
        table[x=model, y expr={\thisrow{zkml} > 17 ? 17 : \thisrow{zkml}}]{\relativedata};

    % ezkl
    \addplot[fill=blue!50, draw=none]
        table[x=model, y expr={\thisrow{ezkl} > 17 ? 17 : \thisrow{ezkl}}]{\relativedata};

    % --- GROUP 3: WARM COLORS ---
    % deep-prove
    \addplot[fill=orange!70!black!70, draw=none]
        table[x=model, y expr={\thisrow{deep-prove} > 17 ? 17 : \thisrow{deep-prove}}]{\relativedata};

    % zkCNN
    \addplot[fill=red!60!black!60, draw=none]
        table[x=model, y expr={\thisrow{zkcnn} > 17 ? 17 : \thisrow{zkcnn}}]{\relativedata};

    \legend{Bionetta(UltraGroth), \textsf{keras2circom}, ZKML, \textsf{ezkl}, \textsf{deep-prove}, \texttt{zkCNN}}
    \end{axis}

    % -------- UPPER AXIS (50–600) --------
    \begin{axis}[
        width=450pt,
        height=150pt,
        ybar,
        bar width=12pt,
        symbolic x coords={ProofSize, VKeySize, ProvingTime, VerifyingTime},
        xtick=data,
        ymin=50, ymax=600,
        ymajorgrids=true,
        enlarge x limits=0.15,
        axis x line=none,
        xticklabel=\empty,
        ytick={100, 200, 300, 400, 500, 600},
        y axis line style={line width=0.75pt},
        ticklabel style={font=\small},
        at={(0,5.75cm)},
    ]

    % same order & colors as lower axis
    \addplot[fill=green!75!black!75, draw=none]
        table[x=model, y expr={\thisrow{bionetta_ug} < 50 ? nan : \thisrow{bionetta_ug}}]{\relativedata};

    \addplot[fill=green!50!black!50, draw=none]
        table[x=model, y expr={\thisrow{keras2circom} < 50 ? nan : \thisrow{keras2circom}}]{\relativedata};

    \addplot[fill=blue!75!black!75, draw=none]
        table[x=model, y expr={\thisrow{zkml} < 50 ? nan : \thisrow{zkml}}]{\relativedata};

    \addplot[fill=blue!50, draw=none]
        table[x=model, y expr={\thisrow{ezkl} < 50 ? nan : \thisrow{ezkl}}]{\relativedata};

    % deep-prove clipped to 600
    \addplot[fill=orange!70!black!70, draw=none]
        table[x=model, y expr={\thisrow{deep-prove} < 50 ? nan : (\thisrow{deep-prove} > 600 ? 600 : \thisrow{deep-prove})}]{\relativedata};

    % zkCNN normally (rarely goes above)
    \addplot[fill=red!60!black!60, draw=none]
        table[x=model, y expr={\thisrow{zkcnn} < 50 ? nan : \thisrow{zkcnn}}]{\relativedata};

    \end{axis}

    % Scaling markers
    \node[draw=none, fill=white] at (2.8, 7) {\footnotesize$\times 3200$};
    \node[draw=none, fill=white] at (5.6, 7) {\footnotesize$\times 700$};

    \end{tikzpicture}

    \caption{Relative resources overhead of existing zkML frameworks compared to Bionetta (UltraGroth). Lower values are better.}
    \label{fig:relative_overhead_final}
\end{figure}

% --- Table of Contents ---
\tableofcontents
\newpage

\section{Introduction}

Zero-knowledge proof systems have been extensively used in various applications,
ranging from blockchain scalability solutions to privacy-preserving identity
protocols, such as Rarimo \cite{snark-survey,rarimo}. Currently, the most widely
adopted type of zero-knowledge proof system for native Ethereum verifications is
zk-SNARKs. In particular, the Groth16 proving system \cite{groth16} allows a
prover to demonstrate the witness knowledge of some statement without revealing
the witness itself using just three pairing operations
\cite{snark-stark-comparison}. The main advantages of zk-SNARKs include:
\begin{itemize}[label={\ding{51}}]
    \item zk-SNARK proofs are \textit{succinct}: the proof size is typically
    polylogarithmic in the size of the witness, and the verification time is
    polylogarithmic in the size of the arithmetical circuit. This makes zk-SNARK
    systems perfect for blockchain applications, where the verification process
    must be done on-chain and thus should be as efficient as possible.
    \item zk-SNARK proofs are \textit{zero-knowledge}: the proof does not reveal
    any information about the witness corresponding to the statement being
    proven. While such property might not be crucial for scalability solutions,
    it is essential for privacy-preserving protocols, which is the primary
    use case of our research.
\end{itemize}

Yet, as of today, the majority of zk-SNARK applications are focused on
cryptographic primitives, such as proving the knowledge of the pre-image of the
hash digest \cite{posseidon}, elliptic curve operations \cite{ec_r1cs}, pairings
\cite{pairings_r1cs} etc. While these applications are sufficient for most
blockchain use cases, many applications still require more complex statements to
be proven. One such application is the \textit{machine learning model proof of
inference} or, in our particular case, the \textit{liveness check} and
\textit{biometric proximity proof}. 

In this report, we introduce \textbf{Bionetta} --- the R1CS-based proving
framework with the primary focus on Groth16 derivatives for effective EVM on-chain 
verification. The main advantages of our approach combined with \textit{UltraGroth} are the following:
\begin{itemize}[label={\ding{51}}]
    \item We inherit all the benefits of Groth16 protocol: constant-sized proof
    (\textbf{320B}), constant verification time (four pairing operations), and
    small verification key size.
    \item The proving time \textit{depends only on the number of non-linearity
    calls in the neural network}. All \textit{linear operations} (such as matrix
    multiplications) are \textbf{free}.
    \item Overall, we claim that \textit{R1CS arithmetization greatly
    outperforms any $\mathcal{P}$lon$\mathcal{K}$ish arithmetization-based
    approaches} for client-side proving.
\end{itemize}

However, in contrast to modern ZKML schemes, our methodology is not universal.
In particular, we set the following requirements:
\begin{itemize}[label={\ding{227}}]
    \item Neural network weights and architecture \textbf{must be public} (which
    is a must for any decentralized identity protocol); we do not support
    private model weights.
    \item Currently, we are supporting a limited yet growing set of neural
    network building blocks. For instance, we can prove
    \textit{MobileNetV2}~\cite{mobilenetv2} or \textit{ResNet18}~\cite{resnet}
    architectures, but we \textbf{do not support general \texttt{onnx} models}.
\end{itemize}

\section{Related Approaches}

Among the existing zkML tools and libraries, we choose the five most notable for
benchmarking: \texttt{keras2circom} \cite{keras2circom}, \texttt{ddkang/zkml}
\cite{ddkang-zkml,ddkang-zkml-github}, \texttt{ezkl} \cite{ezkl}, \texttt{zkCNN} \cite{zk-cnn} and recent
\texttt{deep-prove} \cite{deepprove}, claiming to boost the proving time by
roughly $\times 158$\footnote{Based on the original post
\url{https://x.com/lagrangedev/status/1899853995109396618}.}. We are also aware
of the recent \texttt{zkPyTorch} \cite{zkpytorch} framework and look forward to
testing it in the future, yet currently it is not available for public use (as
of December 2025). Below, we provide an informal overview of the differences
between our approach and those of the aforementioned open-sourced frameworks.

\textcolor{blue!50!black}{\textbf{keras2circom.}} As the name suggests,
\texttt{keras2circom} translates the Keras-based models into Circom circuits.
While our approach also utilizes Circom and Groth-like proving system,
\texttt{keras2circom} cannot handle deep networks: in fact, as the original
repository suggests, on average, the scaling factor of each weight cannot exceed
$10^{76/\ell}$, where $\ell$ is the network depth. Since the number of layers
can easily exceed $30$, scaling by the factor of roughly $10^{76/30} \approx
340$ would introduce a drastic precision loss. Besides that,
\texttt{keras2circom} provides much less optimized circuits for the case when
model weights are public.

\textcolor{blue!50!black}{\textbf{ezkl and ddkang/zkml.}} Meanwhile,
\texttt{ezkl} and \texttt{zkml} use \textrm{Halo2} \cite{halo1,halo2} as their
proving system. Since \textrm{Halo2} relies on the
$\mathcal{P}$lon$\mathcal{K}$ish arithmetization instead of R1CS, each addition
imposes additional gates to the resulting circuit. For that reason, if the
neural network consists of $n$ parameters, $L$ non-linearities and works over
the field $\mathbb{F}$ with the bit-size $b$, the size of the circuit in Halo2
is roughly $\mathcal{O}(n)$. In contrast, our approach reduces this complexity
down to $\mathcal{O}(2^{w} + \frac{b}{w}\cdot L)$, where $w$ is the limb size
used in the lookup argument, This is mentioned in more detail in
\Cref{section:efficiency-analysis-for-bionetta}. Assymptotically, this
reduces the circuit complexity down to $\mathcal{O}(N/\log N)$ for $N=Lb$.

\begin{example}
    Consider the simple fully-connected neural network, inputting $r=2000$
    neurons and outputting $r'=100$ neurons, governed by equation $f(\mathbf{x})
    = \textrm{ReLU}(W\mathbf{x})$ for some matrix $W \in \mathbb{R}^{r' \times
    r}$. While \textrm{Halo2} circuit would roughly consist of $rr' \approx
    200000$ gates, the corresponding Circom circuit over BN254 (so that $b=254$)
    with $w=8$ can be reduced down to $2^{9} + \frac{254}{8}r' \approx 3700$
    constraints, giving $\times 54$ boost over the existing approach.
\end{example}

In addition to smaller circuit sizes, Groth16 provides much better verification
key sizes. While the verification key for our tested models takes only 3KB, the
verification key of \texttt{ddkang/zkml} can reach 650KB and \texttt{ezkl} up to
4.2MB.

\textcolor{blue!50!black}{\textbf{deep-prove.}}
\texttt{deep-prove}\cite{deepprove} is based on a sum-check-based protocol
called \textrm{Ceno} \cite{ceno}. Although we do confirm that the proving time
of \texttt{deep-prove} surpasses one provided by \texttt{ezkl}, the proof sizes
are \textit{prohibitively} large: even for small-sized neural networks (see
subsequent sections for more details), the proof size can easily reach several
megabytes, making \texttt{deep-prove} impractical in on-chain applications.

\textcolor{blue!50!black}{\textbf{zkCNN.}} Notice that previous frameworks fall
into two categories: GKR-based approaches and Halo2-based approaches. zkCNN,
similarly to deep-prove, falls into the category of GKR-based approaches.
Consequently, it requires large proof sizes and costly verifier with the
asymptotic complexity of $O(\sqrt{\eta})$ (where $\eta$ is the circuit size).

\begin{table}[H]
    \centering
    \begin{tabular}{c|cccc}
        \Xhline{3.5\arrayrulewidth}
        \makecell{\textbf{zkML}\\\textbf{Framework}} & \makecell{Provable on \\ mobile device$^{\text{\ding{61}}}$} & \makecell{Effective EVM \\ verification$^{\text{\ding{62}}}$} & \makecell{Supports generic \\ architectures$^{\text{\ding{63}}}$} & \makecell{Low numerical \\ error$^{\text{\ding{80}}}$} \\
        \hline\textbf{Bionetta} & \tick  &  \tick \tick   & \tick        & \tick  \\
        \textbf{keras2circom}   & \cross &  \tick \tick   & \cross       & \cross \\
        \textbf{ddkang/zkml}    & \tick  &  \tick         & \tick \tick  & \tick  \\
        \textbf{EZKL}           & \cross &  \cross \cross & \tick \tick  & \tick  \\
        \textbf{deep-prove}     & \tick  &  \cross \cross & \cross       & \cross \\
        \textbf{zkCNN}          & \tick  &  \cross \cross & \cross       & \tick  \\
        \Xhline{3.5\arrayrulewidth}
    \end{tabular}
    \caption{Comparison of capabilities of different zkML frameworks (in the setting of the client-side proving)}
    \label{tab:comparison}
\end{table}
\vspace{-30px}
\par\noindent\rule{0.6\textwidth}{0.4pt}\\
\footnotesize$^{\text{\ding{61}}}$ $28 \times 28$ MobileNetV2 can be proven in
less than 30 seconds with up to 3GB RAM consumption\\
$^{\text{\ding{62}}}$ Verification procedure is doable on EVM blockchains:
reasonable verification and proof sizes (under 10KB), as well as verifying
function complexity (no need for recursive wrapping of the proving procedure) \\
$^{\text{\ding{63}}}$ Copes well with the medium-sized models not specifically designed for zero-knowledge circuits: for example, general \texttt{.onnx} support\\
$^{\text{\ding{80}}}$ The output of the circuit does not significantly differ from the real output. For example, running the circuits on top of \texttt{deep-prove} results in a significant error while \texttt{keras2circom} can handle only a small number of layers\normalsize\\

\textbf{Summary.} In \Cref{tab:comparison}, we provide the high-level overview
of differences between Bionetta and other prior zkML solutions. In
\Cref{table:zkml-complexity-comparison}, we specify concretely efficiency
metrics of Bionetta, and compare them with previously considered
state-of-the-art zkML solutions\footnote{\texttt{keras2circom} was not
considered due to the large size of circuits.}. Based on the results, Bionetta
demonstrates the best asymptotics of all the existing solutions. We first note
that GKR-based systems show similar results to Bionetta in terms of proving
efficiency. Although Halo2-based approaches have slightly worse prover
asymptotics than Bionetta, they show poor results in practice, with proving
taking more than 20 minutes for even relatively small models. Regarding
verification time and proof size, Bionetta exhibits constant asymptotic
behavior. In contrast, GKR approaches require $O(\sqrt{\eta})$, which is
resource-intensive for on-chain verification. Other approaches similarly exhibit
worse verification and proof size asymptotic behavior than Bionetta. Moreover,
the proof of Halo2 framework reached the size of more than 70 GB.

\newpage

\begin{table}[t!]
    \centering
    \caption{Complexity comparison of Bionetta with several state-of-the-art zkML frameworks. $E$ denotes 
    the pairing cost, $H$ denotes the hashing cost.} 
    \vspace{5px}
    \normalsize

    \resizebox{\textwidth}{!}{
    \renewcommand{\arraystretch}{1.15}
    \begin{tabular}{lllllll}
        \Xhline{2\arrayrulewidth}
        \textbf{zkML} & \textbf{zkSNARK} & \textbf{Setup} & & \textbf{Prover} & \textbf{Verifier} & \textbf{Proof Size} \\
        \hline
        \multirow{12}{*}{Bionetta}
        & \multirow{3}{*}{\makecell[l]{Groth16~\cite{groth16}}}
        & & $\mathbb{G}_1$ & $O(\ell\lambda)$ & \cellcolor{green!25} & \cellcolor{green!25}$2$ \\
        & & \textcolor{purple!60!white}{\textit{C}} & $\mathbb{G}_2$ & $O(\ell\lambda)$ & \multirow{-2}{*}{\cellcolor{green!25}$3 \, E + O(|\mathbbm{x}|)$} & \cellcolor{green!25}$1$ \\
        & & & $\mathbb{F}$ & $O(\ell\lambda \log \ell\lambda)$ & --- & --- \\
        &&&&& \\
        & 
        & & $\mathbb{G}_1$ & \cellcolor{green!25}$O(\ell\lambda/\log \ell\lambda)$ & \cellcolor{green!25} & \cellcolor{green!25}$3$ \\
        & \makecell[l]{UltraGroth\\\textit{(this paper)}} & \textcolor{purple!60!white}{\textit{C}} & $\mathbb{G}_2$ & \cellcolor{green!25}$O(\ell\lambda/\log \ell\lambda)$ & \multirow{-2}{*}{\cellcolor{green!25}$4 \, E + O(|\mathbbm{x}|)$} & \cellcolor{green!25}$1$ \\
        & & & $\mathbb{F}$ & \cellcolor{green!25}$O(\ell\lambda)$ & \cellcolor{green!25}$1 \, H$ & \cellcolor{green!25}$1$ \\
        &&&&& \\
        & 
        \makecell[l]{Spartan~\cite{spartan} \\ \& Virgo~\cite{virgo}} & \textcolor{green!60!black}{\textit{T}} & $\mathbb{F}$ & $O(\ell\lambda\log \ell\lambda)$ & $O(\log^2 \ell\lambda) \, H$ & $O(\log^2\ell\lambda)$ \\
        &&&&& \\
        & \multirow{2}{*}{\makecell[l]{Spartan~\cite{spartan} \\ \& vSQL~\cite{vsql}}} 
        & \multirow{2}{*}{\textcolor{blue!70!black}{\textit{U}}} & $\mathbb{G}$ & $O(\ell\lambda)$ & $O(\log^2 \ell\lambda)$ & $O(\log^2\ell\lambda)$ \\
        & &  & $\mathbb{F}$ & $O(\ell\lambda)$ & $O(\log^2 \ell\lambda)$ & $O(\log^2\ell\lambda)$ \\
        \hdashline
        
        \multirow{2}{*}{zkCNN~\cite{zk-cnn}} & \multirow{2}{*}{\makecell[l]{GKR~\cite{gkr} \\ \& Hyrax~\cite{hyrax}}} 
        & \multirow{2}{*}{\textcolor{green!60!black}{\textit{T}}} & $\mathbb{G}$ &  $O(n+\ell\rho)$ & \cellcolor{red!25}$O(\sqrt{n+\ell\rho})$ & \cellcolor{red!25}$O(\sqrt{n+\ell\rho})$ \\
        & & & $\mathbb{F}$ & $O(\eta+\ell\rho)$ & $O(\log(\eta+\ell\rho))$ & $O(\log(\eta+\ell\rho))$ \\
        \hdashline
        
        \multirow{2}{*}{zkGPT~\cite{zk-gpt}} & \makecell[l]{GKR~\cite{gkr} \\ \& Hyrax~\cite{hyrax}} 
        & \multirow{2}{*}{\textcolor{green!60!black}{\textit{T}}} & $\mathbb{G}$ & \cellcolor{green!25}$O(\eta)$ & \cellcolor{red!25}$O(\sqrt{\eta})$ & \cellcolor{red!25}$O(\sqrt{\eta})$ \\
        & \& Lasso~\cite{lasso} & & $\mathbb{F}$ & \cellcolor{green!25}$O(2^{\rho}+\eta)$ & $O(\log \eta)$ & $O(\log \eta)$ \\
        \hdashline
        
        EZKL~\cite{ezkl} & \multirow{2}{*}{\makecell[l]{Halo2~\cite{halo2} \\ \& IPA~\cite{ipa}}} & 
        \multirow{2}{*}{\textcolor{green!60!black}{\textit{T}}} & $\mathbb{G}$ &  $O(\eta^*)$ & $O(\log\eta^*)$ & $O(\log\eta^*)$ \\
        ZKML~\cite{ddkang-zkml} & & & $\mathbb{F}$ & $O(\eta^*\log\eta^*)$ & $O(\log\eta^*) \, H$ & $O(\log\eta^*)$ \\
        &&&&\multicolumn{3}{l}{where $\eta^* = \eta+2^{\rho}$}\\
        \Xhline{2\arrayrulewidth}
    \end{tabular}
    }

    \small
    \begin{gather*}
        \text{$\ell$: \# of non-linear calls (e.g., ReLU)} \quad \text{$\lambda$: finite field bit-size ($\log_2|\mathbb{F}|$)} \quad \text{$\rho$: precision parameter} \\
        \text{$n$: \# of model parameters} \quad \text{$\eta=\mathsf{poly}(n)$: total \# of addition and multiplication gates} \\
        \text{\textcolor{purple!60!white}{\textit{C}}: per-circuit setup} \quad \text{\textcolor{blue!70!black}{\textit{U}}: universal setup} \quad \text{\textcolor{green!60!black}{\textit{T}}: transparent setup}
    \end{gather*}
    %} 
    \label{table:zkml-complexity-comparison}
\end{table}

\section{Proving System}
In this section, we introduce one of the main ideas behind our approach. We
start by recapping what R1CS is in \Cref{section:r1cs}, then we proceed to
introducing the primary optimization called \textit{circuit-embedded weights} in
\Cref{section:circuit-embedded-weights}, then our quantization scheme in
\Cref{section:arithmetization}, and additionally we show how to implement the
highly-efficient custom neural network layers in
Appendix~\ref{section:r1cs-friendly-architectures}. Finally, the
\Cref{section:ultragroth} wraps up the discussion by significantly reducing the
non-linearity cost from $b$ to $b/w$ constraints.

\subsection{R1CS Arithmetization}\label{section:r1cs}

The R1CS arithmetization is a well-known technique for converting any
computational program encoding the NP statement into a set of quadratic checks
over the finite field $\mathbb{F}$, which are then converted to a set of
polynomial checks. More formally, any program can be encoded as the following
set of quadratic constraints:
\begin{equation*}
    \langle \boldsymbol{\ell}_i, \boldsymbol{z} \rangle \cdot \langle \boldsymbol{r}_i, \boldsymbol{z} \rangle = \langle \boldsymbol{o}_i, \boldsymbol{z} \rangle, \; \forall i \in [m],
\end{equation*} 
where $\boldsymbol{z} \in \mathbb{F}^n$ is the witness, roughly speaking
representing the intermediate calculations results; $\boldsymbol{\ell}_i,
\boldsymbol{r}_i, \boldsymbol{o}_i \in \mathbb{F}^n$ are the constant
\textit{left}, \textit{right} and \textit{output} vectors, respectively.
Finally, we denote the number of constraints by $m$. 

To shorten the notation, vectors $\boldsymbol{\ell}_i, \boldsymbol{r}_i,
\boldsymbol{o}_i$ can be represented by the constant matrices $L, R, O \in
\mathbb{F}^{m \times n}$, respectively. This way, the whole R1CS instance can be
simply written as $L\boldsymbol{z} \odot R\boldsymbol{z} = O\boldsymbol{z}$ and
the goal of languages such as Circom \cite{circom} is to generate the matrices
$L, R, O$ for a given program. 

\begin{example}
    Suppose the program consists in computing the expression $y=x_1^3 + x_2^2$. 

    % Define circle styles and colors
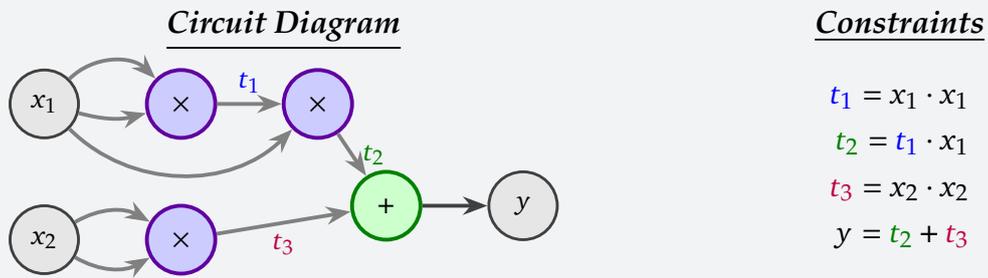
\colorlet{circle edge}{gray!50!black}
\colorlet{circle area}{gray!20}
\colorlet{gate1 edge}{green!50!black}
\colorlet{gate1 area}{green!20}
\colorlet{gate2 edge}{blue!50!purple}
\colorlet{gate2 area}{blue!20!white}
\colorlet{gate3 edge}{blue!50!black}
\colorlet{gate3 area}{blue!20}

\tikzset{
    var/.style={circle, draw=circle edge, fill=circle area, very thick, minimum size=1cm, text centered},
    gate1/.style={circle, draw=gate1 edge, fill=gate1 area, ultra thick, minimum size=1cm, text centered},
    gate2/.style={circle, draw=gate2 edge, fill=gate2 area, ultra thick, minimum size=1cm, text centered},
    gate3/.style={circle, draw=gate3 edge, fill=gate3 area, ultra thick, minimum size=1cm, text centered},
    arrow/.style={-Stealth, ultra thick}
}

\vspace{10px}
\begin{minipage}[t]{0.65\textwidth}
\begin{center}
    \underline{\textbf{Circuit Diagram}}

    \scalebox{0.9}{
        \begin{tikzpicture}
            % Nodes
            \node[var] (x1) at (0, 1) {$x_1$};
            \node[gate2] (x1_x1) at (2, 1) {$\times$};
            \node[gate2] (x1_x1_x1) at (4, 1) {$\times$};

            \node[var] (x2) at (0, -1) {$x_2$};
            \node[gate2] (x2_x2) at (2, -1) {$\times$};

            \node[gate1] (plus) at (5.0, -0.5) {$+$};

            % x1**3
            \draw[arrow,gray] (x1) to [bend left=45] (x1_x1);
            \draw[arrow,gray] (x1) to [bend right=15] (x1_x1);
            \draw[arrow,gray] (x1_x1) -- node[midway, above] {\textcolor{blue}{$t_1$}} (x1_x1_x1);
            \draw[arrow,gray] (x1) to [bend right=45] (x1_x1_x1);

            % x2**2
            \draw[arrow,gray] (x2) to [bend left=30] (x2_x2);
            \draw[arrow,gray] (x2) to [bend right=30] (x2_x2);

            % Summation
            \draw[arrow,gray] (x1_x1_x1) -- node[midway, right] {\textcolor{green!50!black}{$t_2$}} (plus);
            \draw[arrow,gray] (x2_x2) -- node[midway, below] {\textcolor{purple}{$t_3$}} (plus);

            % Result
            \node[var] (q) at (7.0, -0.5) {$y$};
            \draw[arrow,gray!50!black] (plus) -- (q);
        \end{tikzpicture}
    }
    \end{center}
\end{minipage}%
\hfill
\begin{minipage}[t]{0.3\textwidth}
    \begin{center}
        \underline{\textbf{Constraints}}
        \begin{align*}
            \textcolor{blue}{t_1} = x_1 \cdot x_1 \\
            \textcolor{green!50!black}{t_2} = \textcolor{blue}{t_1} \cdot x_1 \\
            \textcolor{purple}{t_3} = x_2 \cdot x_2 \\
            y = \textcolor{green!50!black}{t_2} + \textcolor{purple}{t_3}
        \end{align*}
    \end{center}
\end{minipage}

\vspace{10px}

In this case, the witness looks as $\boldsymbol{z} =
(1,x_1,x_2,t_1,t_2,t_3,y)$\footnote{To account for a constant term in constraints, we
put $1$ as the first element of the witness $\boldsymbol{z}$.}. The corresponding
matrices are:
\begin{equation*}
    L = \begin{bmatrix}
        0 & 1 & 0 & 0 & 0 & 0 & 0\\
        0 & 0 & 0 & 1 & 0 & 0 & 0 \\
        0 & 0 & 1 & 0 & 0 & 0 & 0 \\
        0 & 0 & 0 & 0 & 1 & 1 & 0
    \end{bmatrix}, \; R = \begin{bmatrix}
        0 & 1 & 0 & 0 & 0 & 0 & 0\\
        0 & 1 & 0 & 0 & 0 & 0 & 0 \\
        0 & 0 & 1 & 0 & 0 & 0 & 0 \\
        1 & 0 & 0 & 0 & 0 & 0 & 0
    \end{bmatrix}, \; O = \begin{bmatrix}
        0 & 0 & 0 & 1 & 0 & 0 & 0\\
        0 & 0 & 0 & 0 & 1 & 0 & 0 \\
        0 & 0 & 0 & 0 & 0 & 1 & 0 \\
        0 & 0 & 0 & 0 & 0 & 0 & 1
    \end{bmatrix}
\end{equation*}
\end{example}

Now, the main reason R1CS is so attractive for neural network inference is that
the R1CS arithmetization is very friendly to the \textit{linear} operations. In
particular, consider the following proposition.

\begin{proposition}
    The main advantage of the R1CS arithmetization is 
that any addition or multiplication by constants can be done completely for free! Indeed, consider 
the previous example. Since it contains an addition gate, we can simplify R1CS down to three constraints:
\begin{align*}
    \textcolor{blue}{t_1} = x_1 \cdot x_1 \\
    \textcolor{green!50!black}{t_2} = \textcolor{blue}{t_1} \cdot x_1 && \textcolor{gray}{\text{// These two constraints remain the same}} \\
    y - \textcolor{green!50!black}{t_2} = x_2^2, && \textcolor{gray}{\text{// Combined last two constraints}}
\end{align*}
\end{proposition}

\subsubsection{Proving System Choice}

Having the R1CS instance, we can further use a variety of zero-knowledge
protocols, including Spartan \cite{spartan}, Hyrax \cite{hyrax}, Aurora
\cite{aurora}, Fractal \cite{fractal}, or WHIR \cite{whir} to generate the proof
and the verifier instance (such as the smart-contract). We formally encapsulate
this idea in the following definition.
\begin{definition}
    \textbf{Proving System} is a tuple of the following three algorithms:
    \begin{itemize}
        \item $\mathrm{Setup}(1^{\lambda},\mathcal{R}) \to (\mathsf{pp},
        \mathsf{vp})$: the \textit{setup} algorithm takes as input the security
        parameter $\lambda \in \mathbb{N}$ and the relation $\mathcal{R}$ and
        outputs the public \textit{prover parameters} $\mathsf{pp}$ and the
        \textit{verifier parameters} $\mathsf{vp}$.
        \item $\mathrm{Prove}(\mathsf{pp}, \mathbbm{x}, \mathbbm{w}) \to \pi$:
        the \textit{proving} algorithm takes as input the prover parameters
        $\mathsf{pp}$, the public input $\mathbbm{x}$ and the witness
        $\mathbbm{w}$ and outputs the \textit{proof} $\pi$ that proves the
        knowledge of the witness $\mathbbm{w}$ for the statement $\mathbbm{x}$
        such that $(\mathbbm{x}, \mathbbm{w}) \in \mathcal{R}$.
        \item $\mathrm{Verify}(\mathsf{vp}, \mathbbm{x}, \pi) \to \{0, 1\}$: the
        \textit{verification} algorithm takes as input the verifier parameters
        $\mathsf{vp}$, the public input $\mathbbm{x}$ and the proof $\pi$ and
        outputs $1$ if the proof $\pi$ is valid for the statement $\mathbbm{x}$
        and $0$ otherwise.
    \end{itemize}
\end{definition}

\textbf{Remark.} Further, we assume that all proving systems in use are
\textit{complete} (verification of the valid proof always succeeds),
\textit{sound} (verification of an invalid proof fails with overwhelming
probability), \textit{zero-knowledge} (the proof does not reveal any information
about the witness), non-interactive, and, finally, \textit{succinct} (the proof
size $|\pi|$ is (poly)logarithmetically small in the size of the circuit
$\mathtt{C}$, encoding the relation $\mathcal{R}$).

In our particular case, we base our proving system on \textbf{Groth16} since it
is the most widely adopted proving system for zk-SNARKs with the best
infrastructure support. Besides, it has the smallest proof and verification key
size, so it is a perfect fit for our use case. However, \textit{note that we are
not limited to Groth16 and can easily switch to any other proving system}. This
is illustrated in the \Cref{fig:architecture}. Specifically, we employed
\textit{UltraGroth}, which enables lookup tables in Groth16 with the tiny
verification overhead (see \Cref{section:ultragroth}). This significantly
reduced the overhead when proving the execution of non-linear layers.

% Drawing the diagram of architecture in tikz
\tikzstyle{startstop} = [rectangle, rounded corners, minimum width=2cm, minimum height=1.2cm,text centered, draw=black, fill=gray!10, align=center, ultra thick]
\tikzstyle{process} = [rectangle, minimum width=3cm, minimum height=1.2cm, text centered, draw=green!40!black, fill=green!20, align=center, ultra thick]
\tikzstyle{output} = [rectangle, rounded corners, minimum width=3.2cm, minimum height=1.2cm, text centered, draw=blue!40!black, fill=blue!20!white, align=center, ultra thick]
\tikzstyle{dashedbox} = [draw=black, dashed, thick, inner sep=0.25cm, rectangle, rounded corners, align=center, ultra thick]
\tikzstyle{arrow} = [ultra thick,->,>=stealth]
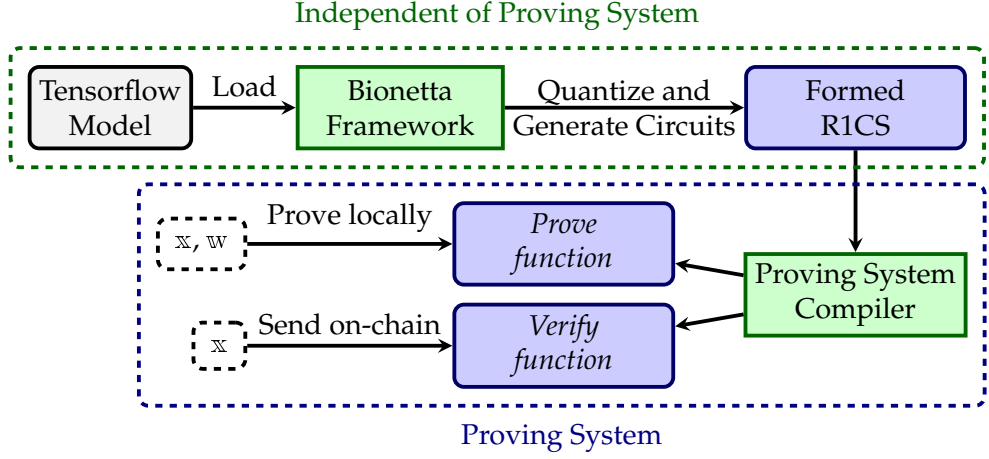
\begin{figure}[H]
    \centering
    \scalebox{0.9}{
    \begin{tikzpicture}[node distance=1.6cm and 1cm]
        % Nodes
        \node (model) [startstop] {Tensorflow\\Model};
        \node (framework) [process, right=1.5cm of model] {Bionetta\\ Framework};
        \node (r1cs) [output, right=3.5 cm of framework] {Formed\\R1CS};
        \node (proving-system) [process, below=1.5cm of r1cs] {Proving System \\ Compiler};
        \node (prove) [output, left=of proving-system, yshift=0.75cm] {\textit{Prove} \\ \textit{function}};
        \node (verify) [output, left=of proving-system, yshift=-0.75cm] {\textit{Verify} \\ \textit{function}};
        \node (prove-inputs) [dashedbox, left=3.0cm of prove] {$\mathbbm{x}, \mathbbm{w}$};
        \node (verify-inputs) [dashedbox, left=3.0cm of verify] {$\mathbbm{x}$};
        % \node (quantized) [output, right=of quant] {Quantized\\Model};
        % \node (zkp) [output, right=of circuit] {ZKP Engine\\(Expander)};
        
        % Arrows
        \draw [arrow] (model) -- node[midway, above, align=center] {Load} (framework);
        \draw [arrow] (framework) -- node[midway, align=center] {Quantize and \\ Generate Circuits} (r1cs);
        \draw [arrow] (r1cs) -- (proving-system);

        \draw [arrow] (proving-system) -- (prove);
        \draw [arrow] (proving-system) -- (verify);

        \draw [arrow] (prove-inputs) -- node[midway, above, align=center] {Prove locally} (prove);
        \draw [arrow] (verify-inputs) -- node[midway, above, align=center] {Send on-chain} (verify);
        %\draw [arrow] (quant) -- (circuit);
        %\draw [arrow] (circuit) -- (zkp);
        
        % Dashed box
        \node[fit={(model) (framework) (r1cs)}, dashedbox, draw=green!40!black, name=groupboxproving] {};
        \node[above=2.75pt of groupboxproving] {\textcolor{green!40!black}{Independent of Proving System}};
        \node[fit={(proving-system) (prove) (verify) (prove-inputs) (verify-inputs)}, dashedbox, draw=blue!50!black, name=groupboxverify] {};
        \node[below=2.75pt of groupboxverify] {\textcolor{blue!50!black}{Proving System}};
    \end{tikzpicture}}
    \caption{Architecture of the Bionetta framework}
    \label{fig:architecture}
\end{figure}

\subsection{Circuit-Embedded Weights}\label{section:circuit-embedded-weights}

The previous idea of free linear operations is the basis of our core
optimization --- \textit{circuit-embedded weights}. To demonstrate its essence,
consider the simple example of a \textbf{linear regression model}. Such model is
governed by the equation:
\begin{equation*}
    f(\textcolor{green!50!black}{\mathbf{x}}; \textcolor{blue}{\boldsymbol{\theta}}) := \langle \textcolor{blue}{\boldsymbol{\theta}}, \textcolor{green!50!black}{\mathbf{x}} \rangle + \textcolor{blue}{\theta_0}, \quad \textcolor{green!50!black}{\mathbf{x}} \in \mathbb{R}^n, \textcolor{blue}{\boldsymbol{\theta}} \in \mathbb{R}^{n+1}
\end{equation*}

Currently, the majority of zkML frameworks frames the relation\footnote{To
further encode the relation into the zero-knowledge circuit, it must operate
over the elements from the finite structure (such as finite field $\mathbb{F}$).
For simplicity, our current discussion regards all signals as the real field
$\mathbb{R}$ elements, but \Cref{section:arithmetization} will specify how to
operate with them over the prime field $\mathbb{F}_p$.} to be proven as:
\begin{equation*}
    \mathcal{R}^{(f)}_{\text{ZKML}} = \left\{ \begin{matrix}
        \mathbbm{x} = (y,\textcolor{blue}{\boldsymbol{\theta}}) \in \mathbb{R} \times \mathbb{R}^{n+1} \\
        \mathbbm{w} = (\textcolor{green!50!black}{\mathbf{x}}) \in \mathbb{R}^n
    \end{matrix} \;\middle|\; y \approx f(\textcolor{green!50!black}{\mathbf{x}};\textcolor{blue}{\boldsymbol{\theta}}) \right\}
\end{equation*}

Such relation is quite natural: in case we re-train the model (thus, this
changes the weights $\textcolor{blue}{\boldsymbol{\theta}}$), we simply pass the
new weights to the prover. However, this approach has a significant drawback:
since in such case weights are signals, the prover needs to impose constraints
on each term $\textcolor{blue}{\theta_i} \cdot \textcolor{green!50!black}{x_i}$
for each $i \in [n]$, even though $\textcolor{blue}{\boldsymbol{\theta}}$ is the
part of the public protocol. This means that the prover computes the scalar
product $\langle \textcolor{blue}{\boldsymbol{\theta}},
\textcolor{green!50!black}{\mathbf{x}} \rangle$ in $\mathcal{O}(n)$ constraints.
While such complexity is acceptable for small-sized models (especially for such
simple cases as the linear regression), when the model becomes deeper, the
number of constraints significantly increases with each such linear operation.

In contrast, we propose to treat the weights
$\textcolor{blue}{\boldsymbol{\theta}}$ as constants and not as signals.
Formally, instead of building the relation $\mathcal{R}^{(f)}_{\text{ZKML}}$
over the public signals $\textcolor{blue}{\boldsymbol{\theta}}$, we
\textit{parameterize} the relation $\mathcal{R}^{(f)}_{\text{Bionetta}}$ over
the \textit{constant} weights $\textcolor{blue}{\boldsymbol{\theta}}$:
\begin{equation*}
    \mathcal{R}^{(f)}_{\text{Bionetta}}(\textcolor{blue}{\boldsymbol{\theta}}) = \left\{ \begin{matrix}
        \mathbbm{x} = (y) \in \mathbb{R} \\
        \mathbbm{w} = (\textcolor{green!50!black}{\mathbf{x}}) \in \mathbb{R}^n
    \end{matrix} \;\middle|\; y \approx f(\textcolor{green!50!black}{\mathbf{x}};\textcolor{blue}{\boldsymbol{\theta}}) \right\}
\end{equation*}

This way, we first train the neural network, then we extract the weights
$\textcolor{blue}{\boldsymbol{\theta}}$, and finally setup the proving system via
$\mathrm{Setup}(1^{\lambda},\mathcal{R}^{(f)}_{\text{Bionetta}}(\textcolor{blue}{\boldsymbol{\theta}}))$. 
In engineering terms, this means that we hardcode the weights
$\textcolor{blue}{\boldsymbol{\theta}}$ right into the code. This is illustrated 
below.

\begin{minipage}[t]{0.48\textwidth}
%\begin{algorithm}
\rule{\linewidth}{1.25pt}
\footnotesize\textbf{Algorithm 1(a)}. Pseudocode of the \textit{classical ZKML method} of implementing the linear regression model.\normalsize

\rule{\linewidth}{1.25pt}\vspace{-5px}
\begin{lstlisting}[basicstyle=\scriptsize\ttfamily]
circuit LinearRegression(n: int):
    |\textcolor{red!50!black}{$\gg$}| public signal input |$\boldsymbol{\theta}$|[n+1];
    private signal input x[n];
    public signal output y;
    y <== 0;

    # Scalar product |$\langle \boldsymbol{\theta}, \boldsymbol{x} \rangle$|
    for i in 1..=n:
        y += |$\boldsymbol{\theta}$|[i] * x[i];

    # Adding the bias |$\theta_0$|
    y <== y + |$\boldsymbol{\theta}$|[0];
\end{lstlisting}
%\end{algorithm}
\end{minipage}
\hfill \hspace{10px}
\begin{minipage}[t]{0.48\textwidth}
%\begin{algorithm}
\rule{\linewidth}{1.25pt}
{\footnotesize\textbf{Algorithm 1(b)}. Pseudocode of the \textit{Bionetta method} of implementing the linear regression model.}
\rule{\linewidth}{1.25pt}\vspace{-5px}
\begin{lstlisting}[basicstyle=\scriptsize\ttfamily]
# Hardcode the weights
|\textcolor{green!50!black}{$\gg$}| const |$\boldsymbol{\theta}$|[n+1] = [0x642, 0x123, |$\dots$|];

circuit LinearRegression(n: int):
    private signal input x[n];
    public signal output y;
    y <== 0;

    # Scalar product |$\langle \boldsymbol{\theta}, \boldsymbol{x} \rangle$|: now costs 0
    for i in 1..=n:
        y += |$\boldsymbol{\theta}$|[i] * x[i];

    # Adding the bias |$\theta_0$|
    y <== y + |$\boldsymbol{\theta}$|[0];
\end{lstlisting}
%\end{algorithm}
\end{minipage}

\begin{proposition}
    Any matrix-matrix or matrix-vector multiplications in Bionetta where the
    matrix is fixed \textbf{can be implemented in $0$ constraints} and the only
    computational overhead is computing such products over the finite field for
    a proper witness generation. Consequently, the vast majority of classical
    machine learning algorithms such as \textit{PCA/LDA},
    \textit{Linear/Logistic Regression}, \textit{linear SVM} can be implemented
    in $0$ constraints.
\end{proposition}

\subsubsection{Linear Regression In-Depth Analysis}

In this section, we specifically show the difference in both the witness size and 
the size of R1CS matrices for both approaches. The reader is encouraged to skip this
section if he/she is not interested in the low-level details.

With that said, let us write down the $L$, $R$, and $O$ matrices for the linear
regression model. In case of the classical approach, we need to introduce the
witness of total size $3n+3$. Namely, our witness $\boldsymbol{z}$ consists of
the following elements:
\begin{equation*}
    \boldsymbol{z} = (1,
    \textcolor{blue}{\underbrace{\theta_0,\theta_1,\dots,\theta_n}_{\text{Weights $\boldsymbol{\theta}$}}},
    \textcolor{green!50!black}{\underbrace{x_1,\dots,x_n}_{\text{Inputs $\boldsymbol{x}$}}},
    \textcolor{purple}{\underbrace{t_1,\dots,t_n}_{\substack{
        \text{Intermediate} \\
        \text{variables } \boldsymbol{t} = \boldsymbol{x} \odot \boldsymbol{\theta}\texttt{[1:]}
      }}},
    y
    ) \in \mathbb{F}^{3n+3}
\end{equation*}

Now, we write down $n+1$ constraints: first $n$ constraints will correspond to
the multiplication checks $\textcolor{purple}{t_i} =
\textcolor{green!50!black}{x_i} \cdot \textcolor{blue}{\theta_i}$, while the
last one will assert $\textcolor{blue}{\theta_0}+\sum_{i=1}^n
\textcolor{purple}{t_i} = y$. One can verify that the corresponding checks can
be encoded as follows\footnote{Note that we can reduce this down to $n$
constraints by combining the last two constraints into one. However, this is not 
crucial for our discussion.}:
\begin{align*}
    L &= \begin{bmatrix}
        \mathbf{0}_n & \textcolor{blue}{\mathbf{0}_n} & \textcolor{blue}{E_{n \times n}} & \textcolor{green!50!black}{\mathbf{0}_{n \times n}} & \textcolor{purple}{\mathbf{0}_{n \times n}} & \mathbf{0}_{n} \\
        1 & \textcolor{blue}{0} & \textcolor{blue}{\mathbf{0}_{n}^{\top}} & \textcolor{green!50!black}{\mathbf{0}_{n}^{\top}} & \textcolor{purple}{\mathbf{0}_{n}^{\top}} & 0 \\
    \end{bmatrix} \in \mathbb{F}^{(n+1) \times (3n+3)}, && \textcolor{gray!60!black}{\texttt{\scriptsize // Encodes left inputs: $L\boldsymbol{z} = \begin{bmatrix}
        \textcolor{blue}{\theta_1} \\ \vdots \\ \textcolor{blue}{\theta_n} \\ 1
    \end{bmatrix}$}} \\
    R &= \begin{bmatrix}
        \mathbf{0}_n & \textcolor{blue}{\mathbf{0}_n} & \textcolor{blue}{\mathbf{0}_{n \times n}} & \textcolor{green!50!black}{E_{n \times n}} & \textcolor{purple}{\mathbf{0}_{n \times n}} & \mathbf{0}_{n} \\
        0 & \textcolor{blue}{1} & \textcolor{blue}{\mathbf{0}_{n}^{\top}} & \textcolor{green!50!black}{\mathbf{0}_{n}^{\top}} & \textcolor{purple}{\mathbf{1}_{n}^{\top}} & 0 \\
    \end{bmatrix} \in \mathbb{F}^{(n+1) \times (3n+3)}, && \textcolor{gray!60!black}{\texttt{\scriptsize // Encodes right inputs: $R\boldsymbol{z} = \begin{bmatrix}
        \textcolor{green!50!black}{x_1} \\ \vdots \\ \textcolor{green!50!black}{x_n} \\ \textcolor{blue}{\theta_0} + \sum_{i=1}^n \textcolor{purple}{t_i}
    \end{bmatrix}$}} \\
    O &= \begin{bmatrix}
        \mathbf{0}_n & \textcolor{blue}{\mathbf{0}_n} & \textcolor{blue}{\mathbf{0}_{n \times n}} & \textcolor{green!50!black}{\mathbf{0}_{n \times n}} & \textcolor{purple}{E_{n \times n}} & \mathbf{0}_{n} \\
        0 & \textcolor{blue}{0} & \textcolor{blue}{\mathbf{0}_{n}^{\top}} & \textcolor{green!50!black}{\mathbf{0}_{n}^{\top}} & \textcolor{purple}{\mathbf{0}_{n}^{\top}} & 1 \\
    \end{bmatrix} \in \mathbb{F}^{(n+1) \times (3n+3)}, && \textcolor{gray!60!black}{\texttt{\scriptsize // Encodes outputs: $O\boldsymbol{z} = \begin{bmatrix}
        \textcolor{purple}{t_1} \\ \vdots \\ \textcolor{purple}{t_n} \\ y
    \end{bmatrix}$}}
\end{align*}

However, if we do not pass the weights $\textcolor{blue}{\boldsymbol{\theta}}$ directly 
into the circuit, we can easily use just a single constraint! Note that our optimized witness 
$\boldsymbol{z}$ now consists of $n+2$ elements:
\begin{equation*}
    \boldsymbol{z} = (1,
    \textcolor{green!50!black}{\underbrace{x_1,\dots,x_n}_{\text{Inputs $\boldsymbol{x}$}}},
    y
    ) \in \mathbb{F}^{n+2}
\end{equation*}

So that the corresponding matices $L(\textcolor{blue}{\boldsymbol{\theta}})$, $R(\textcolor{blue}{\boldsymbol{\theta}})$, 
and $O(\textcolor{blue}{\boldsymbol{\theta}})$ (notice that now they depend on the weights chosen!) are 
$L(\textcolor{blue}{\boldsymbol{\theta}}) \equiv [1,\mathbf{0}_{n+1}^{\top}]$, $R(\textcolor{blue}{\boldsymbol{\theta}})=
[0,\textcolor{blue}{\boldsymbol{\theta}}^{\top},0]$, and $O(\textcolor{blue}{\boldsymbol{\theta}}) \equiv [\mathbf{0}_{n+1}^{\top},1]$.

\textbf{Conclusion.} Using the circuit-embedded weights, we can reduce the
witness size from $3n+3$ to $n+2$ and the number of constraints from $n+1$ to
$1$. Similar optimizations can be applied to any other linear operation. For
example, the PCA algorithm involves running inference over
$f(\textcolor{green!50!black}{\mathbf{x}}; \textcolor{blue}{W},
\textcolor{blue}{\boldsymbol{\beta}}) =
\textcolor{blue}{W}\textcolor{green!50!black}{\mathbf{x}} +
\textcolor{blue}{\boldsymbol{\beta}}$ which is also constraint-free in
\textit{Bionetta}.

\section{Float Quantization}\label{section:arithmetization}

The next section is dedicated to how exactly we encode the real numbers
$\mathbb{R}$ arithmetic into the finite field $\mathbb{F}$ arithmetic. The short
answer is the following: to encode $x \in \mathbb{R}$, we multiply it by a large
power of two, round it, and then convert to the finite field element
$\mathbb{F}_p$. However, the devil is in the details: in such scenario, it is
unclear how to handle overflows for deep enough neural networks --- the issue
that \texttt{keras2circom} \cite{keras2circom}, for instance, has not resolved.

\subsection{Quantization Scheme Definition}

Here and hereafter suppose that the finite field $\mathbb{F}$ is prime, that is
$\mathbb{F} = \mathbb{F}_p$ for a $b$-bit prime $p$. The first question that
arises is \textit{how to handle a sign} of some $x \in \mathbb{F}_p$: after all,
any element of $\mathbb{F}_p$ must be in range $[0,p)$ and thus cannot be
naturally negative. The common approach is to treat the sign of $x \in
\mathbb{F}_p$ as follows:
\begin{equation*}
    \mathsf{sign}(x) = \begin{cases}
        1, & \text{if } 0 \leq x < \frac{p-1}{2} \\
        -1, & \text{if } \frac{p-1}{2} \leq x < p
    \end{cases}
\end{equation*}

We propose to simplify such definition a bit: instead of setting the
``threshold'' for negativity as $\frac{p-1}{2}$, we set it to $2^{b-1}$ where
$b$ is the bitsize of the prime $p$. This admittedly narrows down the range of
possible positive values, yet practically this does not introduce any
significant overhead. The primary reason for such choice is (a) convenience, (b)
better performance in certain cases. 

With such sign handling, it is natural to define the following conversion from the integers $\mathbb{Z}$ to 
the finite field $\mathbb{F}_p$ (which we call $\mathbf{F}_p$) and back (we call it $\mathbf{Z}$):
\begin{equation*}
    \mathbf{F}_p(x) = \begin{cases}
        x, & \text{if } x > 0 \\
        p + x, & \text{if } x < 0 \\
    \end{cases}, \quad \mathbf{Z}(x) = \begin{cases}
        x, & \text{if } 0 \leq x < 2^{b-1} \\
        x - p, & \text{if } 2^{b-1} \leq x < p
    \end{cases}
\end{equation*}

Now, we are ready to define the quantization process.

\begin{definition}
    The \textbf{Bionetta Quantization Scheme} consists of two algorithms:
    quantize $Q_{\rho}: \mathbb{R} \to \mathbb{F}_p$ and dequantize $D_{\rho}:
    \mathbb{F}_p \to \mathbb{R}$. They are defined as follows:
    \begin{equation*}
        Q_{\rho}(x) = \mathbf{F}_p([2^{\rho}x]), \quad D_{\rho}(\widehat{x}) = 2^{-\rho}\mathbf{Z}(\widehat{x})
    \end{equation*}

    We call $\rho \in \mathbb{N}$ the \textbf{precision parameter}. In practice,
    the larger $\rho$ is, the more precise the quantization is, yet the trickier
    it is to handle the overflow: see \Cref{section:overflow}.
\end{definition}

\begin{example}
    Suppose we want to multiply $x=0.2$ and $y=-0.3$ to get $z=xy$ using the Bionetta
    quantization scheme with a precision $\rho=5$ and prime field
    $\mathbb{F}_{9967}$. The quantized values are:
    \begin{align*}
        \widehat{x} &:= Q_5(0.2) = \mathbf{F}_p([2^5 \times 0.2]) = \mathbf{F}_p([6.4]) = \mathbf{F}_p(6) = 6 \\
        \widehat{y} &:= Q_5(-0.3) = \mathbf{F}_p([2^5 \times -0.3]) = \mathbf{F}_p([-9.6]) = \mathbf{F}_p(-10) = 9957
    \end{align*}

    Then, we multiply quantized values over $\mathbb{F}_{9967}$ and get
    $\widehat{z}=9907$. Finally, we dequantize the result to get
    \begin{equation*}
        z = D_{10}(9907) = \mathbf{Z}(9907)/2^{10} = -60/1024 \approx -0.059
    \end{equation*}

    Note that this differs from the real value $z=-0.06$ by a relatively
    negligible amount.
\end{example}

This example shows an essential observation: when multiplying two quantized
values with precisions $\rho$ and $\rho'$, the dequantization precision must be
set to $\rho+\rho'$ (roughly speaking, since $[2^{\rho}x][2^{\rho'}y] \approx
[2^{\rho+\rho'}xy]$). To prove that such dequantization of product indeed leads
to the negligible error, we provide the following theorem.

\begin{theorem}
    Let $\widehat{x} := Q_{\rho}(x)$ and $\widehat{y} := Q_{\rho}(y)$ be the
    quantized values of $x$ and $y$, respectively. Then, if we define circuit
    $\mathtt{C}_f(\widehat{x},\widehat{y}) = \widehat{x}\cdot \widehat{y}$ for
    the multiplication $f(x,y)=x \cdot y$ and assume $\beta := 2\max\{|x|,|y|\}$
    to be relatively small, the following relation for an error
    $\varepsilon_{\rho}$ holds:
    \begin{equation*}
        \varepsilon_{\rho} := \left| D_{2\rho}(\mathtt{C}_f(\widehat{x},\widehat{y})) - f(x,y) \right| \leq 2^{-\rho} \beta + 2^{-2\rho}
    \end{equation*}

    Consequently, the error $\varepsilon_{\rho}$ is negligible in $\rho$.
\end{theorem}

We leave the proof in Appendix \ref{section:quant-proof-appendix} for those
interested in specifics.

\subsection{Overflow Handling}\label{section:overflow}

The careful reader might notice the primary issue of the quantization scheme:
the overflow. Indeed, suppose we want to encode the function $f(x) = x^n$ in the
quantized form for fairly large $n \in \mathbb{N}$. The most straightforward way
to quantize this expression is to quantize $x$ first (to get $\widehat{x} :=
Q_{\rho}(x)$), then compute $\widehat{x}^n$ and finally dequantize the result with
precision $\rho' := n\rho$. While this might perfectly fine for small $n$'s
(such as $2$ or $3$), for larger $n$'s the result might easily overflow the
field $\mathbb{F}_p$. 

To handle this issue, we propose the following solution. Suppose we have
conducted $4$ multiplications in the quantization steps and got the quantized
value $\widehat{x}^4$ with precision $4\rho < b-1$. Then, we can divide the
result by $2^{3\rho}$ to get the value of $x^4$, but now with precision $\rho$.
This way, we can handle the overflow issue by dividing the intermediate values
by $2^{\ell\rho}$ (right shift by $\ell\rho$) for suitable $\ell \in \mathbb{N}$
which we call the \textbf{precision cut operation}.

\begin{example}
Suppose we have the BN254 prime field $\mathbb{F}_p$ (so that $b=254$) and we
have set the precision parameter $\rho := 80$. We want to compute the value of
$x^5$ for $x=0.2$. Note that $\widehat{x}^5$ (for $\widehat{x} := Q_{\rho}(x)$)
yields the value with the precision $5\rho=400$, which is too large for the
field $\mathbb{F}_p$. We propose to build the following circuit $\mathtt{C}_f:
\mathbb{F}_p \to \mathbb{F}_p$:
\begin{equation*}
    \mathtt{C}_f(\widehat{x}) = ((\widehat{x}^3 \gg (2\rho)) \times \widehat{x}^2) \gg (2\rho)
\end{equation*}
In this case, $D_{\rho}(\mathtt{C}_f(\widehat{x}))$ would yield the value of
$x^5$ with negligible error. Indeed, $\widehat{x}^3 \gg (2\rho)$ yields $x^3$ with
precision $\rho$, then multiplication by $\widehat{x}^2$ rises precision to $3\rho$
which we finally cut to $\rho$ by the right shift by $2\rho$.
\end{example}

\textbf{Note:} \textit{$\widetilde{\mathtt{C}}_f(\widehat{x})=(\widehat{x}^3 \gg
(2\rho)) \times \widehat{x}^2$ is also a valid circuit, but the dequantization
must be done with precision $3\rho$ so that
$D_{3\rho}(\widetilde{\mathtt{C}}_f(\widehat{x})) \approx f(x)$.}

\subsection{Activation Functions}

Now, with the quantization scheme in mind, we can proceed to explaining how to
implement activations in Bionetta. We already know that the multiplication by
matrix $W \in \mathbb{R}^{m \times n}$ and the addition of the bias vector
$\boldsymbol{\beta} \in \mathbb{R}^m$ are free in R1CS, so the linear layer
$\mathbf{y} = W\mathbf{x} + \boldsymbol{\beta}$ is the most efficient layer in
terms of constraints.

However, we cannot simply stack linear layers to form the architecture. Indeed,
the composition of linear layers is still a linear layer, so we need to
introduce the non-linear activation functions. Unfortunately, many common
activations such as \textbf{Sigmoid} (corresponding to $\sigma(x) =
\frac{1}{1+e^{-x}}$) or \textbf{Tanh} (corresponding to $\tanh(x) = \frac{1 -
e^{-2x}}{1 + e^{-2x}}$) contain exponentials and thus become problematic when it
comes to the R1CS. While there are approaches to handle such activations using
polynomial approximations \cite{activation-polynomial-approximation}, we find it
more convenient to simply use the activations that can be precisely implemented
in R1CS, specified in \Cref{table:activations}. Besides, the large number of 
modern neural network architectures use such activations, so 
we do not lose much in terms of expressiveness.

\begin{table}
\begin{tabular}{>{\raggedright\arraybackslash}m{2.5cm} >{\raggedright\arraybackslash}m{4cm} >{\centering\arraybackslash}m{6cm} >{\centering\arraybackslash}m{2.5cm}}
    \toprule
    \textbf{Name} & \textbf{Function} & \textbf{Sketch} & \textbf{\#Constraints} \\
    \midrule
    % ReLU
    \textbf{ReLU} & $\max\{0,x\}$ & \begin{tikzpicture}[scale=0.5]
        \begin{axis}[
            axis lines = middle,
            axis line style={very thick},
            xlabel = \Large$x$,
            ymin = -1, ymax = 5,
            xmin = -4, xmax = 4,
            grid = both,
            minor tick num=1,
            major grid style={line width=.2pt,draw=gray!50},
            minor grid style={line width=.1pt,draw=gray!20},
            samples=100,
            domain=-4:4,
            width=10cm,
            height=7cm,
        ]
        \addplot[blue, line width=3pt] {max(0,x)};
        \end{axis}
        \end{tikzpicture} & $\approx b$ \\
    
    % LeakyReLU
    \textbf{LeakyReLU} & $\max\{2^{-\ell}x,x\}$ & \begin{tikzpicture}[scale=0.5]
        \begin{axis}[
            axis lines = middle,
            axis line style={very thick},
            xlabel = \Large$x$,
            ymin = -1, ymax = 5,
            xmin = -4, xmax = 4,
            grid = both,
            minor tick num=1,
            major grid style={line width=.2pt,draw=gray!50},
            minor grid style={line width=.1pt,draw=gray!20},
            samples=100,
            domain=-4:4,
            width=10cm,
            height=7cm,
        ]
        \addplot[blue, line width=3pt] { (0.25*x > x) ? 0.25*x : x };
        \end{axis}
    \end{tikzpicture} & $\approx b$ \\
    \textbf{ReLU6} & $\min\{6,\max\{0,x\}\}$ & \begin{tikzpicture}[scale=0.5]
        \begin{axis}[
            axis lines = middle,
            axis line style={very thick},
            xlabel = \Large$x$,
            ymin = -2, ymax = 8,
            xmin = -2, xmax = 8,
            grid = both,
            minor tick num=1,
            major grid style={line width=.2pt,draw=gray!50},
            minor grid style={line width=.1pt,draw=gray!20},
            samples=100,
            domain=-4:10,
            width=10cm,
            height=7cm,
        ]
        \addplot[blue, line width=3pt] { (x >= 6) ? 6 : ((x < 0) ? 0 : x) };
        \end{axis}
    \end{tikzpicture} & $\approx 2b$ \\
    \textbf{Hard Sigmoid} & $\frac{1}{6}\text{ReLU6}(x+3)$ & \begin{tikzpicture}[scale=0.5]
        \begin{axis}[
            axis lines = middle,
            axis line style={very thick},
            xlabel = \Large$x$,
            ymin = -1, ymax = 1.5,
            xmin = -4, xmax = 4,
            grid = both,
            minor tick num=1,
            major grid style={line width=.2pt,draw=gray!50},
            minor grid style={line width=.1pt,draw=gray!20},
            samples=100,
            domain=-6:6,
            width=10cm,
            height=7cm,
        ]
        \addplot[blue, line width=3pt] { (x >= 3) ? 1 : ((x < -3) ? 0 : ((x+3)/6)) };
        \end{axis}
    \end{tikzpicture} & $\approx 2b$ \\
    \textbf{Hard Swish} & $\frac{1}{6}x\text{ReLU6}(x+3)$ & \begin{tikzpicture}[scale=0.5]
        \begin{axis}[
            axis lines = middle,
            axis line style={very thick},
            xlabel = \Large$x$,
            ymin = -1, ymax = 1.5,
            xmin = -4, xmax = 4,
            grid = both,
            minor tick num=1,
            major grid style={line width=.2pt,draw=gray!50},
            minor grid style={line width=.1pt,draw=gray!20},
            samples=100,
            domain=-6:6,
            width=10cm,
            height=7cm,
        ]
        \addplot[blue, line width=3pt] { ((x >= 3) ? 1 : ((x < -3) ? 0 : ((x+3)/6)))*x };
        \end{axis}
    \end{tikzpicture} & $\approx 2b$ \\
    \bottomrule
\end{tabular}
\caption{The list of activation functions supported by \textit{Bionetta} and
their R1CS cost. According to the \textit{MobileNetV3} paper \cite{mobilenetv3}, 
we also include effective implementations of sigmoid and swish functions which 
do not require calculating exponents in-circuit.}
\label{table:activations}
\end{table}

The question now is how many constraints do we need to implement, say, 
\textbf{ReLU} or \textbf{LeakyReLU} in R1CS. 

\begin{proposition}
    The $\mathsf{ReLU}$ and $\mathsf{LeakyReLU}$ activations can be implemented in
    R1CS with $b+1$ constraints.
\end{proposition}

\textbf{Reasoning.} Suppose we are implementing the $\mathsf{ReLU}(x) =
\max\{0,x\}$. To compare the field element $x$ with $0$, we need to check the
$(b-1)$th bit. This requires decomposing $x$ into the binary form: $x =
\sum_{i=0}^{b-1}x_i2^i$ where we need to constraint each $x_i$ using quadratic
check $x_i(1-x_i) = 0$. This requires $b$ constraints. Finally, the result is
simply $x_{b} \cdot x$, requiring one more quadratic constraint.

One more interesting feature with $\mathsf{ReLU}$-based activations is that we
can \textit{cut the precision for free} (recall that such operation consists in
computing $x \gg \rho\ell$).

\begin{proposition}
    Computing $\mathsf{ReLU}(\widehat{x}) \gg \rho\ell$ for some fixed $\ell \in \mathbb{N}$
    costs $b+1$ constraints.
\end{proposition}
Indeed, since we decompose $x$ into the binary form to check the sign when
computing $\mathsf{ReLU}$, we can simply drop the last $\ell\rho$ bits of the
decomposed $x$ and get the value of $\mathsf{ReLU}(x)$ with precision cut by
$\ell\rho$ for free.

Now, this observation is \textit{crucial}. Indeed, the careful reader could have
asked a reasonable question: why don't we use polynomial approximations of
ReLU/polynomial activation functions (as deeply researched in $\Pi$-nets: see
\cite{pi-nets})? The reason is that polynomial approximations cause the
significant raise in precision: although computing $Q(\widehat{\mathbf{x}})$ for
$d$-degree polynomial $Q \in \mathbb{F}[x]$ requires only $d$ constraints per
value, such computation increases the precision of $\widehat{\mathbf{x}}$ by
$\rho d$ bits. Consequently, even moderately deep networks would require cutting
precision of a large number of values, each costing roughly $b$ constraints. 
This way, the benefit of $d$ constraints per value is lost in the overhead of
precision cuts.

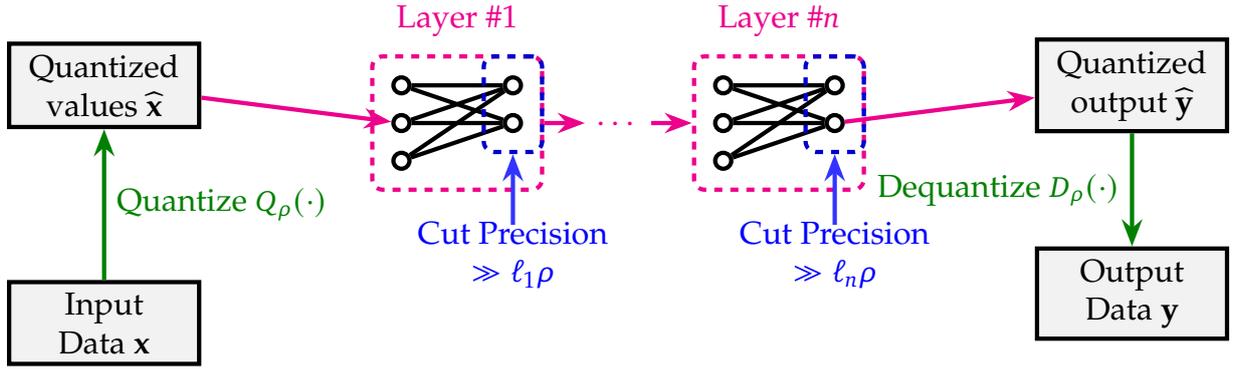
\begin{figure}[H]
\begin{tikzpicture}[
    neuron/.style={circle, draw, fill=white, minimum size=6pt, inner sep=0pt, ultra thick},
    box/.style={draw, fill=gray!10, minimum width=2.5cm, minimum height=1cm, align=center, ultra thick},
    arrow/.style={ultra thick, ->, >=Stealth},
    qarrow/.style={arrow, draw=green!50!black, ultra thick},
    narrow/.style={arrow, draw=magenta, ultra thick},
    cutarrow/.style={arrow, draw=blue!80, ultra thick},
    scale=0.8
]

% Input and quantization
\node[box] (input) {Input\\Data $\mathbf{x}$};
\node[box, above=2cm of input] (quant) {Quantized\\values $\widehat{\mathbf{x}}$};

% Layer 1
\foreach \i in {1,2,3} {
    \node[neuron, right=2.5cm of quant, yshift=0.5cm-\i*0.5cm] (in1\i) {};
}
\foreach \i in {1,2} {
    \node[neuron, right=1.2cm of in11, yshift=0.5cm-\i*0.5cm] (out1\i) {};
}
% Connections layer 1
\foreach \i in {1,2,3} {
    \foreach \j in {1,2} {
        \draw[ultra thick] (in1\i) -- (out1\j);
    }
}

\node[fit={(in11) (in12) (in13) (out11) (out12)}, dashedbox, draw=magenta, name=layer1] {};
\node[fit={(out11) (out12)}, dashedbox, draw=blue!80!black, name=layer1out] {};
\node[above=2.75pt of layer1] {\textcolor{magenta}{Layer \#1}};

% Cut precision after layer 1
\draw[cutarrow] ($(layer1out) + (0, -2.0cm)$) -- (layer1out) node[blue,align=center] at ($(layer1out) + (0, -2.5cm)$) {Cut Precision \\ $\gg \ell_1\rho$};

% Layer 2
\foreach \i in {1,2,3} {
    \node[neuron, right=2.5cm of out11, yshift=0.5cm-\i*0.5cm] (in2\i) {};
}
\foreach \i in {1,2} {
    \node[neuron, right=1.2cm of in21, yshift=0.5cm-\i*0.5cm] (out2\i) {};
}
% Connections layer 2
\foreach \i in {1,2,3} {
    \foreach \j in {1,2} {
        \draw[ultra thick] (in2\i) -- (out2\j);
    }
}
\node[fit={(in21) (in22) (in23) (out21) (out22)}, dashedbox, draw=magenta, name=layer2] {};
\node[fit={(out21) (out22)}, dashedbox, draw=blue!80!black, name=layer2out] {};
\node[above=2.75pt of layer2] {\textcolor{magenta}{Layer \#$n$}};
% Cut precision after layer 2
\draw[cutarrow] ($(layer2out) + (0, -2.0cm)$) -- (layer2out) node[blue, align=center] at ($(layer2out) + (0, -2.5cm)$) {Cut Precision \\ $\gg \ell_n\rho$};

% Draw the text node right between layer1 and layer2 in the middle (x coordinate)
\node[magenta, name=dots] at ($(layer1)!0.5!(layer2)$) {$\dots$};
\draw[narrow] (layer1) -- (dots);
\draw[narrow] (dots) -- (layer2);

% Output quantized and real
\node[box, right=2.5cm of out21] (quantout) {Quantized\\output $\widehat{\mathbf{y}}$};
\node[box, below=1.5cm of quantout] (output) {Output\\Data $\mathbf{y}$};

% Arrows for quantize and dequantize
\draw[qarrow] (input) -- node[green!50!black, midway, right, align=center] {Quantize $Q_{\rho}(\cdot)$} (quant);
\draw[narrow] (quant) -- (in12);
\draw[narrow] (out22) -- (quantout);
\draw[qarrow] (quantout) -- node[green!50!black, midway, left, align=center] {Dequantize $D_{\rho}(\cdot)$} (output);

\end{tikzpicture}
\caption{The precision cut operation in the Bionetta system. The dashed
\textcolor{magenta}{magenta} boxes represent the layers of the neural network.
The dashed \textcolor{blue!80!black}{blue} box represents the output of the
layer $i$, which is cut by $\ell_i\rho$ bits. Finally,
\textcolor{green!50!black}{green} arrows represent the pre- and post-processing
of the data done outside the circuit.}
\label{fig:cutting-precisions}
\end{figure}

This motivates us to use the $\mathsf{ReLU}$-based activations where we can 
not only compute the activation function with $b+1$ constraints, but also cut
the precision for free. This idea is illustrated in \Cref{fig:cutting-precisions}.

\section{UltraGroth: Lookup Tables in R1CS}\label{section:ultragroth}

\subsection{Quadratic Arithmetic Programs (QAP)}

Here we recap the construction of the original Groth16 proving system
\cite{groth16}. From our previous discussion, we know that the R1CS encodes the
program in the form $L\boldsymbol{z} \odot R\boldsymbol{z} = O\boldsymbol{z}$
for the given fixed matrices $L,R,O \in \mathbb{F}^{m \times n}$ with
$\boldsymbol{z} \in \mathbb{F}^n$. To make the proof succinct, we need to encode
this check in the form of a polynomial identity. To do this, we build $3n$
polynomials $\{\ell_i(X)\}_{i \in [n]},\{r_i(X)\}_{i \in [n]},\{o_i(X)\}_{i \in
[n]} \subseteq \mathbb{F}^{(\leq m)}[X]$ which interpolate the columns of the
corresponding matrices:
\begin{equation*}
    \ell_i(\omega^j) = L_{i,j}, \; r_i(\omega^j) = R_{i,j}, \; o_i(\omega^j) = O_{i,j}, \quad i \in [n], j \in [m]
\end{equation*}

Now, the R1CS check can be expressed as the following polynomial identity: the
polynomial $m(X) := \sum_{i\in [n]}z_i\ell_i(X) \cdot \sum_{i\in
[n]}z_ir_i(X) - \sum_{i\in [n]}z_io_i(X)$ has roots in the domain $\Omega
= \{\omega^j\}_{j \in [m]}$. Equivalently, the vanishing polynomial
$t(X) = \prod_{\alpha \in \Omega}(X-\alpha)$ must divide the polynomial
$m(X)$. With the suitable choice of $\Omega$ (namely, primitive roots of $2^t$
for suitable $t$), the expression can be simplified down to $t(X) = X^m
- 1$. 

All in all, we can define the \textbf{Quadratic Arithmetic Program} (QAP)
relation as follows:
\begin{equation*}
    \textcolor{blue!80!black}{\mathcal{R}_{\text{QAP}} = \left\{ 
        \begin{matrix} 
            \mathbbm{x} = \{z_i\}_{i \in \mathcal{I}_{X}} \in \mathbb{F}^{l} \\
            \mathbbm{w} = \{z_i\}_{i \in \mathcal{I}_{W}} \in \mathbb{F}^{n-l}
        \end{matrix}
        \; \middle| \; 
        \begin{matrix}\sum_{i \in [n]}z_i\ell_i(X) \cdot \sum_{i \in [n]}z_ir_i(X) = \sum_{i \in [n]}z_io_i(X) + t(X)h(X) \\
        \text{for some polynomial} \; h(X) \in \mathbb{F}[X] \; \text{and} \; z_0:=1\end{matrix}
    \right\}}
\end{equation*}

Here, $l$ is the number of \textit{public inputs} with indices $\mathcal{I}_X
\subseteq [n]$ and $n$, as previously discussed, is the total solution witness
size. We denote the set of indices of the \textit{private inputs} as
$\mathcal{I}_W := [n] \setminus \mathcal{I}_X$. 

\subsection{Groth16 Proof Construction}

The Groth16 proving system operates over the bilinear group $\mathcal{G} =
(\mathbb{G}_1,\mathbb{G}_2,\mathbb{G}_T,e)$ with the pairing $e: \mathbb{G}_1
\times \mathbb{G}_2 \to \mathbb{G}_T$. All groups are cyclic and let
corresponding generators be $g_1$, $g_2$ and $g_T$, respectively. The field
$\mathbb{F}$ is prime: $\mathbb{F} = \mathbb{F}_p$ and the security parameter is
$\lambda := p$. Now, we specify the implementation of three key procedures of
the proving system: \textcolor{green!50!black}{$\mathsf{Setup}(1^{\lambda})$},
\textcolor{blue!80!black}{$\mathsf{Prove}(\mathsf{pp},\mathbbm{x},\mathbbm{w})$}
and \textcolor{magenta}{$\mathsf{Verify}(\mathsf{pp},\mathbbm{x},\pi)$}.

\textcolor{green!50!black}{$\mathsf{Setup}(1^{\lambda})$}. Denote by $\zeta_i(X) :=
\beta \ell_i(X) + \alpha r_i(X) + o_i(X)$. The trusted setup generates the toxic
waste $\alpha,\beta,\gamma,\delta,\tau \xleftarrow{R} \mathbb{F}^{\times}$ and
computes the following common parameters:
\begin{align*}
    \mathsf{pp} &= \left( g_1^{\alpha},g_1^{\beta},g_1^{\delta},g_2^{\beta}, g_2^{\gamma}, g_2^{\delta}, \left\{g_1^{\tau^i}, g_2^{\tau^i}, g_1^{\tau^it(\tau)/\delta}\right\}_{i \in [n]}, \left\{g_1^{\zeta_i(\tau)/\gamma}\right\}_{i \in \mathcal{I}_X}, \left\{g_1^{\zeta_i(\tau)/\delta}\right\}_{i \in \mathcal{I}_W}\right) \\
    \mathsf{vp} &= \left( g_1,g_2, g_2^{\gamma}, g_2^{\delta}, g_T^{\alpha\beta}, \left\{g_1^{\zeta_i(\tau)/\gamma}\right\}_{i \in \mathcal{I}_{X}} \right)
\end{align*}

\textcolor{blue!80!black}{$\mathsf{Prove}(\mathsf{pp},\mathbbm{x},\mathbbm{w})$}.
The prover samples random scalars $r,s \xleftarrow{R} \mathbb{F}$ and outputs
the proof $\pi=(g_1^{a(\tau)}, g_1^{c(\tau)}, g_2^{b(\tau)})$ where:
\begin{gather*}
    a(X) = \alpha + \sum_{i \in [n]}z_i\ell_i(X) + r\delta, \quad b(X) = \beta + \sum_{i \in [n]}z_ir_i(X) + s\delta, \\
    c(X) = \frac{1}{\delta}\left(\sum_{i \in \mathcal{I}_W}z_i\zeta_i(X) + h(X)t(X)\right) + a(X)s + b(X)r - rs\delta
\end{gather*}

\textcolor{magenta}{$\mathsf{Verify}(\mathsf{pp},\mathbbm{x},\pi)$}. The
verifier parses the proof $\pi = (\pi_A, \pi_C, \pi_B) \in \mathbb{G}_1^2 \times
\mathbb{G}_2$ and accepts the proof if and only if the following check holds:
\begin{equation*}
    e(\pi_A, \pi_B) = e(g_1^{\alpha}, g_2^{\beta}) \cdot e(\pi_{\text{IC}}, g_2^{\gamma}) \cdot e(\pi_C, g_2^{\delta}).
\end{equation*}

Here, we denote the public input commitment part as $\pi_{\text{IC}} :=
g_1^{\text{IC}(\tau)} \in \mathbb{G}_1$ which the verifier computes where
$\text{IC}(X) := \sum_{i \in \mathcal{I}_X}z_i \cdot \frac{\zeta_i(X)}{\gamma}$.

\subsubsection{Efficiency Analysis}

We provide the following proposition.

\begin{proposition}
    Denote by $E_1$ and $E_2$ the cost of exponentiation over $\mathbb{G}_1$ and
    $\mathbb{G}_2$, respectively. Suppose pairing costs $E$. Then, the Groth16
    proving system complexity is following:
    \begin{itemize}
        \item The proving procedure
        $\textcolor{green!50!black}{\mathsf{Prove}(\mathsf{pp},\mathbbm{x},\mathbbm{w})}$
        runs in $\mathcal{O}(n \cdot E_1 + n \cdot E_2)$.
        \item The verification procedure
        $\textcolor{magenta}{\mathsf{Verify}(\mathsf{vp},\mathbbm{x},\pi)}$
        requires a time $3E + \mathcal{O}(l \cdot E_1)$, while the corresponding
        verification parameters require $\mathcal{O}(l)$ elements from
        $\mathbb{G}_1$. 
        \item The proof $\pi$ consists of $3$ group elements: two from
        $\mathbb{G}_1$ and one from $\mathbb{G}_2$.
    \end{itemize}
\end{proposition}

Note that in our case, when the input are private, while the weights are public,
the vast majority of signals are private, thus $l \ll n$. For that reason, since
verification procedure requires only $\mathcal{O}(l)$ elements from
$\mathbb{G}_1$, the Groth16 verification is effective.

\subsection{Lookup Tables}

The main idea of the UltraGroth system is to utilize the idea of the lookup
tables, which became ``native'' to PlonKish based proving systems
\cite{plookup}. Suppose the values in the lookup tables are $\{t_i\}_{i \in
[d]}$ while the part of the witness which we want to check is $\{z_i\}_{i \in
[n]}$: that is, we check whether $\{z_i\}_{i \in [n]} \subseteq \{t_i\}_{i \in
[d]}$. In our particular case, we will want to have $t_i=i$ for $d=2^w$ where 
$w$ is the bit-width of the input. The lookup table will be used to check
whether the input is in the range $[0,2^w)$. 

One of the most optimal checks was provided in \cite{logarithmic-derivatives}. We 
briefly restate the key result of the paper. 

\begin{theorem}[Following \cite{logarithmic-derivatives}]
    Given two sequences of elements $\{t_i\}_{i \in [d]}$ and $\{z_i\}_{i \in
    [n]}$ from field $\mathbb{F}$, the inclusion check $\{z_i\}_{i \in [n]}
    \subseteq \{t_i\}_{i \in [d]}$ is satisfied if and only if there exists the
    set of multiplicities $\{\mu_i\}_{i \in [d]}$ where $\mu_i = \#\{j \in [n]:
    z_j = t_i\}$ such that:
    \begin{equation*}
        \sum_{i \in [n]} \frac{1}{X+z_i} = \sum_{i \in [d]} \frac{\mu_i}{X+t_i}
    \end{equation*}

    In particular, checking such equality at a random point from $\mathbb{F}$
    results in the soundness error of up to $(n+d)/|\mathbb{F}|$, which
    becomes negligible for large enough $|\mathbb{F}|$.
\end{theorem}

\textbf{Note.} The second part of the theorem comes from the fact that we can
multiply both sides of the equation by the common denominator $\prod_{i \in
[n]}(X+t_i)\prod_{i \in [d]}(X+z_i)$. This way, both sides of the equation
become polynomials of degree approximately $n+d$. Thus, applying the
Schwartz-Zippel lemma, we obtain the soundness of $1-(n+d)/|\mathbb{F}|$.

Now, the primary issue why this approach is not used in the modern Groth16-based
constructions is that there is no efficient way to sample the randomness inside
the circuit. Indeed, constructions such as Fiat-Shamir heuristic requires the
verifier to hash the transcript, however this has to be done in a ``clever
way'': for instance, calculating the hash in-circuit would require an
overwhelming number of constraints. 

\subsection{UltraGroth Construction}

Suppose besides the regular constraints, there is a need to apply the lookup
technique to $d$ subsets of the witness $\{\mathcal{I}_W^{\langle i
\rangle}\}_{i \in [d]} \subseteq \mathcal{I}_W$. We denote by
$\mathcal{I}_W^{\langle d \rangle}$ the remaining part of the witness
$\mathcal{I}_W \setminus \bigcup_{i \in [d]}\mathcal{I}_W^{\langle i \rangle}$.
We call such subsets \textbf{rounds}: so in total we have $d+1$ rounds.

\begin{example}
    In our particular case, $d=1$. This way, we have two rounds in total:
    \begin{itemize}
        \item The round $\mathcal{I}_W^{\langle 0 \rangle}$ consists of the
        outputs of the $\mathsf{ReLU}$ function splitted into chunks.
        \item The round $\mathcal{I}_W^{\langle 1 \rangle}$ consists of all
        remaining checks.
    \end{itemize}
\end{example}

Now, we extend the witness $\boldsymbol{z}$ to include the indices of sampled
randomnesses. Suppose $\{\mathcal{I}_R^{\langle i \rangle}\}_{i \in [d]}$ are
such indices for each round (note that we do not need any randomness in the
$d^{\text{th}}$ round). We additionally note that in practice each
$\mathcal{I}_{R}^{\langle i \rangle}$ is commonly a single index with the
sampled randomness $\alpha_i \in \mathbb{F}_p$, but to make the protocol more
general, assume that $\mathcal{I}_R^{\langle i \rangle}$ corresponds to the
multivariate randomness $\boldsymbol{\alpha}_i \in \mathbb{F}_p^{\ell_i}$ of
size $\ell_i$. Surely, extending the witness by $\sum_{i \in [d]}\ell_i$ scalars
is negligible since this number is typically approximately $d$ (in the
particular case of Bionetta, it is simply $1$).

\begin{figure}[H]
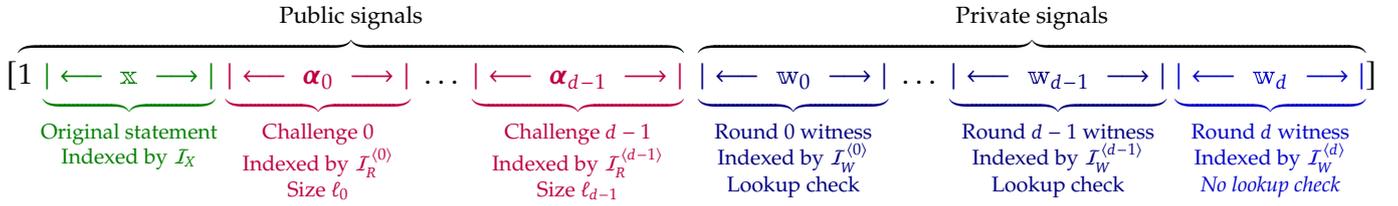

    \newcommand*{\horzbar}{\rule[.5ex]{2.5ex}{0.5pt}}
    \begin{equation*}
        \small\hspace{-30px}
        [
            \overbrace{1 \; 
            \textcolor{green!50!black}{\underbrace{|\longleftarrow \; \mathbbm{x} \; \longrightarrow|}_{
                \substack{
                    \text{Original statement} \\ 
                    \text{Indexed by} \; \mathcal{I}_X
                }
            }} \; 
            \textcolor{purple}{\underbrace{|\longleftarrow \; \boldsymbol{\alpha}_0 \; \longrightarrow|}_{
                \substack{
                    \text{Challenge $0$} \\ 
                    \text{Indexed by} \; \mathcal{I}_R^{\langle 0 \rangle} \\
                    \text{Size $\ell_0$}
                }
            }} \;
            \dots \; 
            \textcolor{purple}{\underbrace{|\longleftarrow \; \boldsymbol{\alpha}_{d-1} \; \longrightarrow|}_{
                \substack{
                    \text{Challenge $d-1$} \\ 
                    \text{Indexed by} \; \mathcal{I}_R^{\langle d-1 \rangle} \\
                    \text{Size $\ell_{d-1}$}
                }
            }}}^{\text{Public signals}} \;
            \overbrace{\textcolor{blue!50!black}{\underbrace{|\longleftarrow \; \mathbbm{w}_0 \; \longrightarrow|}_{
                \substack{
                    \text{Round $0$ witness} \\ 
                    \text{Indexed by} \; \mathcal{I}_W^{\langle 0 \rangle} \\
                    \text{Lookup check}
                }
            }} \;
            \dots \; 
            \textcolor{blue!50!black}{\underbrace{|\longleftarrow \; \mathbbm{w}_{d-1} \; \longrightarrow|}_{
                \substack{
                    \text{Round $d-1$ witness} \\ 
                    \text{Indexed by} \; \mathcal{I}_W^{\langle d-1 \rangle} \\
                    \text{Lookup check}
                }
            }} \;
            \textcolor{blue!80!black}{\underbrace{|\longleftarrow \; \mathbbm{w}_{d} \; \longrightarrow|}_{
                \substack{
                    \text{Round $d$ witness} \\ 
                    \text{Indexed by} \; \mathcal{I}_W^{\langle d \rangle} \\
                    \textit{No lookup check}
                }
            }}}^{\text{Private signals}}
        ]
    \end{equation*}
    \caption{The witness includes three parts: the
    \textcolor{green!50!black}{public part $\mathcal{I}_X$}, the
    \textcolor{blue!50!black}{private part $\mathcal{I}_W$} split into $d+1$
    rounds $\{\textcolor{blue!50!black}{\mathcal{I}_W^{\langle i \rangle}}\}_{i
    \in [d+1]}$ and the \textcolor{purple}{sampled challenges
    $\{\mathcal{I}_R^{\langle i \rangle}\}_{i \in [d]}$}.}
    \label{fig:ultragroth-witness-structure}
\end{figure}

In summary, we depict the witness structure in \Cref{fig:ultragroth-witness-structure}.

Besides the new witness structure, since we are going to apply the Fiat-Shamir
heuristic, we use the transformation $\mathcal{H}: \mathbb{F}_p \times
\mathbb{G}_1 \to \mathbb{F}_p$ modeled as a random oracle. 

\textcolor{green!50!black}{$\mathsf{Setup}(1^{\lambda})$}. The setup procedure
now consists in sampling additional $d+1$ scalars $\delta_0,\dots,\delta_d
\xleftarrow{R} \mathbb{F}^{\times}$. As in the previous section, we denote the
polynomial linear combination by $\zeta_i(X) := \beta \ell_i(X) + \alpha r_i(X)
+ o_i(X)$. The modified common parameters are\footnote{To simplify the notation,
we denote $\{\{a_{i,j}\}_{j \in \mathcal{I}_i}\}_{i \in [n]}$ by
$\{a_{i,j}\}_{i,j \in [n] \times \mathcal{I}_i}$.}:
\begin{gather*}
    \textcolor{gray}{\mathsf{pp} = \left( g_1^{\alpha},g_1^{\beta},\textcolor{blue}{\{g_1^{\delta_i}\}_{i \in [d]}},g_2^{\beta}, g_2^{\gamma}, \textcolor{blue}{\{g_2^{\delta_i}\}_{i \in [d]}}, \left\{g_1^{\tau^i}, g_2^{\tau^i}, g_1^{\tau^it(\tau)/\delta}\right\}_{i \in [n]}, \left\{g_1^{\zeta_i(\tau)/\gamma}\right\}_{i \in \mathcal{I}_X}, \textcolor{blue!80!black}{\left\{g_1^{\zeta_j(\tau)/\delta_i}\right\}_{(i,j) \in [d] \times \mathcal{I}_W^{(i)}}}\right)} \\
    \textcolor{gray}{\mathsf{vp} = \left( g_1,g_2, g_2^{\gamma}, g_2^{\delta}, g_T^{\alpha\beta}, \left\{g_1^{\zeta_i(\tau)/\gamma}\right\}_{i \in \mathcal{I}_{X}} \right)}
\end{gather*}

\textcolor{blue!80!black}{$\mathsf{Prove}(\mathsf{pp},\mathbbm{x},\mathbbm{w})$}.
The prover conducts the steps specified in Algorithm \ref{alg:ultragroth-prove}.

\begin{center}
    \begin{minipage}{0.85\linewidth}
        \begin{algorithm}[H]
            \SetAlgoLined

            Initialize accumulator $a_0 := \mathcal{H}(\mathsf{pp})$

            \Comment{\textbf{Step 1.} Sample challenges for in-circuit lookup check}
            \For{round $i$ in $[d]$}{
                $r_i \xleftarrow{R} \mathbb{F}_p$ \Comment*[r]{Sampling a random scalar}
                
                \Comment{Compute commitment}
                
                $\pi_C^{\langle i \rangle} \gets g_1^{c_i(\tau)}$, $c_i(X) := \sum_{j \in \mathcal{I}_W^{\langle i \rangle}} z_j \frac{\zeta_j(X)}{\delta_i} + r_i \delta_d$
                
                $a_{i+1} \gets \mathcal{H}(a_i, \pi_C^{\langle i \rangle})$ \Comment*[r]{Compute accumulator}
                
                \Comment{Iteratively fill the $i^{\text{th}}$ challenge $\boldsymbol{\alpha}_i \in \mathbb{F}_p^{\ell_i}$}
                
                \For{$j \in \mathcal{I}_R^{\langle i \rangle}$}{
                    $z_j \gets \mathcal{H}(a_{i+1}, g_1^j)$
                }
            }

            \Comment{\textbf{Step 2.} Generate the proof}
            Compute the polynomial $h(X)$ as usual from the QAP check $\sum_{i \in [n]}z_i\ell_i(X) \cdot \sum_{i \in [n]}z_ir_i(X) = \sum_{i \in [n]}z_io_i(X) + t(X)h(X)$
            
            Sample random $r,s \xleftarrow{R} \mathbb{F}_p$ and compute $\pi_A \gets g_1^{a(\tau)}$, $\pi_B \gets g_2^{b(\tau)}$ and 
            the last commitment $\pi_C^{\langle d \rangle} \gets g_1^{c_d(\tau)}$ where:
            \begin{align*}
                a(X) &= \alpha + \sum_{i \in [n]}z_i\ell_i(X) + r\textcolor{blue}{\delta_d}, \\
                b(X) &= \beta + \sum_{i \in [n]}z_ir_i(X) + s\textcolor{blue}{\delta_d}, \\
                c_d(X) &= \frac{1}{\delta_d}\left(\sum_{i \in \mathcal{I}_W^{\langle d \rangle}}z_i\zeta_i(X) + h(X)t(X)\right) + a(X)s + b(X)r \textcolor{blue}{- \sum_{i \in [d]}r_i\delta_i - rs\delta_d}
            \end{align*}

            \Output{Proof $\pi=(\pi_A,\pi_B,\pi_C^{\langle 0 \rangle},\dots,\pi_C^{\langle d \rangle}) \in \mathbb{G}_1 \times \mathbb{G}_2 \times \mathbb{G}_1^{d+1}$}

            \caption{The $\textcolor{blue!80!black}{\mathsf{Prove}(\mathsf{pp},\mathbbm{x},\mathbbm{w})}$ procedure.}
            \label{alg:ultragroth-prove}
        \end{algorithm}
    \end{minipage}
\end{center}

\textcolor{magenta}{$\mathsf{Verify}(\mathsf{pp},\mathbbm{x},\pi)$}. The verifier 
first recomputes the challenges sampled by the prover $\{\boldsymbol{\alpha}_i\}_{i \in [d]}$
and verifies that the corresponding public signals are correct. Then, it checks: 
\begin{equation*}
    e(\pi_A,\pi_B) = e(g_1^{\alpha}, g_2^{\beta}) \cdot e(\pi_{\text{IC}}, g_2^{\gamma}) \cdot \prod_{i \in [d+1]}e(\pi_C^{\langle i \rangle}, g_2^{\delta_i})
\end{equation*}

Finally, the following theorem states the security of the UltraGroth
protocol.

\begin{theorem}\label{theorem:ugroth-security} In the generic group model (GGM)
    and random oracle model (ROM), UltraGroth is complete, sound, and
    zero-knowledge.
\end{theorem}

\textbf{Proof.} See Appendix~\ref{section:ultragroth-proof-appendix}.

\subsubsection{Efficiency Analysis for Bionetta}\label{section:efficiency-analysis-for-bionetta}
Let us now analyze the efficiency of the UltraGroth system. As can be seen,
introducing the lookup check costs only 1 additional pairing and an additional
hashing operation. The prover complexity in turn is reduced to
$\mathcal{O}(2^{w+1} + \frac{bL}{w} + 4L)$ where $L$ is the number of range checks
and $b$ is the field bitsize. Asymptotically, this gives $\mathcal{O}(N/\log N)$
complexity for $N=bL$.

\begin{figure}[H]
\begin{tikzpicture}[scale=0.9]
    \begin{axis}[
      width=12cm,
      height=8cm,
      domain=1:30,
      samples=300,
      xlabel={Chunk size $w$},
      ylabel={Complexity $c(w;L) = 2^{w+1} + \frac{bL}{w}$},
      grid=major,
      thick,
      ymin=0,
      xmin=7,
      xmax=25,
      title={Prover complexity vs chunk size $w$},
    ]
    \addplot[red, ultra thick] {2^(x+1) + (254 * 2^21) / x};
    \addlegendentry{$L = 2^{21}$}
    
    \addplot[blue, ultra thick] {2^(x+1) + (254 * 2^20) / x};
    \addlegendentry{$L = 2^{20}$}

    \addplot[green!50!black, ultra thick] {2^(x+1) + (254 * 2^19) / x};
    \addlegendentry{$L = 2^{19}$}

    \addplot[orange!90!black, ultra thick] {2^(x+1) + (254 * 2^18) / x};
    \addlegendentry{$L = 2^{18}$}

    \addplot[magenta, ultra thick] {2^(x+1) + (254 * 2^17) / x};
    \addlegendentry{$L = 2^{17}$}

    \end{axis}
\end{tikzpicture}
\caption{The prover complexity depending on the chunk size $w$. The different colors 
represent different values of $L$.}
\label{fig:chunk-size}
\end{figure}
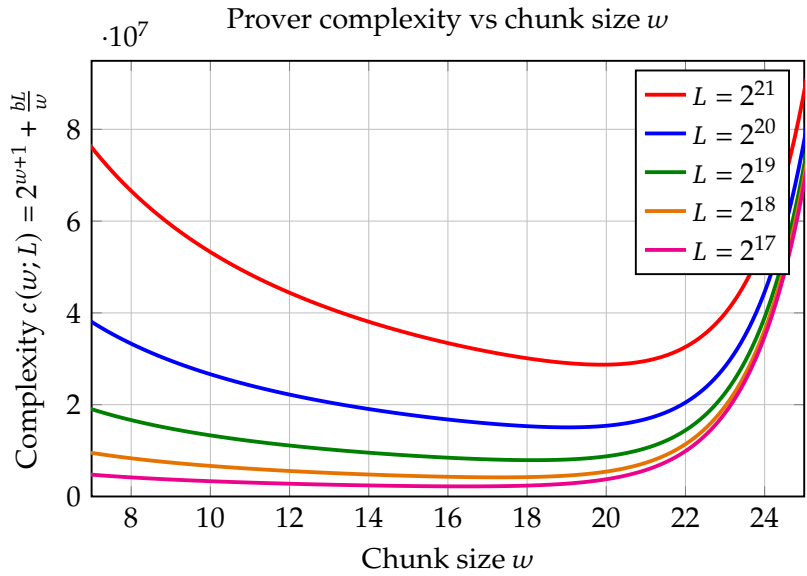

The question, though, is which chunk parameter $w$ to choose. In other words,
which value of $w$ minimizes the cost function $c(w;L) = 2^{w+1} + \frac{bL}{w}$
for the given integer $L$? The plot itself is visualized in
\Cref{fig:chunk-size}. It seems that the best choice would be to pick $w$ around
$18$ for the given $L$ values. However, we give a rigorous statement.

\begin{proposition}
    The optimal chunk size $\widehat{w}$ must be closest to the solution 
    to the given equation:
    \begin{equation*}
        2^{w}w^2 = \frac{Lb}{2\log 2}
    \end{equation*}
\end{proposition}

\textbf{Reasoning.} Simply find the derivative of the function $c(w;L)$ and set
it to $0$. The derivative is $\partial_w c(w;L) = 2^{w+1}\log 2
- \frac{Lb}{w^2}$. Equating this to zero gives the equation above. It is also
easy to check that it is indeed a minimum: notice that the second derivative is
positive: $\partial^2_wc(w;L)=2^{w+1}\log^22 +
\frac{2Lb}{w^3}$, thus $\partial^2_w c(\widehat{w};L) > 0$.

Finally, we provide the following proposition that gives the straightforward
interpretation of complexity of UltraGroth.

\begin{proposition}
    The function $c(\widehat{w}; N) = \mathcal{O}(N/\log N)$ if $\widehat{w}$ is
    the optimal chunk size.
\end{proposition}

\section{Experiments}\label{section:experiments}

\subsection{Setup}

To evaluate the performance of aforementioned
tools, we benchmark each framework over a set of five neural networks. The first
neural network model (which we call the MNIST model) is built with a large
number of standard \textit{Dense} (fully connected) and \textit{ReLU} layers to
measure baseline zkML frameworks' overhead. This models takes a flattened $28
\times 28$ MNIST image as an input and produces a 10-class
classification. The architecture consits of two million parameters and 6K
activations. We additionally benchmark results on \textit{LeNet5}~\cite{lenet}, VGG11-mini\footnote{
We reduced VGG11 for benchmarking since \texttt{zkml} and EZKL frameworks ran out of RAM
memory for original architecture.}~\cite{vgg11}, \textit{MobileNetV2}~\cite{mobilenetv2} and
\textit{ResNet18}~\cite{resnet} models to test how Bionetta performs on
real-world models.

We conducted tests with the latest versions of each library from their GitHub repositories
\cite{ezkl,ddkang-zkml-github,keras2circom,deepprove,zk-cnn}.

We expect Bionetta to perform better than existing methods when the model has a
lot of linear components. However, it turns out that even with the large number
of activations, Bionetta performs better than existing approaches.

\textcolor{blue!50!black}{\textbf{System Settings.}} We use TensorFlow 2.12
\cite{tensorflow2015-whitepaper} for Bionetta and change TensorFlow/PyTorch
version based on the proving system.

\textbf{Hardware:} Intel Xeon E5-2665 CPU (16 threads @ 3.10GHz) and 350GB of RAM

\textbf{Software:} Bionetta uses custom fork of Rapidsnark prover that supports
Groth16 and $1$QAP UltraGroth.

\subsection{Results}

\begin{figure*}
    \centering
    \begin{tabular}{cc}
        \begin{tikzpicture}
            \begin{axis}[
                title={\textbf{Proving Time}},
                xlabel={Model Size ($\times 10^6$)},
                ylabel={Time (seconds)},
                width=0.4\textwidth, height=4.5cm,
                grid=both,
                minor tick num=1,
                xticklabel style={/pgf/number format/1000 sep=},
                ymin=0.1, ymax=15,
                unbounded coords=jump,
                font=\small
            ]
                \addplot[color=blue, mark=*, dotted, very thick, mark size=2pt, solid] table[x expr={\thisrow{DenseSize}*794/1000000}, y=UG_Prove]{\benchmarks};
                \addplot[color=red, mark=square*, dotted, very thick, mark size=2pt, solid] table[x expr={\thisrow{DenseSize}*794/1000000}, y=EZKL_Prove]{\benchmarks};
                \addplot[color=orange, mark=triangle*, dotted, very thick, mark size=2pt, solid] table[x expr={\thisrow{DenseSize}*794/1000000}, y=ZKML_Prove]{\benchmarks};
                \addplot[color=teal, mark=diamond*, dotted, very thick, mark size=2pt, solid] table[x expr={\thisrow{DenseSize}*794/1000000}, y=DP_Prove]{\benchmarks};
                \addplot[color=violet, mark=square*, dotted, very thick, mark size=2pt, solid] table[x expr={\thisrow{DenseSize}*794/1000000}, y=zkCNN_Prove]{\benchmarks};
                
                \node[anchor=north west, align=left, font=\scriptsize, fill=white, fill opacity=0.8, draw=black!20, inner sep=2pt] at (rel axis cs:0.02,0.98) {
                    \textcolor{red}{\ding{110}}\textcolor{orange}{\ding{115}} up to $\times 800$ of\\
                    Bionetta's proving time
                };
            \end{axis}
        \end{tikzpicture}
        &
        % --- Top-Right Plot: Verification Time ---
        \begin{tikzpicture}
            \begin{axis}[
                title={\textbf{Verification Time}},
                xlabel={Model Size ($\times 10^6$)},
                ylabel={Time (seconds, log scale)},
                ymode=log,
                width=0.4\textwidth, height=4.5cm,
                grid=both,
                minor tick num=1,
                xticklabel style={/pgf/number format/1000 sep=},
                ymin=0.01, ymax=20,
                unbounded coords=jump,
                font=\small
            ]
                \addplot[color=blue, mark=*, dotted, very thick, mark size=2pt, solid] table[x expr={\thisrow{DenseSize}*794/1000000}, y=UG_Verify]{\benchmarks};
                \addplot[color=red, mark=square*, dotted, very thick, mark size=2pt, solid] table[x expr={\thisrow{DenseSize}*794/1000000}, y=EZKL_Verify]{\benchmarks};
                \addplot[color=orange, mark=triangle*, dotted, very thick, mark size=2pt, solid] table[x expr={\thisrow{DenseSize}*794/1000000}, y=ZKML_Verify]{\benchmarks};
                \addplot[color=teal, mark=diamond*, dotted, very thick, mark size=2pt, solid] table[x expr={\thisrow{DenseSize}*794/1000000}, y=DP_Verify]{\benchmarks};
                \addplot[color=violet, mark=square*, dotted, very thick, mark size=2pt, solid] table[x expr={\thisrow{DenseSize}*794/1000000}, y=zkCNN_Verify]{\benchmarks};
            \end{axis}
        \end{tikzpicture}
        \\
        % --- Bottom-Left Plot: RAM Usage ---
        \begin{tikzpicture}
            \begin{axis}[
                title={\textbf{RAM Usage}},
                xlabel={Model Size ($\times 10^6$)},
                ylabel={Size (GB)},
                width=0.4\textwidth, height=4.5cm,
                grid=both,
                minor tick num=1,
                xticklabel style={/pgf/number format/1000 sep=},
                ymin=0.1, ymax=4,
                unbounded coords=jump,
                font=\small
            ]
                \addplot[color=blue, mark=*, dotted, very thick, mark size=2pt, solid] table[x expr={\thisrow{DenseSize}*794/1000000}, y=UG_RAM]{\benchmarks};
                % \addplot[color=red, mark=square*, dotted, very thick, mark size=2pt, solid] table[x expr={\thisrow{DenseSize}*794/1000000}, y=EZKL_RAM]{\benchmarks};
                \addplot[color=orange, mark=triangle*, dotted, very thick, mark size=2pt, solid] table[x expr={\thisrow{DenseSize}*794/1000000}, y=ZKML_RAM]{\benchmarks};
                \addplot[color=teal, mark=diamond*, dotted, very thick, mark size=2pt, solid] table[x expr={\thisrow{DenseSize}*794/1000000}, y=DP_RAM]{\benchmarks};
                \addplot[color=violet, mark=square*, dotted, very thick, mark size=2pt, solid] table[x expr={\thisrow{DenseSize}*794/1000000}, y=zkCNN_RAM]{\benchmarks};

                \node[anchor=north west, align=left, font=\scriptsize, fill=white, fill opacity=0.8, draw=black!20, inner sep=2pt] at (rel axis cs:0.02,0.98) {
                    \textcolor{red}{\ding{110}}\textcolor{orange}{\ding{115}} up to $\times 15$ of\\
                    Bionetta's RAM usage
                };
            \end{axis}
        \end{tikzpicture}
        &
        % --- Bottom-Right Plot: Proof Size ---
        \begin{tikzpicture}
            \begin{axis}[
                title={\textbf{Proof Size}},
                xlabel={Model Size ($\times 10^6$)},
                ylabel={Size (KB, log scale)},
                ymode=log,
                width=0.4\textwidth, height=4.5cm,
                grid=both,
                minor tick num=1,
                xticklabel style={/pgf/number format/1000 sep=},
                ymin=0.7, ymax=7200,
                unbounded coords=jump,
                font=\small
            ]
                \addplot[color=blue, mark=*, dotted, very thick, mark size=2pt, solid] table[x expr={\thisrow{DenseSize}*794/1000000}, y=UG_Proof]{\benchmarks};
                \addplot[color=red, mark=square*, dotted, very thick, mark size=2pt, solid] table[x expr={\thisrow{DenseSize}*794/1000000}, y=EZKL_Proof]{\benchmarks};
                \addplot[color=orange, mark=triangle*, dotted, very thick, mark size=2pt, solid] table[x expr={\thisrow{DenseSize}*794/1000000}, y=ZKML_Proof]{\benchmarks};
                \addplot[color=teal, mark=diamond*, dotted, very thick, mark size=2pt, solid] table[x expr={\thisrow{DenseSize}*794/1000000}, y=DP_Proof]{\benchmarks};
                \addplot[color=violet, mark=square*, dotted, very thick, mark size=2pt, solid] table[x expr={\thisrow{DenseSize}*794/1000000}, y=zkCNN_Proof]{\benchmarks};
            \end{axis}
        \end{tikzpicture}
    \end{tabular}
    
    \caption{Comparison of proving time, verification time, RAM usage, and proof size across different zkML frameworks as a function of the model's size. \textit{Legend:} 
    \textcolor{blue}{\ding{108}}~UltraGroth, 
    \textcolor{red}{\ding{110}}~EZKL, 
    \textcolor{orange}{\ding{115}}~ZKML,
    \textcolor{teal}{\ding{117}}~\texttt{deep-prove}, and
    \textcolor{violet}{\ding{110}}~zkCNN.}
    \label{fig:asymptotics}
\end{figure*}
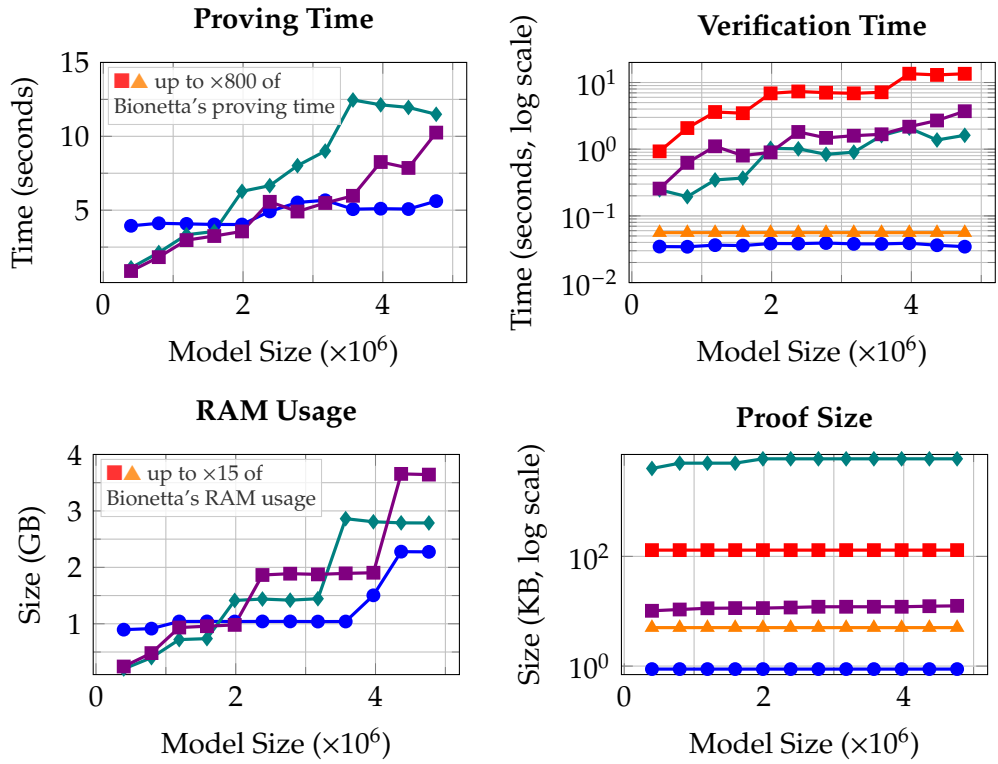

% For Bionetta, we used custom fork of \texttt{rapidsnark}~\cite{rapidsnark} prover that supports $1$QAP UltraGroth.
\textbf{General Benchmarking.} The results are summarized\footnote{Unfortunately, some frameworks like
\texttt{keras2circom}~\cite{keras2circom} or
\texttt{deep-prove}~\cite{deep-prove} do not support complex architectures like
ResNet18 or MobileNetV2, so we do not include them in this table. Yet, we
further include \texttt{deep-prove} in~\Cref{fig:asymptotics}.} in
\Cref{tab:general-benchmarks}, while the summary for the MNIST model is
visualized in \Cref{fig:relative_overhead_final}. As can be seen, Bionetta paired with
UltraGroth greatly outperforms all Halo2-based frameworks in terms of all
metrics, including proof size (surpassing EZKL by a factor of up to $\times
500$), proving and verification key sizes (up to $\times 9000$-fold reduction),
proving (yielding up to a $\times 580$-fold reduction) and verification times
(with $\times 1000$ reduction compared to EZKL). 

Compared to GKR-based approaches (such as \texttt{zkCNN} or
\texttt{deep-prove}), we expectedly achieve better proof sizes ($\times 40$
improvement compared to \texttt{zkCNN} and $\times 4200$ compared to
\texttt{deep-prove}) and verification times ($\times 72.5$ boost). For proving
time, the performance of Bionetta remains on par with such frameworks,
indicating no significant drop while still offering substantial gains elsewhere.
In \Cref{fig:asymptotics}, we consider more systematically how proving time
scales with the model increase.

Moreover, compared to existing solutions, these results indicate that using
Bionetta is practical for many potential blockchain applications: recall that
UltraGroth proof verification requires only $4$ pairing operations while
verification keys are of size up to 5KB.

\begin{table*}

\centering
\small
\setlength{\tabcolsep}{4pt}

\caption{Benchmark results for four frameworks on an Intel Xeon E5-2665 CPU (16
threads @ 3.10GHz) and 350GB of RAM. We present five key metrics: proof size,
proving key (PK) size, verification key (VK) size, proving time, and
verification time. Colors indicate performance ranking:
\colorbox{green!25}{best}, \colorbox{red!20}{worst}.\\
\textbf{Notes:}\\
``\cross'' indicates that the zkML framework does not support all operations used in the given model. \\
$^{\text{\ding{61}}}$\texttt{zkCNN} is GKR-based and therefore does not require proving and verification keys. \\
}
\label{tab:general-benchmarks}
\begin{NiceTabular}{llccccc}[colortbl-like]
\toprule
\textbf{Framework} & \textbf{Metric} & \textbf{MNIST} & \textbf{LeNet5} & \textbf{VGG11-mini} & \textbf{ResNet18} & \textbf{MobileNetV2} \\
\midrule

% --- Bionetta (UltraGroth) ---
\multirow{6}{*}{\centering\arraybackslash\textbf{Bionetta+UltraGroth}} & Proof Size (KB) & \cellcolor{green!25}0.88 & \cellcolor{green!25}0.88 & \cellcolor{green!25}0.88 & \cellcolor{green!25}0.88 & \cellcolor{green!25}0.88 \\
& PK (GB)         & \cellcolor{green!10}0.20 & \cellcolor{green!10}0.22 & \cellcolor{green!10}0.63 & \cellcolor{green!10}1.00 & \cellcolor{green!10}2.41 \\
& VK (MB)         & \cellcolor{green!10}0.004 & \cellcolor{green!10}0.004 & \cellcolor{green!10}0.004 & \cellcolor{green!10}0.004 & \cellcolor{green!10}0.004 \\
& Prove (s)       & \cellcolor{green!25}3.05 & 3.75 & 7.70 & \cellcolor{green!25}14.10 & \cellcolor{green!25}24.50 \\
& RAM (GB)        & \cellcolor{green!25}0.27 & 0.28 & \cellcolor{green!25}0.48 & \cellcolor{green!25}0.75 & \cellcolor{green!25}1.80 \\
& Verify (s)      & \cellcolor{green!25}0.010 & \cellcolor{green!25}0.010 & \cellcolor{green!25}0.020 & \cellcolor{green!25}0.015 & \cellcolor{green!25}0.018 \\
\midrule

% --- EZKL ---
\multirow{6}{*}{\centering\arraybackslash\texttt{ezkl}~\cite{ezkl}} 
 & Proof Size (KB) & \cellcolor{red!20}{127.0} & \cellcolor{red!20}{127.0} & \cellcolor{red!20}{478.0} & \cellcolor{red!20}180.0 & \cellcolor{red!20}{175.0} \\
 & PK (GB)         & 8.30 & 2.10 & 17.00 & \cellcolor{red!20}34.00 & \cellcolor{red!20}76.00 \\
 & VK (MB)         & \cellcolor{red!20}4.10 & \cellcolor{red!20}1.10 & \cellcolor{red!20}8.10 & 17.00 & \cellcolor{red!20}37.00 \\
 & Prove (s)       & \cellcolor{red!20}1310 & \cellcolor{red!20}535 & \cellcolor{red!20}2650 & 6840 & \cellcolor{red!20}14320 \\
 & RAM (GB)        & 21.15 & 5.45 & 41.85 & 103.50 & \cellcolor{red!20}203.95 \\
 & Verify (s)      & \cellcolor{red!20}5.40 & \cellcolor{red!20}2.00 & \cellcolor{red!20}9.70 & 25.30 & \cellcolor{red!20}40.60 \\
\midrule

% --- ZKML ---
\multirow{6}{*}{\centering\arraybackslash\texttt{zkml}~\cite{ddkang-zkml}} 
 & Proof Size (KB) & 5.05 & 6.70 & 6.70 & 6.70 & 5.10 \\
 & PK (GB)         & \cellcolor{red!20}16.10 & \cellcolor{red!20}2.40 & \cellcolor{red!20}19.30 & 19.30 & 17.70 \\
 & VK (MB)         & 2.60 & 0.40 & 3.15 & 3.15 & 2.60 \\
 & Prove (s)       & 1100 & 185 & 1580 & 1660 & 1135 \\
 & RAM (GB)        & \cellcolor{red!20}39.95 & \cellcolor{red!20}6.85 & \cellcolor{red!20}51.95 & 52.30 & 44.05 \\
 & Verify (s)      & 0.012 & 0.013 & 0.022 & 0.023 & 0.020 \\
\midrule

% --- zkCNN ---
\multirow{4}{*}{\centering\arraybackslash\texttt{zkCNN}$^{\text{\ding{61}}}$~\cite{zk-cnn}}
 & Proof Size (KB) & 23.25 & 38.10 & 42.60 & \cross & \cross \\
 & Prove (s)       & 3.45 & \cellcolor{green!25}1.05 & \cellcolor{green!25}4.10 & \cross & \cross \\
 & RAM (GB)        & 1.00 & \cellcolor{green!25}0.15 & 0.80 & \cross & \cross \\
 & Verify (s)      & \cellcolor{red!20}1.00 & \cellcolor{red!20}0.30 & \cellcolor{red!20}1.45 & \cross & \cross \\
\bottomrule
\end{NiceTabular}
\end{table*}

\textbf{Client-side vs Server-side proving.} We examine how optimizations
introduced in \cref{section:circuit-embedded-weights} reduce circuit sizes. We
compiled all aforementioned models and recorded the number of constraints and
proving efficiency. Results are depicted in~\Cref{tab:client-server}.
Expectedly, the decrease in the number of constraints largely depends on the
neural network architecture: for instance, UltraGroth circuit size decreases
$\times 3$ for LeNet5 circuit while for ResNet18 this optimization yields
$\times 32$ improvement. For the latter case, this reduces proving time from 270
seconds down to only 15.

\begin{table}[t]
\centering
\small
\setlength{\tabcolsep}{3pt} % Reduced colsep for single column fit

\caption{Comparison of circuits complexities for client-side (marked as
``Client'' below) and server-side (marked as ``Server'') provings over QAP and
$1$QAP. Values for Server-Side proving time and RAM are estimated based on
constraints number. In \colorbox{orange!20}{orange} we color the most optimal
parameters used in Bionetta.}
\label{tab:client-server}

{\footnotesize$^{\text{\ding{61}}}$Values for server-side proving are estimated based on the
constraints number since Bionetta is primarily designed for client-side proving.}

% Resizebox ensures the wide table fits in the single column
\resizebox{0.8\columnwidth}{!}{
\begin{NiceTabular}{llccccc}[colortbl-like]
\toprule
\textbf{Mode} & \textbf{Metric} & \textbf{MNIST} & \textbf{LeNet5} & \textbf{VGG11} & \textbf{ResNet18} & \textbf{MobNetV2} \\
\midrule

% --- Bionetta (UltraGroth, Client-Side) ---
\rowcolor{orange!20}
\textbf{1QAP} & Constraints ($\times 10^6$) & 0.25 & 0.26 & 0.73 & 1.16 & 2.78 \\
\rowcolor{orange!20}\small Client & Prove (s)   & 3.05 & 3.75 & 7.70 & 14.10 & 24.50 \\
\rowcolor{orange!20}& RAM (GB)    & 0.27 & 0.28 & 0.48 & 0.75 & 1.80 \\
\midrule

% --- Bionetta (UltraGroth, Server-side) ---
\multirow{3}{*}{\centering\arraybackslash\shortstack{\textbf{1QAP}\\\small Server$^{\text{\ding{61}}}$}} 
& Constraints ($\times 10^6$) & 2.19 & 0.65 & 4.17 & 37.85 & 7.97 \\
& Prove (s)   & 15 & 6 & 30 & 270 & 60 \\
& RAM (GB)    & 1.2 & 0.4 & 2.2 & 19.8 & 4.2 \\
\midrule

% --- Bionetta (Groth16, Client-Side) ---
\multirow{3}{*}{\centering\arraybackslash\shortstack{\textbf{QAP}\\\small Client}}
& Constraints ($\times 10^6$) & 1.57 & 1.69 & 7.67 & 12.0 & 31.8 \\
& Prove (s)   & 8.30 & 8.70 & 56.4 & 87.5 & 230.0 \\
& RAM (GB)    & 0.58 & 0.60 & 4.05 & 6.3 & 16.65 \\
\midrule

% --- Bionetta (Groth16, Server-Side) ---
\multirow{3}{*}{\centering\arraybackslash\shortstack{\textbf{QAP}\\\small Server$^{\text{\ding{61}}}$}} 
& Constraints ($\times 10^6$) & 3.51 & 2.08 & 11.1 & 49.0 & 37.8 \\
& Prove (s)   & 25 & 15 & 80 & 350 & 270 \\
& RAM (GB)    & 1.9 & 1.1 & 5.8 & 25.5 & 19.7 \\

\bottomrule
\end{NiceTabular}
}
\end{table}

\textbf{Efficiency Metrics Behaviour.} We additionally provide benchmarks for
dependency of key efficiency metrics from the number of non-linearities in the
neural network. For this purpose, we define the following MNIST model:
\begin{align*}
    & f(\mathbf{x};W_1,W_2) = W_2\text{ReLU}(W_1\mathbf{x}), \\
    & \text{with} \; \mathbf{x} \in \mathbb{R}^{784}, W_1 \in \mathbb{R}^{N \times 784}, W_2
\in \mathbb{R}^{10 \times N},
\end{align*}
where we vary the parameter $N$ --- number of neurons in the hidden layer. The
results for $12$ different values of $N$ in range $500 \leq N \leq 6000$ are
depicted in \Cref{fig:asymptotics}. Bionetta's verification time and proof size
are consistently the best across all values of $N$. For proving time and RAM
consumption with $N<2000$, the Bionetta gives slightly worse performance than
GKR-based approaches. However, as the model becomes larger (starting from
roughly $2$ million parameters), the Bionetta surpasses these frameworks. For
instance, when the model reaches 4.7 million parameters, zkCNN generates proof
in 10 seconds using 3.6~GB of RAM while Bionetta in 5.3 seconds consuming only
2.2~GB of RAM.

% --- DATA DEFINITION ---
% Using the original absolute values
\pgfplotstableread{
Model   DenseSize   UG_Prove    EZKL_Prove  ZKML_Prove  DP_Prove    zkCNN_Prove UG_Verify   EZKL_Verify ZKML_Verify DP_Verify   zkCNN_Verify    UG_RAM  EZKL_RAM    ZKML_RAM    DP_RAM  zkCNN_RAM   UG_Proof    EZKL_Proof  ZKML_Proof  DP_Proof    zkCNN_Proof
1   500 3.9416  199.91  157.7   1.097   0.88047 0.034636    0.92798 0.056   0.246   0.25658 0.89599 2.8485  5.6695  0.20 0.2431  0.882   129.95  5.024   4000.4176   10.188
2   1000    4.1123  365.91  165.2   2.131   1.8192  0.034425    2.0748  0.056   0.194   0.62478 0.91454 5.4744  5.802   0.4 0.4808  0.882   129.99  5.024   5000.1148   10.75
3   1500    4.0719  690.65  168.47  3.337   2.9728  0.036404    3.6127  0.056   0.346   1.1052  1.041   10.568  6.0211  0.72 0.9339  0.882   129.93  5.024   5000.5423   11.312
4   2000    4.0336  711.18  1114.4  3.558   3.2498  0.035856    3.4668  0.056   0.369   0.80104 1.0419  10.384  39.8    0.73711 0.9571  0.882   129.94  5.024   5000.5582   11.406
5   2500    4.0288  1275.2  1108.2  6.275   3.5614  0.038513    6.8764  0.056   1.03    0.89704 1.0405  20.862  39.881  1.4137  0.9805  0.882   129.94  5.024   6000.0006   11.406
6   3000    4.914   1309.9  1157.7  6.659   5.567   0.038373    7.3461  0.056   1.013   1.8138  1.0416  20.784  40.154  1.4387  1.8646  0.882   129.94  5.024   6000.0167   11.688
7   3500    5.5207  1329.9  1110    8.013   4.9078  0.039146    7.0409  0.056   0.837   1.48    1.0418  20.532  40.199  1.4185  1.8878  0.882   129.91  5.024   6000.0004   12.062
8   4000    5.6704  1333.2  1157.5  9.0 5.4854  0.038026    6.8832  0.056   0.902   1.5931  1.0393  21.081  40.35   1.4443  1.8755  0.882   129.99  5.024   6000.0109   12.062
9   4500    5.0672  1375    1151    12.459  5.9728  0.037965    7.1369  0.056   1.598   1.6738  1.0403  21.106  40.419  2.8626  1.8932  0.882   129.95  5.024   6000.5042   12.062
10  5000    5.1006  2530    1142.5  12.132  8.26    0.038902    13.554  0.056   2.071   2.1755  1.5043  41.331  40.593  2.8075  1.9045  0.882   129.97  5.024   6000.4877   12.062
11  5500    5.0758  2538.4  1142.5  11.954  7.8587  0.036275    13.005  0.056   1.374   2.6936  2.278   41.892  40.479  2.788   3.6561  0.882   129.94  5.024   6000.4871   12.344
12  6000    5.613   2607.5  1143.4  11.491  10.251  0.034534    13.523  0.056   1.616   3.7119  2.2744  41.884  40.87   2.7865  3.6433  0.882   130.02  5.024   6000.4708   12.531
}\benchmarks

\textbf{Efficiency on Mobile Devices.} To further demonstrate that Bionetta is
capable of proving neural networks on lightweight client devices, we conduct an
experiment where we launch all five compiled models on \textit{iPhone 14 Pro}
(A16 Bionic chip, 6‑core CPU with 2 performance and 4 efficiency cores, 5‑core
GPU, and 6 GB of RAM). We depict the results in \Cref{tab:benchmarks}. For
Groth16, we use both the \texttt{rapidsnark}~\cite{rapidsnark} and recently
published mobile-first proving framework \texttt{imp1}~\cite{imp1} by Ingonyama.
As can be seen, the MobileNetV2 neural network generates the proof in 23 seconds
while ResNet requires only 14 seconds. While the peak RAM consumption is
considerable (approximately 1.7GB), it is still a doable number for modern edge
devices. 

\begin{table}
\centering
\caption{Bionetta Benchmarks (proving time and peak RAM consumption) on iPhone 14 Pro for both UltraGroth and Groth16. ``\cross'' mark indicates that too much RAM was required to launch the proving procedure. For Groth16, we specify both \texttt{rapidsnark}~\cite{rapidsnark} and \texttt{imp1}~\cite{imp1} benchmarks.}
\label{tab:benchmarks}
\resizebox{0.8\textwidth}{!}{%
\begin{tikzpicture}
\node[inner sep=0pt] {
\begin{tabular}{
  l
  S[table-format=2.2] % UltraGroth Time
  S[table-format=4.0] % UltraGroth RAM
  c % Groth16 Time (two values)
  c % Groth16 RAM (two values)
}
\toprule
& \multicolumn{2}{c}{\textbf{UltraGroth}} & \multicolumn{2}{c}{\textbf{Groth16}} \\
& \multicolumn{2}{c}{} & \multicolumn{2}{c}{(\texttt{rapidsnark}/\texttt{imp1})} \\
\cmidrule(lr){2-3} \cmidrule(lr){4-5}

\textbf{Model} & {\textbf{Time (s)}} & {\textbf{RAM (MB)}} & {\textbf{Time (s)}} & {\textbf{RAM (MB)}} \\
\midrule

MNIST        & 2.80 & 230  & 8.2/5.0     & 570/540   \\
LeNet5       & 3.55 & 145  & 7.7/4.80    & 600/560   \\
VGG11-mini   & 7.65 & 465  & 52.5/36.5   & 3800/3550 \\
ResNet18     & 13.60 & 700 & \cross/\cross & \cross/\cross \\
MobileNetV2  & 22.96 & 1700 & \cross/\cross & \cross/\cross \\
\bottomrule
\end{tabular}
};
\end{tikzpicture}
}
\label{tab:mobile-benches}
\end{table}

\textbf{Accuracy.} Finally, we estimate the accuracy of Bionetta. We launch
every model on $10^5$ randomly generated inputs (for example, by sampling each
input from standard normal distribution) both in TensorFlow and through
arithmetical circuit. Then, we dequantize the circuit result and compare it to
one obtained in TensorFlow. Assuming that we use precision $\rho$, the output of
the TensorFlow is $\mathbf{y}$ and dequantized result is $\mathbf{y}_{\rho}$, we
measure the relative error as $\varepsilon_{\rho} = \|\mathbf{y} -
\mathbf{y}_{\rho}\|/\|\mathbf{y}\|$. After running experiments on each model,
the maximal observed relative errors among all models are $\varepsilon_{15}
\approx 5.2 \cdot 10^{-3}$, $\varepsilon_{30} \approx 3.7 \cdot 10^{-5}$,
$\varepsilon_{45} \approx 2.0 \cdot 10^{-6}$, $\varepsilon_{60} \approx 9.4
\cdot 10^{-7}$, which can be considered negligible in practical applications.

\section{Conclusion}

In this paper, we have presented a new framework for zero-knowledge machine
learning, called \textit{Bionetta}. We have experimentally shown that our
framework is capable of proving and verifying machine learning models with high
accuracy and low overhead. We have compared our framework with existing
frameworks, such as \texttt{ddkang/zkml}, \texttt{ezkl}, \texttt{keras2circom},
\texttt{deep-prove} and \texttt{zkCNN}, and shown that our framework outperforms them in terms of
proof size, verification key size, and proving time, while providing a
reasonable verification procedure, fully compatible with Ethereum smart
contracts.

In addition, we implemented the custom modification of the Groth16 protocol to
significantly boost the performance of the proving procedure with minimal
overhead for the verification stage. Finally, we plan to make the Bionetta
framework open-sourced and available for public use, allowing researchers and
developers to build upon our work and contribute to the development of
zero-knowledge machine learning.

\section{Authors}

\noindent
\begin{minipage}[t]{0.3\textwidth}
\textcolor{blue!50!black}{\textbf{Cryptography}} \\
Dmytro Zakharov\\
Oleksandr Kurbatov\\
Artem Sdobnov\\
Lev Soukhanov\\
Yevhenii Sekhin\\
Vitalii Volovyk
\end{minipage}
\hspace{0.05\textwidth}
\begin{minipage}[t]{0.3\textwidth}
    \textcolor{green!50!black}{\textbf{Machine Learning}} \\
    Mykhailo Velykodnyi \\
    Mark Cherepovskyi \\
    Kyrylo Baibula
\end{minipage}
\hspace{0.05\textwidth}
\begin{minipage}[t]{0.3\textwidth}
    \textcolor{purple!80!white}{\textbf{Management, Visioning, and Guidance}}\\
    Lasha Antadze\\
    Pavlo Kravchenko\\ 
    Volodymyr Dubinin\\ 
    Yaroslav Panasenko
\end{minipage}

\printbibliography

@inproceedings{groth16,
  address   = {Berlin, Heidelberg},
  author    = {Groth, Jens},
  booktitle = {Advances in Cryptology -- EUROCRYPT 2016},
  editor    = {Fischlin, Marc and Coron, Jean-Sebastien},
  isbn      = {978-3-662-49896-5},
  pages     = {305--326},
  publisher = {Springer Berlin Heidelberg},
  title     = {On the Size of Pairing-Based Non-interactive Arguments},
  year      = {2016}
}

@article{snark-survey,
  author        = {Morais, Eduardo and Koens, Tommy and van Wijk, Cees and Koren, Aleksei},
  date          = {2019/07/31},
  doi           = {10.1007/s42452-019-0989-z},
  id            = {Morais2019},
  isbn          = {2523-3971},
  journal       = {SN Applied Sciences},
  number        = {8},
  pages         = {946},
  title         = {A survey on zero knowledge range proofs and applications},
  url           = {https://doi.org/10.1007/s42452-019-0989-z},
  volume        = {1},
  year          = {2019}
}

@article{circom,
  author   = {Bellés-Muñoz, Marta and Isabel, Miguel and Muñoz-Tapia, Jose Luis and Rubio, Albert and Baylina, Jordi},
  journal  = {IEEE Transactions on Dependable and Secure Computing},
  title    = {Circom: A Circuit Description Language for Building Zero-Knowledge Applications},
  year     = {2023},
  volume   = {20},
  number   = {6},
  pages    = {4733-4751},
  doi      = {10.1109/TDSC.2022.3232813}
}

@article{snark-stark-comparison,
  author     = {Oude Roelink, Bjorn and El-Hajj, Mohammed and Sarmah, Dipti},
  doi        = {https://doi.org/10.1002/spy2.401},
  eprint     = {https://onlinelibrary.wiley.com/doi/pdf/10.1002/spy2.401},
  journal    = {SECURITY AND PRIVACY},
  number     = {5},
  pages      = {e401},
  title      = {Systematic review: Comparing zk-SNARK, zk-STARK, and bulletproof protocols for privacy-preserving authentication},
  url        = {https://onlinelibrary.wiley.com/doi/abs/10.1002/spy2.401},
  volume     = {7},
  year       = {2024}
}

@inproceedings{pairings_r1cs,
  address   = {Cham},
  author    = {Housni, Youssef El},
  booktitle = {Applied Cryptography and Network Security},
  editor    = {Tibouchi, Mehdi and Wang, XiaoFeng},
  isbn      = {978-3-031-33488-7},
  pages     = {339--362},
  publisher = {Springer Nature Switzerland},
  title     = {Pairings in Rank-1 Constraint Systems},
  year      = {2023}
}

@inproceedings{posseidon,
  title     = {Poseidon: A New Hash Function for Zero-Knowledge Proof Systems},
  author    = {Lorenzo Grassi and Dmitry Khovratovich and Christian Rechberger and Arnab Roy and Markus Schofnegger},
  booktitle = {USENIX Security Symposium},
  year      = {2021},
  url       = {https://api.semanticscholar.org/CorpusID:221069468}
}

@article{ec_r1cs,
  author     = {Aranha, Diego F. and El Housni, Youssef and Guillevic, Aurore},
  title      = {A survey of elliptic curves for proof systems},
  year       = {2022},
  issue_date = {Nov 2023},
  publisher  = {Kluwer Academic Publishers},
  address    = {USA},
  volume     = {91},
  number     = {11},
  issn       = {0925-1022},
  url        = {https://doi.org/10.1007/s10623-022-01135-y},
  doi        = {10.1007/s10623-022-01135-y},
  journal    = {Des. Codes Cryptography},
  month      = dec,
  pages      = {3333–3378},
  numpages   = {46}
}

@inproceedings{ddkang-zkml,
  author    = {Chen, Bing-Jyue and Waiwitlikhit, Suppakit and Stoica, Ion and Kang, Daniel},
  title     = {ZKML: An Optimizing System for ML Inference in Zero-Knowledge Proofs},
  year      = {2024},
  isbn      = {9798400704376},
  publisher = {Association for Computing Machinery},
  address   = {New York, NY, USA},
  url       = {https://doi.org/10.1145/3627703.3650088},
  doi       = {10.1145/3627703.3650088},
  booktitle = {Proceedings of the Nineteenth European Conference on Computer Systems},
  pages     = {560–574},
  numpages  = {15},
  location  = {Athens, Greece},
  series    = {EuroSys '24}
}

@misc{ezkl,
  title         = {Verifiable evaluations of machine learning models using zkSNARKs},
  author        = {Tobin South and Alexander Camuto and Shrey Jain and Shayla Nguyen and Robert Mahari and Christian Paquin and Jason Morton and Alex 'Sandy' Pentland},
  year          = {2024},
  eprint        = {2402.02675},
  archiveprefix = {arXiv},
  primaryclass  = {cs.LG},
  url           = {https://arxiv.org/abs/2402.02675}
}

@misc{deep-prove,
	author       = {Lagrange Labs},
	title        = {deep-prove},
	year         = {2025},
	publisher    = {GitHub},
	journal      = {GitHub repository},
	howpublished = {\url{https://github.com/Lagrange-Labs/deep-prove}}
}

@article{mobilenetv2,
  title   = {MobileNetV2: Inverted Residuals and Linear Bottlenecks},
  author  = {Mark Sandler and Andrew G. Howard and Menglong Zhu and Andrey Zhmoginov and Liang-Chieh Chen},
  journal = {2018 IEEE/CVF Conference on Computer Vision and Pattern Recognition},
  year    = {2018},
  pages   = {4510-4520},
  url     = {https://api.semanticscholar.org/CorpusID:4555207}
}

@misc{mobilenetv3,
  title         = {Searching for MobileNetV3},
  author        = {Andrew Howard and Mark Sandler and Grace Chu and Liang-Chieh Chen and Bo Chen and Mingxing Tan and Weijun Wang and Yukun Zhu and Ruoming Pang and Vijay Vasudevan and Quoc V. Le and Hartwig Adam},
  year          = {2019},
  eprint        = {1905.02244},
  archiveprefix = {arXiv},
  primaryclass  = {cs.CV},
  url           = {https://arxiv.org/abs/1905.02244}
}

@misc{activation-polynomial-approximation,
  title         = {Highly Accurate CNN Inference Using Approximate Activation Functions over Homomorphic Encryption},
  author        = {Takumi Ishiyama and Takuya Suzuki and Hayato Yamana},
  year          = {2020},
  eprint        = {2009.03727},
  archiveprefix = {arXiv},
  primaryclass  = {cs.LG},
  url           = {https://arxiv.org/abs/2009.03727}
}

@inproceedings{pi-nets,
  title={P-nets: Deep polynomial neural networks},
  author={Chrysos, Grigorios G and Moschoglou, Stylianos and Bouritsas, Giorgos and Panagakis, Yannis and Deng, Jiankang and Zafeiriou, Stefanos},
  booktitle={Proceedings of the IEEE/CVF Conference on Computer Vision and Pattern Recognition},
  pages={7325--7335},
  year={2020}
}

@misc{resnet,
  title         = {Deep Residual Learning for Image Recognition},
  author        = {Kaiming He and Xiangyu Zhang and Shaoqing Ren and Jian Sun},
  year          = {2015},
  eprint        = {1512.03385},
  archiveprefix = {arXiv},
  primaryclass  = {cs.CV},
  url           = {https://arxiv.org/abs/1512.03385}
}

@misc{vgg11,
	author       = "Karen Simonyan and Andrew Zisserman",
	title        = "Very Deep Convolutional Networks for Large-Scale Image Recognition",
	booktitle    = "International Conference on Learning Representations",
	year         = "2015",
}

@misc{lenet,
	author={Lecun, Y. and Bottou, L. and Bengio, Y. and Haffner, P.},
	journal={Proceedings of the IEEE}, 
	title={Gradient-based learning applied to document recognition}, 
	year={1998},
	volume={86},
	number={11},
	pages={2278-2324},
	doi={10.1109/5.726791}
}

@inproceedings{zk-cnn,
	author = {Liu, Tianyi and Xie, Xiang and Zhang, Yupeng},
	title = {zkCNN: Zero Knowledge Proofs for Convolutional Neural Network Predictions and Accuracy},
	year = {2021},
	isbn = {9781450384544},
	publisher = {Association for Computing Machinery},
	address = {New York, NY, USA},
	url = {https://doi.org/10.1145/3460120.3485379},
	doi = {10.1145/3460120.3485379},
	booktitle = {Proceedings of the 2021 ACM SIGSAC Conference on Computer and Communications Security},
	pages = {2968–2985},
	numpages = {18},
	location = {Virtual Event, Republic of Korea},
	series = {CCS '21}
}

@misc{ddkang-zkml-github,
  author       = {Daniel Kang},
  title        = {ddkang/zkml Framework},
  year         = {2022},
  publisher    = {GitHub},
  journal      = {GitHub repository},
  howpublished = {\url{https://github.com/ddkang/zkml}}
}

@misc{keras2circom,
  author       = {drCathieSo.eth},
  title        = {keras2circom},
  year         = {2022},
  publisher    = {GitHub},
  journal      = {GitHub repository},
  howpublished = {\url{https://github.com/socathie/keras2circom}}
}

@misc{deepprove,
  author       = {Lagrange Labs},
  title        = {deep-prove},
  year         = {2025},
  publisher    = {GitHub},
  journal      = {GitHub repository},
  howpublished = {\url{https://github.com/Lagrange-Labs/deep-prove}}
}

@misc{zkpytorch,
  author = {Tiancheng Xie and Tao Lu and Zhiyong Fang and Siqi Wang and Zhenfei Zhang and Yongzheng Jia and Dawn Song and Jiaheng Zhang},
  title = {{zkPyTorch}: A Hierarchical Optimized Compiler for Zero-Knowledge Machine Learning},
  howpublished = {Cryptology {ePrint} Archive, Paper 2025/535},
  year = {2025},
  url = {https://eprint.iacr.org/2025/535}
}

@misc{rarimo,
  author = {Rarimo},
  title = {Rarimo},
  publisher    = {GitHub},
  journal      = {GitHub repository},
  year = {2025},
  howpublished = {\url{https://github.com/rarimo}}
}

@misc{ceno,
  author       = {Tianyi Liu and Zhenfei Zhang and Yuncong Zhang and Wenqing Hu and Ye Zhang},
  title        = {Ceno: Non-uniform, Segment and Parallel Zero-knowledge Virtual Machine},
  howpublished = {Cryptology {ePrint} Archive, Paper 2024/387},
  year         = {2024},
  url          = {https://eprint.iacr.org/2024/387}
}

@misc{plookup,
  author = {Ariel Gabizon and Zachary J.  Williamson},
  title = {plookup: A simplified polynomial protocol for lookup tables},
  howpublished = {Cryptology {ePrint} Archive, Paper 2020/315},
  year = {2020},
  url = {https://eprint.iacr.org/2020/315}
}

@inproceedings{se-networks,
  title={Squeeze-and-excitation networks},
  author={Hu, Jie and Shen, Li and Sun, Gang},
  booktitle={Proceedings of the IEEE conference on computer vision and pattern recognition},
  pages={7132--7141},
  year={2018}
}

@misc{vit,
  title={An Image is Worth 16x16 Words: Transformers for Image Recognition at Scale}, 
  author={Alexey Dosovitskiy and Lucas Beyer and Alexander Kolesnikov and Dirk Weissenborn and Xiaohua Zhai and Thomas Unterthiner and Mostafa Dehghani and Matthias Minderer and Georg Heigold and Sylvain Gelly and Jakob Uszkoreit and Neil Houlsby},
  year={2021},
  eprint={2010.11929},
  archivePrefix={arXiv},
  primaryClass={cs.CV},
  url={https://arxiv.org/abs/2010.11929}, 
}

@misc{logarithmic-derivatives,
  author = {Ulrich Haböck},
  title = {Multivariate lookups based on logarithmic derivatives},
  howpublished = {Cryptology {ePrint} Archive, Paper 2022/1530},
  year = {2022},
  url = {https://eprint.iacr.org/2022/1530}
}

@misc{whir,
    author = {Gal Arnon and Alessandro Chiesa and Giacomo Fenzi and Eylon Yogev},
    title = {WHIR: Reed–Solomon Proximity Testing with Super-Fast Verification},
    howpublished = {Cryptology {ePrint} Archive, Paper 2024/1586},
    year = {2024},
    url = {https://eprint.iacr.org/2024/1586}
}

@misc{tensorflow2015-whitepaper,
  title  = {{TensorFlow}: Large-Scale Machine Learning on Heterogeneous Systems},
  url    = {https://www.tensorflow.org/},
  note   = {Software available from tensorflow.org},
  author = {
            Mart\'{i}n~Abadi and
            Ashish~Agarwal and
            Paul~Barham and
            Eugene~Brevdo and
            Zhifeng~Chen and
            Craig~Citro and
            Greg~S.~Corrado and
            Andy~Davis and
            Jeffrey~Dean and
            Matthieu~Devin and
            Sanjay~Ghemawat and
            Ian~Goodfellow and
            Andrew~Harp and
            Geoffrey~Irving and
            Michael~Isard and
            Yangqing Jia and
            Rafal~Jozefowicz and
            Lukasz~Kaiser and
            Manjunath~Kudlur and
            Josh~Levenberg and
            Dandelion~Man\'{e} and
            Rajat~Monga and
            Sherry~Moore and
            Derek~Murray and
            Chris~Olah and
            Mike~Schuster and
            Jonathon~Shlens and
            Benoit~Steiner and
            Ilya~Sutskever and
            Kunal~Talwar and
            Paul~Tucker and
            Vincent~Vanhoucke and
            Vijay~Vasudevan and
            Fernanda~Vi\'{e}gas and
            Oriol~Vinyals and
            Pete~Warden and
            Martin~Wattenberg and
            Martin~Wicke and
            Yuan~Yu and
            Xiaoqiang~Zheng},
  year   = {2015}
}

@article{mnist,
  title     = {The mnist database of handwritten digit images for machine learning research},
  author    = {Deng, Li},
  journal   = {IEEE Signal Processing Magazine},
  volume    = {29},
  number    = {6},
  pages     = {141--142},
  year      = {2012},
  publisher = {IEEE}
}

@misc{spartan,
  author       = {Srinath Setty},
  title        = {Spartan: Efficient and general-purpose {zkSNARKs} without trusted setup},
  howpublished = {Cryptology {ePrint} Archive, Paper 2019/550},
  year         = {2019},
  url          = {https://eprint.iacr.org/2019/550}
}

@INPROCEEDINGS{vsql,
	author = { Zhang, Yupeng and Genkin, Daniel and Katz, Jonathan and Papadopoulos, Dimitrios and Papamanthou, Charalampos },
	booktitle = { 2017 IEEE Symposium on Security and Privacy (SP) },
	title = {{ vSQL: Verifying Arbitrary SQL Queries over Dynamic Outsourced Databases }},
	year = {2017},
	volume = {},
	ISSN = {2375-1207},
	pages = {863-880},
	doi = {10.1109/SP.2017.43},
	url = {https://doi.ieeecomputersociety.org/10.1109/SP.2017.43},
	publisher = {IEEE Computer Society},
	address = {Los Alamitos, CA, USA},
	month = May
}

@article{virgo,
	title={Transparent Polynomial Delegation and Its Applications to Zero Knowledge Proof},
	author={Jiaheng Zhang and Tiancheng Xie and Yupeng Zhang and Dawn Xiaodong Song},
	journal={2020 IEEE Symposium on Security and Privacy (SP)},
	year={2020},
	pages={859-876},
	url={https://api.semanticscholar.org/CorpusID:209467198}
}

@article{gkr,
	author = {Goldwasser, Shafi and Kalai, Yael Tauman and Rothblum, Guy N.},
	title = {Delegating Computation: Interactive Proofs for Muggles},
	year = {2015},
	issue_date = {August 2015},
	publisher = {Association for Computing Machinery},
	address = {New York, NY, USA},
	volume = {62},
	number = {4},
	issn = {0004-5411},
	url = {https://doi.org/10.1145/2699436},
	doi = {10.1145/2699436},
	journal = {J. ACM},
	month = sep,
	articleno = {27},
	numpages = {64}
}

@InProceedings{lasso,
	author="Setty, Srinath
	and Thaler, Justin
	and Wahby, Riad",
	editor="Joye, Marc
	and Leander, Gregor",
	title="Unlocking the Lookup Singularity with Lasso",
	booktitle="Advances in Cryptology -- EUROCRYPT 2024",
	year="2024",
	publisher="Springer Nature Switzerland",
	address="Cham",
	pages="180--209",
	isbn="978-3-031-58751-1"
}

@inproceedings{ipa,
	author = {Bunz, Benedikt and Maller, Mary and Mishra, Pratyush and Tyagi, Nirvan and Vesely, Psi},
	title = {Proofs for Inner Pairing Products and Applications},
	year = {2021},
	isbn = {978-3-030-92077-7},
	publisher = {Springer-Verlag},
	address = {Berlin, Heidelberg},
	url = {https://doi.org/10.1007/978-3-030-92078-4_3},
	doi = {10.1007/978-3-030-92078-4_3},
	booktitle = {Advances in Cryptology – ASIACRYPT 2021: 27th International Conference on the Theory and Application of Cryptology and Information Security, Singapore, December 6–10, 2021, Proceedings, Part III},
	pages = {65–97},
	numpages = {33},
	location = {Singapore, Singapore}
}

@article{zk-gpt,
	title={zkGPT: An Efficient Non-interactive Zero-knowledge Proof Framework for LLM Inference},
	author={Qu, Wenjie and Sun, Yijun and Liu, Xuanming and Lu, Tao and Guo, Yanpei and Chen, Kai and Zhang, Jiaheng},
	journal={USENIX Security 2025, Prepublication}
}

@misc{hyrax,
  author       = {Riad S.  Wahby and Ioanna Tzialla and abhi shelat and Justin Thaler and Michael Walfish},
  title        = {Doubly-efficient {zkSNARKs} without trusted setup},
  howpublished = {Cryptology {ePrint} Archive, Paper 2017/1132},
  year         = {2017},
  url          = {https://eprint.iacr.org/2017/1132}
}

@misc{aurora,
  author       = {Eli Ben-Sasson and Alessandro Chiesa and Michael Riabzev and Nicholas Spooner and Madars Virza and Nicholas P.  Ward},
  title        = {Aurora: Transparent Succinct Arguments for {R1CS}},
  howpublished = {Cryptology {ePrint} Archive, Paper 2018/828},
  year         = {2018},
  url          = {https://eprint.iacr.org/2018/828}
}

@misc{fractal,
  author       = {Alessandro Chiesa and Dev Ojha and Nicholas Spooner},
  title        = {Fractal: Post-Quantum and Transparent Recursive Proofs from Holography},
  howpublished = {Cryptology {ePrint} Archive, Paper 2019/1076},
  year         = {2019},
  url          = {https://eprint.iacr.org/2019/1076}
}

@misc{rapidsnark,
  author = {iden3},
  title = {rapidsnark},
  publisher    = {GitHub},
  journal      = {GitHub repository},
  year = {2025},
  howpublished = {\url{https://github.com/iden3/rapidsnark}}
}

@misc{imp1,
  author = {Ingonyama},
  title = {imp1},
  publisher    = {GitHub},
  journal      = {GitHub repository},
  year = {2025},
  howpublished = {\url{https://github.com/ingonyama-zk/imp1}}
}

@misc{halo1,
    author = {Sean Bowe and Jack Grigg and Daira Hopwood},
    title = {Recursive Proof Composition without a Trusted Setup},
    howpublished = {Cryptology {ePrint} Archive, Paper 2019/1021},
    year = {2019},
    url = {https://eprint.iacr.org/2019/1021}
}

@misc{halo2,
  author = {zcash},
  title = {halo2},
  publisher    = {GitHub},
  journal      = {GitHub repository},
  year = {2025},
  howpublished = {\url{https://zcash.github.io/halo2/}}
}

\newpage

% Begin appendix
\begin{appendices}

\section{UltraGroth Security Proof}\label{section:ultragroth-proof-appendix}

First, let us give the definitions of properties that the UltraGroth must
satisfy in order to be a secure zk-SNARK protocol.

\begin{definition}
    A zk-SNARK protocol $\Pi = (\mathsf{Setup}, \mathsf{Prove}, \mathsf{Verify})$ may 
    have the following properties:
    \begin{enumerate}
        \item \underline{Perfect Completeness}. $\Pi$ is complete if for any efficient relation $\mathcal{R}$
        the valid proof $\pi$ is always accepted by the verifier:
        \begin{equation*}
            \text{Pr}\left[ \mathsf{Verify}(\mathsf{vp}, \mathbbm{x}, \pi) = 1 \; \middle| \begin{matrix}
                (\mathsf{pp}, \mathsf{vp}) \gets \mathsf{Setup}(1^{\lambda}, \mathcal{R}) \\
                \pi \gets \mathsf{Prove}(\mathsf{pp}, \mathbbm{x}, \mathbbm{w}) \\
                (\mathbbm{x}, \mathbbm{w}) \in \mathcal{R}
            \end{matrix} \right] = 1
        \end{equation*}
        \item \underline{Perfect Zero-Knowledge.} $\Pi$ provides perfect
        zero-knowledge if for any efficient relation $\mathcal{R}$, there exists
        PPT simulator $\mathsf{Sim}$ such that for any $(\mathbbm{x},
        \mathbbm{w}) \in \mathcal{R}$, the distribution of a real proof is
        identical to the distribution of a simulated proof. Formally, after a
        trusted setup $(\mathsf{pp}, \mathsf{vp}) \gets
        \mathsf{Setup}(1^{\lambda}, \mathcal{R})$, the following two
        distributions are identical:
        \begin{equation*}
            \left\{ \pi \gets \mathsf{Prove}(\mathsf{pp}, \mathbbm{x}, \mathbbm{w}) : \pi \right\} \approx_C \left\{ \pi' \gets \mathsf{Sim}(\mathsf{vp}, \mathbbm{x}) : \pi' \right\}
        \end{equation*}
        \item \underline{Computational Knowledge Soundness}. $\Pi$ is a
        computational proof of knowledge if for any efficient relation
        $\mathcal{R}$ and for any PPT malicious prover $\mathcal{P}^*$, there
        exists a PPT extractor $\mathcal{E}$ which can ``extract'' the secret
        witness $\mathbbm{w}$ from $\mathcal{P}^*$. Formally, if a prover
        $\mathcal{P}^*$ can produce a valid proof $\pi$ for a statement
        $\mathbbm{x}$ with non-negligible probability, then the extractor (given
        oracle access to $\mathcal{P}^*$) can produce a valid witness
        $\mathbbm{w}^*$ for that statement:
        \begin{equation*}
            \Pr \left[ \begin{matrix}
                \mathsf{Verify}(\mathsf{vp}, \mathbbm{x}, \pi) = 1 \\ (\mathbbm{x}, \mathbbm{w}^*) \in \mathcal{R}
            \end{matrix} \; \middle| \; \begin{matrix}
                (\mathsf{pp}, \mathsf{vp}) \gets \mathsf{Setup}(1^{\lambda}, \mathcal{R}); \\
                \pi \gets \mathcal{P}^*(\mathsf{pp}, \mathbbm{x}); \\
                \mathbbm{w}^* \gets \mathcal{E}^{\mathcal{P}^*}(\mathsf{pp}, \mathbbm{x})
            \end{matrix} \right] \ge 1 - \mathsf{negl}(\lambda)
        \end{equation*}
    \end{enumerate}
\end{definition}

\begin{theorem}
    UltraGroth protocol is a secure zk-SNARK protocol, i.e., it satisfies
    perfect completeness, perfect zero-knowledge, and computational knowledge
    soundness.
\end{theorem}

\textbf{Proof.} \underline{\textit{Completeness}} is easy to show. Recall that the verifier has to check whether
\begin{equation*}
    e(\pi_A,\pi_B) = e(g_1^{\alpha}, g_2^{\beta}) \cdot e(\pi_{\text{IC}}, g_2^{\gamma}) \cdot \prod_{i \in [d+1]}e(\pi_C^{\langle i \rangle}, g_2^{\delta_i})
\end{equation*}

Consider the expression $m_C := \prod_{i \in [d+1]}e(\pi_C^{\langle i \rangle},
g_2^{\delta_i})$. Let $g_T := e(g_1,g_2)$, then:
\begin{align*}
    m_C=\prod_{i \in [d+1]}e(\pi_C^{\langle i \rangle}, g_2^{\delta_i}) 
    &= g_T^{\sum_{i \in [d]}\delta_ic_i(\tau) + \delta_dc_d(\tau)} \\
    &= g_T^{\sum_{i \in [d]}\delta_i\left(\sum_{j \in \mathcal{I}_W^{\langle i \rangle}} z_j \frac{\zeta_j(\tau)}{\delta_i} + r_i \delta_d\right) + \delta_dc_d(\tau)} \\
    &= g_T^{\sum_{i \in [d]}\sum_{j \in \mathcal{I}_W^{\langle i \rangle}}z_j\zeta_j(\tau) + \delta_d\sum_{i \in [d]}\delta_ir_i + \delta_dc_d(\tau)}
\end{align*}

Now, notice that $\sum_{i \in [d]}\sum_{j \in \mathcal{I}_W^{\langle i
\rangle}}$ is essentially summing over index array $\bigcup_{i \in
[d]}\mathcal{I}_W^{\langle i \rangle}$, which in turn is the set $\mathcal{I}_W
\setminus \mathcal{I}_W^{\langle d \rangle}$. Additionally, we expand
$c_d(\tau)$:
\begin{align*}
    m_C &= g_T^{\sum_{j \in \mathcal{I}_W \setminus \mathcal{I}_W^{\langle d \rangle}} z_j \zeta_j(\tau) + \cancel{\delta_d\sum_{i \in [d]}\delta_ir_i} + \sum_{i \in \mathcal{I}_W^{\langle d \rangle}}z_i\zeta_i(\tau) + h(\tau)t(\tau) + \delta_da(\tau)s + \delta_db(\tau)r - \cancel{\delta_d\sum_{i \in [d]}r_i\delta_i} - rs\delta_d^2}
\end{align*}

Note that $\sum_{j \in \mathcal{I}_W \setminus \mathcal{I}_W^{\langle d
\rangle}} + \sum_{j \in \mathcal{I}_W^{\langle d \rangle}} = \sum_{j \in
\mathcal{I}_W}$. Hence, we further simplify the expression:
\begin{align*}
    m_C &= g_T^{\sum_{j \in \mathcal{I}_W} z_j\zeta_j(\tau) + h(\tau)t(\tau) + \delta_da(\tau)s + \delta_db(\tau)r  - rs\delta_d^2}
\end{align*}

But notice that this expression is nothing but
\begin{align*}
    m_C = e\left(g_1^{\widetilde{c}_d(\tau)}, g_2^{\delta_d}\right), \quad \widetilde{c}_d(X) = \frac{1}{\delta_d}\left(\sum_{j \in \mathcal{I}_W} z_j\zeta_j(X) + h(X)t(X)\right) + a(X)s + b(X)r - rs\delta_d
\end{align*}

This is exactly the expression that we had in the original Groth16's verifier 
equation, but instead of a single $\delta$, only the last coefficient
$\delta_d$ is used. Hence, the correctness of the UltraGroth verifier is
guaranteed by the correctness of the original Groth16 verifier.

Now we show \underline{\textit{perfect zero-knowledge}}. Note that the pairing
check verifies whether $a(\tau)b(\tau) = \alpha\beta + \mathsf{IC}(\tau)\gamma +
\sum_{i \in [d+1]}\delta_ic_i(\tau)$. For that reason, we propose the following
simulator $\mathsf{Sim}$ that outputs a valid proof $\boldsymbol{\pi}$ that is
indistinguishable from the real proof.

\textcolor{blue!80!black}{\underline{\textbf{Simulator $\mathsf{Sim}(\mathsf{vp}, \mathbbm{x}) \to \boldsymbol{\pi}$}}}:
\vspace{-20px}
\begin{enumerate}
    \item \textit{Sample uniformly random $a',b',\{c_i'\}_{i \in [d]} \xleftarrow{R}
    \mathbb{F}$, corresponding to the exponents of corresponding proof elements.} 
    \item \textit{Output the proof $\boldsymbol{\pi} = (g_1^{a'}, g_2^{b'}, \{g_1^{c_i'}\}_{i \in [d+1]})$ where:}
    \begin{equation*}
        c_d' = \frac{a'b' - \alpha\beta - \sum_{j \in \mathcal{I}_X}z_j\zeta_j(\tau) - \sum_{i \in [d]}c_i'\delta_i}{\delta_d}
    \end{equation*}
\end{enumerate}

Let us show why it works. The proof correctness follows directly from the
verification equation. Note that first $d+2$ components of the proof are
uniformly distributed groups elements (since the corresponding powers are
uniformly distributed) while the last term $c_d'$ is computed uniquely from
other coefficients; thus it is also uniformly random. Hence, the UltraGroth
protocol is perfectly zero-knowledge.

Finally, we show \underline{\textit{computational knowledge soundness}}. We
generalize the results closely following the original Groth16 paper
\cite{groth16}. Recall that Groth16 is a pairing-based NIZK argument for QAP
derived from NILP. UltraGroth, in turn, can also be viewed as an extension of
NILP. That said, we simply have to show that any affine strategy that succeeds
in non-negligible probability can be used to extract the witness. Recall that in
the affine strategy, the prover computes the proof by applying the linear
transformation on the common reference string: $\boldsymbol{\pi} = \Pi
\boldsymbol{\sigma} = (a(\tau), b(\tau), \{c_i(\tau)\}_{i \in [d+1]}) \in
\mathbb{F}^{d+3}$. Here,
\begin{equation*}
    \boldsymbol{\sigma} = \left(\alpha,\beta,\gamma,\{\delta_i\}_{i \in [d+1]}, \{\tau^i\}_{i \in [n]}, \left\{\frac{\zeta_i(\tau)}{\gamma}\right\}_{i \in \mathcal{I}_X}, \left\{\frac{\zeta_j(\tau)}{\delta_i}\right\}_{i,j \in [d+1] \times \mathcal{I}_W^{\langle i \rangle}}, \left\{\frac{\tau^it(\tau)}{\delta_d}\right\}_{i \in \mathcal{I}_W^{\langle d \rangle}} \right)
\end{equation*}

Therefore, the prover's strategy for the first part $a$ is following:
\begin{equation*}
    a = a_{\alpha}\alpha + a_{\beta}\beta + a_{\gamma}\gamma + \sum_{i \in [d+1]} a_{\delta}^{\langle i \rangle} \delta_i + a(\tau) + \sum_{i \in \mathcal{I}_X} a_i\frac{\zeta_i(\tau)}{\gamma} + \sum_{i \in [d+1]}\sum_{j \in \mathcal{I}_W^{\langle d \rangle}} a_j\frac{\zeta_j(\tau)}{\delta_i} + a_h(\tau)\frac{t(\tau)}{\delta_d},
\end{equation*}
where field elements $a_{\alpha}, a_{\beta},a_{\gamma},\{a_{\delta}^{\langle i
\rangle}\}_{i \in [d+1]}, \{a_i\}_{i \in \mathcal{I}_W}$ and polynomials
$a(\tau)$ and $a_h(\tau)$ are known and correspond to the first row of the
matrix $\Pi$. The same structure, albeit with different coefficients, holds for
$b$ and $\{c_i\}_{i \in [d+1]}$.

Now we view the verification equation as an equality of multivariate Laurent
polynomials. By the Schwartz-Zippel lemma, the prover has a negligible
probability for making the verification equation hold for randomly selected
indeterminates $\alpha, \beta, \gamma, \{\delta_i\}_{i \in [d+1]}$ and $\tau$. Now 
recall that the check is following:
\begin{equation*}
    ab = \alpha\beta + \mathsf{IC}(\tau)\gamma + \sum_{i \in [d+1]}\delta_ic_i.
\end{equation*}

Now we are going to derive what are the coefficients of the corresponding
polynomials $a$, $b$, and $\{c_i\}_{i \in [d+1]}$. The terms with indeterminate $\alpha^2$ are
$a_{\alpha}b_{\alpha}\alpha^2=0$ and so $a_{\alpha} = 0$ or $b_{\alpha}=0$.
Without loss of generality, assume $b_{\alpha}=0$. Terms with $\alpha\beta$
gives us $a_{\alpha}b_{\beta} + a_{\beta}b_{\alpha} = a_{\alpha}b_{\beta}=1$. 
Again, without loss of generality, assume $a_{\alpha}=1$ and $b_{\beta}=1$. 
Finally, the indeterminate $\beta^2$ gives $a_{\beta}b_{\beta} = a_{\beta}=0$. 
Therefore, the expressions for $a$ and $b$ are now simplified down to:
\begin{equation*}
    a = \alpha + a_{\gamma}\gamma + \sum_{i \in [d+1]} a_{\delta}^{\langle i \rangle} \delta_i + \dots, \quad b = \beta + b_{\gamma}\gamma + \sum_{i \in [d+1]} b_{\delta}^{\langle i \rangle} \delta_i + \dots
\end{equation*}

Now consider the terms involving $1/\delta_d^2$. We have:
\begin{equation*}
    \left(\sum_{j \in \mathcal{I}_W^{\langle d \rangle}} a_j\zeta_j(\tau) + a_h(\tau)t(\tau)\right)\left(\sum_{j \in \mathcal{I}_W^{\langle d \rangle}} b_j\zeta_j(\tau) + b_h(\tau)t(\tau)\right) = 0
\end{equation*}

Without loss of generality, assume $\sum_{j \in \mathcal{I}_W^{\langle d
\rangle}} a_j\zeta_j(\tau) + a_h(\tau)t(\tau)=0$. However, looking at term
$\frac{\alpha}{\delta_d}\sum_{j \in \mathcal{I}_W^{\langle d \rangle}}
b_j\zeta_j(\tau) + a_h(\tau)t(\tau) = 0$ shows that in fact $\sum_{j \in
\mathcal{I}_W^{\langle d \rangle}} b_j\zeta_j(\tau) + a_h(\tau)t(\tau)=0$ as
well. 

Now, consider the terms with $1/\delta_i^2$ for $i \neq d$. We have:
\begin{equation*}
    \sum_{j \in \mathcal{I}_W^{\langle i \rangle}} a_j\zeta_j(\tau) \cdot \sum_{j \in \mathcal{I}_W^{\langle i \rangle}} b_j\zeta_j(\tau) = 0
\end{equation*}

Without loss of generality assume $\sum_{j \in \mathcal{I}_W^{\langle d
\rangle}} a_j\zeta_j(\tau) = 0$. Again, looking at term
$\frac{\alpha}{\delta_i}\sum_{j \in \mathcal{I}_W^{\langle i \rangle}}
b_j\zeta_j(\tau) = 0$ shows that in fact $\sum_{j \in \mathcal{I}_W^{\langle i
\rangle}} b_j\zeta_j(\tau) = 0$ as well.

Now terms involving $1/\gamma^2$ give us
\begin{equation*}
    \sum_{j \in \mathcal{I}_X} a_j\zeta_j(\tau) \cdot \sum_{j \in \mathcal{I}_X} b_j\zeta_j(\tau) = 0
\end{equation*}

For the same very reason, we derive that both sums are zero. Finally, the terms 
$a_{\gamma}\beta\gamma = 0$ and $b_{\gamma}\alpha\gamma=0$ show that $a_{\gamma}=0$
and $b_{\gamma}=0$. Therefore, we have
\begin{equation*}
    a = \alpha + a(\tau) + \sum_{i \in [d+1]}a_{\delta}^{\langle i \rangle} \delta_i, \quad b = \beta + b(\tau) + \sum_{i \in [d+1]}b_{\delta}^{\langle i \rangle} \delta_i
\end{equation*}

Now, let us write the product $ab$ explicitly:
\begin{align*}
    ab &= \cancel{\alpha\beta} + \alpha b(\tau) + \alpha\sum_{i \in [d+1]}b_{\delta}^{\langle i \rangle} \delta_i \\
        &+ \beta a(\tau) + a(\tau)b(\tau) + a(\tau)\sum_{i \in [d+1]}b_{\delta}^{\langle i \rangle} \delta_i \\
        &+ \sum_{i \in [d+1]}a_{\delta}^{\langle i \rangle} \delta_i\beta + \sum_{i \in [d+1]}a_{\delta}^{\langle i \rangle} \delta_i b(\tau) + \sum_{i \in [d+1]}a_{\delta}^{\langle i \rangle}\delta_i \sum_{j \in [d+1]}b_{\delta}^{\langle j \rangle} \delta_j \\
        &= \cancel{\alpha\beta} + \sum_{j \in \mathcal{I}_X}z_j(\beta \ell_i(\tau) + \alpha r_i(\tau) + o_i(\tau)) + \sum_{i \in [d+1]}\delta_ic_i
\end{align*}

Consider the terms that involve $\alpha$ (but not $\alpha\delta_i$ and such!) and $\beta$.
We have:
\begin{equation*}
    b(\tau) = \sum_{j \in \mathcal{I}_X} z_jr_i(\tau) + \sum_{j \in \mathcal{I}_W} c_jr_j(\tau), \quad a(\tau) = \sum_{j \in \mathcal{I}_X} z_j\ell_i(\tau) + \sum_{j \in \mathcal{I}_W} c_j\ell_j(\tau)
\end{equation*}

Thus, define $z_j := c_j$ for $j \in \mathcal{I}_W$. This way, we have
\begin{equation*}
    a(\tau) = \sum_{j \in [n]} z_j\ell_j(\tau), \quad b(\tau) = \sum_{j \in [n]} z_jr_j(\tau)
\end{equation*}

Finally, looking at the terms involving powers of $\tau$ we have:
\begin{equation*}
    \sum_{j \in [n]}z_j\ell_j(\tau) \cdot \sum_{j \in [n]}z_jr_j(\tau) = \sum_{j \in [n]}z_jo_j(\tau) + c_h(\tau)t(\tau)
\end{equation*}

This finally shows that $\{z_j\}_{j \in \mathcal{I}_W} \equiv \{c_j\}_{j \in
\mathcal{I}_W}$ is a witness for the statement $\mathbbm{x} = \{z_j\}_{j \in
\mathcal{I}_W}$.

\section{R1CS-Friendly Architectures}\label{section:r1cs-friendly-architectures}

\subsection{Encoder-Decoder Layer}

Now, given the fact that the neural network has $m$ non-linear activation calls,
the approximate number of constraints for the whole network is roughly $mb$.
Although this does not seem like an issue at first, this quickly becomes one of
the biggest problems in constructing the circuit. Consider the fairly
simple layer, inputting and outputting $m=1000$ neurons. Then,
$\mathbf{y}=\mathsf{ReLU}(W\mathbf{x})$ would cost $1000b$ constraints, which,
for the BN254 field, equates to \textit{roughly 260k constraints}. That is a lot
for the single layer. 

The primary issue is as follows: assume at some point the output of the layer is
$\boldsymbol{z} \in \mathbb{F}^{m}$ and we want to apply the activation function
$\phi$ element-wise. This way, we are asserting the mapping $\phi(z_1,\dots,z_m)
= \left[\phi(z_1),\dots,\phi(z_m)\right]^{\top}$. Assuming $\phi$ is ReLU-based
activation, we need to \textit{impose $mb$ constraints}. Can we do better? 

The answer is \textit{yes}, albeit with certain nuances. Suppose we have a layer
with $n$ inputs and $m$ outputs. \textit{We inject an additional layer with $k$
hidden neurons and apply the non-linearity $\phi$ here and remove the
non-linearity from the output layer.} This way, the encoder-decoder layer would
first project the input $\mathbf{x} \in \mathbb{R}^{n}$ to the latent space
$\boldsymbol{z} \in \mathbb{R}^{k}$ using the linear (encoder) layer
$\boldsymbol{z} = W_E\mathbf{x}$, then apply the non-linearity $\phi$ to the
latent space $\boldsymbol{z}$ and finally project the result back to the output
space $\mathbf{y} \in \mathbb{R}^{m}$ using the linear (decoder) layer
$\mathbf{y} = W_D\phi(\mathbf{z})$. All in all, we get the forward pass
$\mathbf{y} = W_D \phi(W_E\mathbf{x})$.

\begin{definition}
    The \textbf{Encoder-Decoder Layer} is a layer with $n$ inputs and $m$
    outputs, which consists of two linear layers (encoder and decoder) and a
    non-linear activation function $\phi$ applied to the latent space of the
    \textbf{hidden size} $k$. It is described as follows\footnote{While formally
    the forward propagation equation should include the bias term (so that
    equation becomes $W_D\cdot \phi(W_E\mathbf{x} +
    \boldsymbol{\beta}_E)+\boldsymbol{\beta}_D$), we omit it for the sake of
    simplicity by assuming that activation $\mathbf{x}$ is passed with the ``$1$''
    appended.}:
    \begin{equation*}
        \mathsf{EDLayer}(\mathbf{x}; W_D,W_E) = W_D\phi(W_E\mathbf{x})
    \end{equation*}
    where $W_E \in \mathbb{R}^{k \times n}$ is the encoder matrix, $W_D \in
    \mathbb{R}^{m \times k}$ is the decoder matrix and $\phi$ is the non-linear
    activation function. In case $n=m$, following \cite{resnet}, we additionally
    insert the residual connection between the input and output of the layer, so
    that the final output is
    \begin{equation*}
        \mathsf{EDLayer}(\mathbf{x}; W_D,W_E) = W_D\phi(W_E\mathbf{x}) + \mathbf{x}
    \end{equation*}

    We call the ratio $\gamma := \frac{k}{m} < 1$ the \textbf{squeezing factor} of the ED
    Layer.
\end{definition}

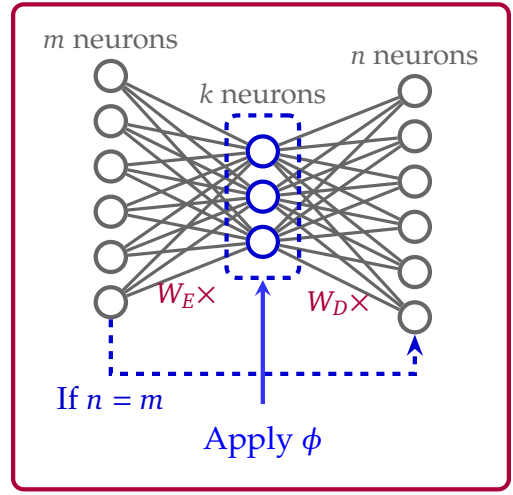
\begin{figure}[H]
% Define colors
\colorlet{neutral-color}{gray!80!black}
\colorlet{activation-color}{blue!80!black}
\colorlet{fc-color}{green!70!black}
\colorlet{ed-color}{purple!90!black}

\centering
\begin{tikzpicture}
    % --- Single-layer fully connected NN ---

    % Input layer
    \foreach \i in {1,2,3,4,5,6}
      \node[circle, draw=neutral-color, ultra thick, minimum size=1pt, name=input\i] at (0,2.5 - \i*0.6) {};
    
    % Output layer
    \foreach \i in {1,2,3,4,5,6}
      \node[circle, draw=activation-color, ultra thick, minimum size=1pt, name=output\i] at (2,2.3 - \i*0.6) {};
    
    % Connections
    \foreach \i in {1,2,3,4,5,6}
      \foreach \j in {1,2,3,4,5,6}
        % Except for \i = \j = 6
        \draw[-, neutral-color, very thick] (input\i) -- (output\j);

    \path (input6) -- (output6) node[midway, below, fc-color] {$W \times$};
    \node[above=-0.5pt of input1] (input1label) {\small \textcolor{neutral-color}{$m$ neurons}}; 
    \node[above=7.5pt of output1] (output1label) {\small \textcolor{neutral-color}{$n$ neurons}}; 

    % Apply activation function
    \node[fit={(output1) (output6)}, dashedbox, draw=blue!80!black, name=fc-out] {};
    \draw[arrow, draw=blue!80, ultra thick] ($(fc-out) + (0, -2.75cm)$) -- (fc-out) node[blue, align=center, name=fc-activation] at ($(fc-out) + (0, -3.25cm)$) {Apply $\phi$};
    
    % Fit box for the regular fully-connected layer
    \node[fit={(input1label) (input1) (input6) (output1) (output1label) (output6) (fc-activation)}, draw=fc-color, thick, inner sep=0.25cm, rectangle, rounded corners, align=center, ultra thick, name=fc] {};
    \node[above=2.75pt of fc, align=center] {\textcolor{fc-color}{Regular Fully-Connected Layer} \\ \textcolor{neutral-color}{$\mathsf{FCLayer}(\mathbf{x};\textcolor{fc-color}{W}) = \textcolor{activation-color}{\phi}(\textcolor{fc-color}{W}\mathbf{x})$}};
    
    % --- Encoder-decoder NN (Right) ---

    % Input layer
    \foreach \i in {1,2,3,4,5,6}
      \node[circle, draw=neutral-color, ultra thick, minimum size=1pt, name=einput\i] at (6,2.5 - \i*0.6) {};
    
    % Latent layer
    \foreach \i in {1,2,3}
      \node[circle, draw=activation-color, ultra thick, minimum size=1pt, name=latent\i] at (8,1.5 - \i*0.6) {};

    % Output layer
    \foreach \i in {1,2,3,4,5,6}
    \node[circle, draw=neutral-color, ultra thick, minimum size=1pt, name=eoutput\i] at (10,2.3 - \i*0.6) {};
    
    % Connections to latent
    \foreach \i in {1,2,3,4,5,6}
        \foreach \j in {1,2,3}
            \draw[-, neutral-color, very thick] (einput\i) -- (latent\j);

    \path (einput6) -- (latent3) node[midway, below, ed-color] {$W_E \times$};
    
    % Connections from latent
    \foreach \i in {1,2,3}
        \foreach \j in {1,2,3,4,5,6}
            \draw[-, neutral-color, very thick] (latent\i) -- (eoutput\j);

    \path (latent3) -- (eoutput6) node[midway, below, ed-color] {$W_D \times$};
    \node[above=-0.5pt of einput1] (einput1label) {\small \textcolor{neutral-color}{$m$ neurons}}; 
    \node[above=7.5pt of latent1] (latent1label) {\small \textcolor{neutral-color}{$k$ neurons}}; 
    \node[above=-0.5pt of eoutput1] (eoutput1label) {\small \textcolor{neutral-color}{$n$ neurons}}; 
    \draw[-{Stealth[length=8pt,open]}, ultra thick, dashed, activation-color] (einput6) |- ($(eoutput6)+(0,-0.75cm)$) node[pos=0.25, below=0.35cm, name=residual] {If $n=m$} -| (eoutput6);

    % Apply activation function
    \node[fit={(latent1) (latent3)}, dashedbox, draw=blue!80!black, name=ed-out] {};
    \draw[arrow, draw=blue!80, ultra thick] ($(ed-out) + (0, -2.75cm)$) -- (ed-out) node[blue, align=center, name=ed-activation] at ($(ed-out) + (0, -3.25cm)$) {Apply $\phi$};

    % Fit box for the encoder-decoder layer
    \node[fit={(einput1) (einput1label) (einput6) (eoutput1) (eoutput1label) (eoutput6) (residual) (ed-activation)}, draw=ed-color, thick, inner sep=0.25cm, rectangle, rounded corners, align=center, ultra thick, name=ed] {};
    \node[above=2.75pt of ed, align=center] {\textcolor{ed-color}{Encoder-Decoder Layer} \\ \textcolor{neutral-color}{$\mathsf{EDLayer}(\mathbf{x};\textcolor{ed-color}{W_E},\textcolor{ed-color}{W_D}) = \textcolor{ed-color}{W_D}\textcolor{activation-color}{\phi}(\textcolor{ed-color}{W_E}\mathbf{x}) + \delta_{nm}\mathbf{x}$}};
    % Residual connection
\end{tikzpicture}
\caption{The visual difference between the regular
\textcolor{fc-color}{fully-connected layer} and the
\textcolor{ed-color}{encoder-decoder layer}. The encoder-decoder layer consists
of two linear layers (encoder \textcolor{ed-color}{$W_E$} and
decoder \textcolor{ed-color}{$W_D$}) and a non-linear activation function
\textcolor{activation-color}{$\phi$} applied to the latent space. The regular
fully-connected layer applies the activation function to the output of the
layer obtained by the linear projection \textcolor{fc-color}{$W$}.}
\label{fig:encoder-decoder}
\end{figure}

Of course, this way some accuracy is lost, but we argue that the accuracy loss is
insignificant for the proper architecture. Such architecture is mainly beneficial 
because of the following fact.

\begin{theorem}
    The encoder-decoder layer $\mathsf{EDLayer}(\mathbf{x}; D,E)$ can be
    implemented in R1CS with roughly $kb$ constraints instead of $mb$, 
    thus giving the speedup of $1/\gamma$ in contrast to the 
    regular fully-connected layer $\mathsf{FCLayer}(\mathbf{x}; W)$.
\end{theorem}

\subsection{Encoder-Decoder Convolutional Layers}

While it is clear how to deal with the fully-connected layers, we need the
analogous construction for the convolutional layers, which operate over images.
Recall that the goal when working with images is to build the mapping from the
volume $\boldsymbol{X} \in \mathbb{R}^{W \times H \times C}$ (stack of $C$
images of size $W \times H$) to another volume $\boldsymbol{Y} \in
\mathbb{R}^{W' \times H' \times C'}$. This is typically done using the $C'$
kernels of size $f \times f$, which we denote by $\{K_c\}_{c \in [C']} \subseteq
\mathbb{R}^{f \times f}$ with stride $s$ which convolves the input image as
follows:
\begin{equation*}
    Y_{i,j,c} = \sum_{k \in [C]}\sum_{u \in [f]}\sum_{v \in [f]} K_{u,v,c} \cdot X_{i\cdot s+u,j\cdot s+v,k}.
\end{equation*}

where the output volume $\boldsymbol{Y}$ is of size $\frac{W}{s} \times
\frac{H}{s} \times C'$. Since such transformation is linear, conducting multiple
convolutions would result in a single analogous linear layer. Therefore, we
introduce the non-linearity similarly to the fully-connected layer: simply apply
the activation (with the offset) to each output latent variable:
\begin{equation*}
    Y_{i,j,c} = \phi\left(\sum_{k \in [C]}\sum_{u \in [f]}\sum_{v \in [f]} K_{u,v,c} \cdot X_{i\cdot s+u,j\cdot s+v,k} + \beta_{i,j,c}\right).
\end{equation*}

Now, the primary issue with this version of convolution is that it requires
computing roughly $H'W'C'$ non-linear activations (essentially, the whole output
volume). To see why that is a problem assume for concreteness that the
intermediate volume size turned out to be $W'=H'=C'=32$, which is not that
uncommon to the modern architectures. In that case, for vanilla Groth16, we
would require approximately $W'H'C'b \approx 8.36 \, \text{mln}$ constraints
(with roughly $420 \, \text{k}$ for the improved UltraGroth protocol). To
mitigate this issue, we need to come up with the way to reduce the number of
non-linear activations.

\subsubsection{Squeeze-and-Excitation Block}

One of such examples in the existing literature is the 
\textit{Squeeze-and-Excitation} (SE) block \cite{se-networks}. The original
network proceeds as follows: given the input volume 
$\boldsymbol{X} \in \mathbb{R}^{W \times H \times C}$, we apply the 
\textit{global average pooling} operation to get the vector $\mathbf{z} \in \mathbb{R}^C$,
where each $z_c$ is the average of the $c$-th channel of the input volume:
\begin{equation*}
    z_c = \frac{1}{W \cdot H} \sum_{i=1}^{W}\sum_{j=1}^{H} X_{i,j,c}, \quad c \in [C]
\end{equation*}

\begin{figure}[H]
\begin{tikzpicture}[scale=1, every node/.style={font=\small}, 
    block/.style={draw, thick, minimum width=1.8cm, minimum height=1cm, align=center},
    volume/.style={fill=gray!20, draw=gray!80!black, ultra thick},
    arrow/.style={-Latex, thick},
    op/.style={draw, thick, ellipse, fill=yellow!30, minimum width=1.5cm, minimum height=0.8cm}
  ]

% Input feature map (3D box)
% Parameters
\def\width{2}
\def\height{2}
\def\depth{0.5}
\def\channels{7}

% Define a list of colors
\definecolor{myred}{RGB}{230, 80, 80}
\definecolor{myorange}{RGB}{255, 165, 0}
\definecolor{myyellow}{RGB}{255, 230, 100}
\definecolor{mygreen}{RGB}{100, 200, 100}
\definecolor{mycyan}{RGB}{100, 220, 220}
\definecolor{myblue}{RGB}{80, 140, 230}
\definecolor{mypurple}{RGB}{180, 100, 220}

\def\colorlist{{"myred","myorange","myyellow","mygreen","mycyan","myblue","mypurple"}}

% Define a list of modified colors
\definecolor{mynewred}{RGB}{200, 90, 100}       % muted, deep red (darker, lower gamma)
\definecolor{myneworange}{RGB}{255, 200, 100}  % soft, creamy orange (higher gamma)
\definecolor{mynewyellow}{RGB}{240, 210, 60}   % golden yellow (darker and warmer)
\definecolor{mynewgreen}{RGB}{160, 240, 160}   % pastel mint green (brighter and lighter)
\definecolor{mynewcyan}{RGB}{80, 160, 190}     % muted teal (cooler and darker)
\definecolor{mynewblue}{RGB}{110, 170, 255}    % vivid sky blue (brighter)
\definecolor{mynewpurple}{RGB}{170, 120, 230}   % deep violet (low gamma, rich tone)

\def\newcolorlist{{"mynewred","myneworange","mynewyellow","mynewgreen","mynewcyan","mynewblue","mynewpurple"}}

% --- Draw the input layer ---
\foreach \i in {0,...,6} {
    \pgfmathsetmacro{\z}{\i*\depth}
    \pgfmathparse{\colorlist[\i]}
    \edef\layercolor{\pgfmathresult}

    % Draw a filled rectangle as a slice of the volume
    \filldraw[fill=\layercolor, opacity=0.9, draw=black, very thick]
    (0,0,\z) -- ++(\width,0,0) -- ++(0,\height,0) -- ++(-\width,0,0) -- cycle;

    % Define a coordinate at the top center of the layer
    \coordinate (layer\i) at ({0.5*\width}, {\height + 0.5}, \z);

    % Optional: draw top edge
    %\draw[black, very thick] (0,0,\z+\depth) -- ++(\width,0,0) -- ++(0,\height,0) -- ++(-\width,0,0) -- cycle;

    % Optional: connect corners to next layer
    \ifnum\i<6
    \foreach \dx/\dy in {0/0, \width/0, \width/\height, 0/\height} {
        \draw[black, very thick] (\dx,\dy,\z) -- (\dx,\dy,\z+\depth);
    }
\fi
}

% --- Draw the output layer ---
\foreach \i in {0,...,6} {
    \pgfmathsetmacro{\z}{\i*\depth}
    \pgfmathparse{\newcolorlist[\i]}
    \edef\layercolor{\pgfmathresult}

    % Draw a filled rectangle as a slice of the volume
    \filldraw[fill=\layercolor, opacity=0.9, draw=black, very thick]
    (10,0,\z) -- ++(\width,0,0) -- ++(0,\height,0) -- ++(-\width,0,0) -- cycle;

    % Define a coordinate at the top center of the layer
    \coordinate (output-layer\i) at ({10 + 0.5*\width}, {\height + 0.5}, \z);

    % Optional: draw top edge
    %\draw[black, very thick] (0,0,\z+\depth) -- ++(\width,0,0) -- ++(0,\height,0) -- ++(-\width,0,0) -- cycle;

    % Optional: connect corners to next layer
    \ifnum\i<6
    \foreach \dx/\dy in {0/0, \width/0, \width/\height, 0/\height} {
        \draw[black, very thick] (10+\dx,\dy,\z) -- (10+\dx,\dy,\z+\depth);
    }
\fi
}

% --- Draw the encoder-decoder architecture ---
% Input layer
\foreach \i in {0,1,2,3,4,5,6}
    \pgfmathparse{\colorlist[\i]}
    \edef\layercolor{\pgfmathresult}
    \node[circle, draw=\layercolor, ultra thick, minimum size=1pt, name=einput\i, scale=0.75] at (3.25,-1.5 - \i*0.4) {};

% Latent layer
\foreach \i in {1,2,3,4}
    \node[circle, draw=gray!50!black, ultra thick, minimum size=0.75pt, name=latent\i, scale=0.75] at (4.5,-1.7 - \i*0.5) {};

% Output layer
\foreach \i in {0,1,2,3,4,5,6}
    \pgfmathparse{\newcolorlist[\i]}
    \edef\layercolor{\pgfmathresult}
    \node[circle, draw=\layercolor, ultra thick, minimum size=0.75pt, name=eoutput\i, scale=0.75] at (5.75,-1.5 - \i*0.4) {};

% Connections to latent
\foreach \i in {0,1,2,3,4,5,6}
    \foreach \j in {1,2,3,4}
        \draw[-, gray!50!black, very thick] (einput\i) -- (latent\j);

% Gated mechanism label below the last latent neuron
\node[below=10pt of latent4, align=center] (gated-mechanism-label) {Gated Mechanism}; 

% Connections from latent
\foreach \i in {1,2,3,4}
    \foreach \j in {0,1,2,3,4,5,6}
        \draw[-, gray!50!black, very thick] (latent\i) -- (eoutput\j);

% Label above the input
\node[above=-0.5cm] at (layer0) {\large Input $\boldsymbol{X}$};
\node[above=-0.5cm] at (output-layer0) {\large Output $\boldsymbol{Y}$};

% Labels with width/height/channels marking
\node at (0, -1.5, 0) {\large $W$};
\node at (-1.5, 0, 0) {\large $H$};
\node at (-0.8, 1.7, 0) {\large $C$};

% --- Draw the arrows to the encoder-decoder layer ---
% Connect each input layer slice to its corresponding neuron
\foreach \i in {0,...,6} {
    \pgfmathsetmacro{\z}{\i*\depth}
    \pgfmathsetmacro{\y}{0.0}
    \pgfmathsetmacro{\x}{\width}

    \pgfmathparse{\colorlist[\i]}
    \edef\arrowcolor{\pgfmathresult}

    % Step 1: Start at the top-center of the rectangle
    % Step 2: Go down
    % Step 3: Go right to the neuron position
    \draw[-Stealth, \arrowcolor, ultra thick]
        (\x, \y, \z) -- ++(0, -0.5, 0) coordinate (midpt) -- (einput\i);
}

% --- Draw arrows from input to output layer ---
% Circle with multiplication sign
\node[draw, circle, minimum size=12.5pt, very thick, inner sep=0pt] (mult) at (7,0) {$\boldsymbol{\times}$};
\node[above=-0.5pt of mult, align=center] (multlabel) {Channel-wise \\ multiplication}; 

% Draw an arrow between the layer and the multiplication sign
\foreach \i in {0,...,6} {
    \pgfmathparse{\newcolorlist[\i]}
    \edef\arrowcolor{\pgfmathresult}

    % Compute the coordinates of the corner (directly below mult)
    \coordinate (midpoint\i) at (7, {-0.5 - \i*0.4});

    % Draw elbow arrow: from eoutput\i to point below mult, then up
    \draw[-Stealth, \arrowcolor, ultra thick]
        (eoutput\i) -- (midpoint\i) -- (mult.south);
}

% Draw an arrow between the layer and multiplication sign
\draw[-Stealth, gray!50!black, ultra thick] (0.85, 0, 0) -- (mult.west);
\draw[-Stealth, gray!50!black, ultra thick] (mult.east) -- (9.85-0.5*\width, 0, 0);

\end{tikzpicture}
\caption{The Squeeze-and-Excitation (SE) block --- one of R1CS-friendly layers}
\label{fig:squeeze-excitation}
\end{figure}
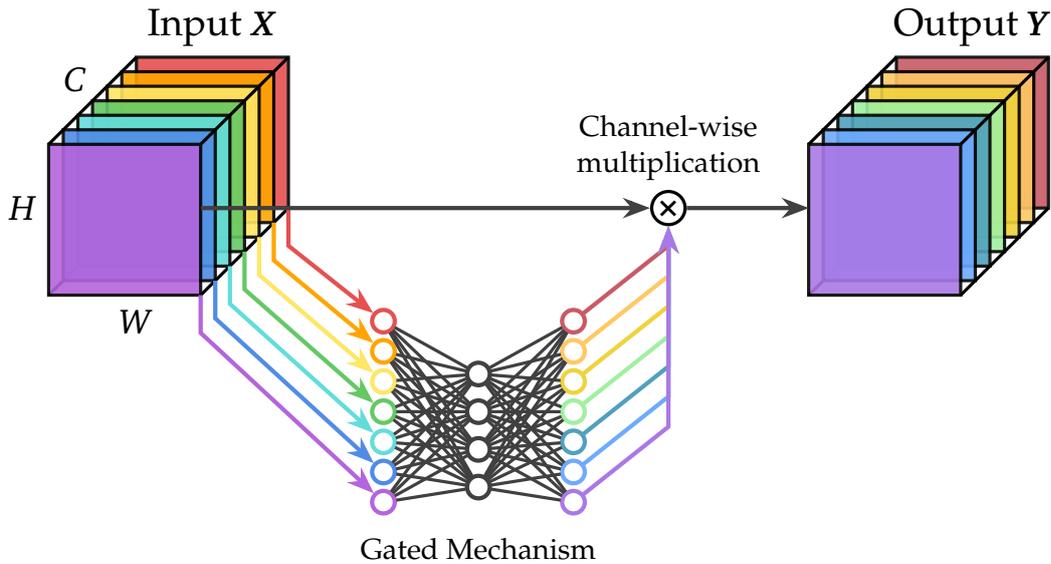

Next, we apply the gating mechanism to the vector $\mathbf{z}$ to set
$\mathbf{s} \in \mathbb{R}^C$, which essentially is the encoder-decoder layer,
but with the non-linearity applied at the end: $\mathbf{s} = \sigma(W_D
\phi(W_E\mathbf{z}))$. Here, the number of neurons in the hidden layer is
$\frac{C}{r}$ for some constant $r$. Finally, we multiply each channel of the
input volume $\boldsymbol{X}$ by the corresponding $s_c$ to get the output
volume: $\boldsymbol{Y}[:,:,c] \gets s_c \cdot \boldsymbol{X}[:,:,c]$. The whole 
computation flow is illustrated in \Cref{fig:squeeze-excitation}.

We give the following proposition to get the number of constraints.

\begin{proposition}
    Given that both $\sigma$ and $\phi$ cost $b$ constraints to verify, the
    total cost of the squeeze-and-excitation block is approximately
    $\left(1+\frac{1}{r}\right)Cb+HWC$. 
\end{proposition}

\textbf{Reasoning.} Since all linearities are free, we only need to (a) check
the activation over the hidden layer, (b) check the activation over the output
layer and (c) check the multiplication of the input volume with the output of
the SE block. The first step costs $\frac{C}{r} \cdot b$ constraints, the second
step costs $Cb$ constraints and the third step costs $HWC$ constraints. The
total cost is therefore $\left(1+\frac{1}{r}\right)Cb + HWC$.

One modification of such block is the following: instead of applying the global
average pooling along the whole volume, we can effectively split it into blocks
(say, forming the grid of size $P_W \times P_H$). This way, we increase the
number of constraints by the factor of $P_WP_H$, but the neural network can
learn more complex representations. 

\subsubsection{Encoder-Decoder Convolutional Layer}

The issue with the Squeeze-and-Excitation Block is that it does not change the
shape of the volume. For that reason, we give the more interesting construction,
which is the \textit{encoder-decoder convolutional layer}. We combine certain
ideas from the Vision Transformer \cite{vit} and the previously mentioned
squeeze-and-excitation block \cite{se-networks}.

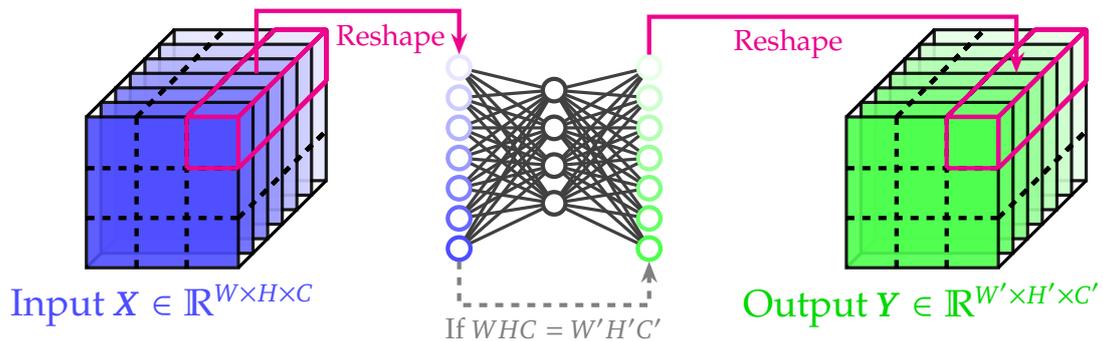
\begin{figure}[H]
\centering
\begin{tikzpicture}[scale=1, every node/.style={font=\small}, 
    block/.style={draw, thick, minimum width=1.8cm, minimum height=1cm, align=center},
    volume/.style={fill=gray!20, draw=gray!80!black, ultra thick},
    arrow/.style={-Latex, thick},
    op/.style={draw, thick, ellipse, fill=yellow!30, minimum width=1.5cm, minimum height=0.8cm}
  ]

% Input feature map (3D box)
% Parameters
\def\width{2}
\def\height{2}
\def\depth{0.5}
\def\channels{7}
\def\offset{10}

% Define a list of colors
\colorlet{blue10}{blue!10}
\colorlet{blue20}{blue!20}
\colorlet{blue30}{blue!30}
\colorlet{blue40}{blue!40}
\colorlet{blue50}{blue!50}
\colorlet{blue60}{blue!60}
\colorlet{blue70}{blue!70}
\colorlet{selected-color}{magenta}

\def\colorlist{{"blue10","blue20","blue30","blue40","blue50","blue60","blue70"}}

% Define a list of colors
\colorlet{green10}{green!10}
\colorlet{green20}{green!20}
\colorlet{green30}{green!30}
\colorlet{green40}{green!40}
\colorlet{green50}{green!50}
\colorlet{green60}{green!60}
\colorlet{green70}{green!70}

\def\outputcolorlist{{"green10","green20","green30","green40","green50","green60","green70"}}

% --- Draw the input layer ---
\foreach \i in {0,...,6} {
    \pgfmathsetmacro{\z}{\i*\depth}
    \pgfmathparse{\colorlist[\i]}
    \edef\layercolor{\pgfmathresult}

    % Draw a filled rectangle as a slice of the volume
    \filldraw[fill=\layercolor, opacity=0.9, draw=black, very thick]
    (0,0,\z) -- ++(\width,0,0) -- ++(0,\height,0) -- ++(-\width,0,0) -- cycle;

    % Define a coordinate at the top center of the layer
    \coordinate (layer\i) at ({0.5*\width}, {\height + 0.5}, \z);

    % Optional: draw top edge
    %\draw[black, very thick] (0,0,\z+\depth) -- ++(\width,0,0) -- ++(0,\height,0) -- ++(-\width,0,0) -- cycle;

    % Optional: connect corners to next layer
    \ifnum\i<6
    \foreach \dx/\dy in {0/0, \width/0, \width/\height, 0/\height} {
        \draw[black, very thick] (\dx,\dy,\z) -- (\dx,\dy,\z+\depth);
    }
    \fi

    % Draw splitting lines
    \draw[black, dashed, ultra thick] (0.0*\width,0.33*\height,3.0) -- (1.0*\width,0.33*\height,3.0);
    \draw[black, dashed, ultra thick] (0.0*\width,0.66*\height,3.0) -- (1.0*\width,0.66*\height,3.0);
    \draw[black, dashed, ultra thick] (0.33*\width,1.0*\height,3.0) -- (0.33*\width,0.0*\height,3.0);
    \draw[black, dashed, ultra thick] (0.33*\width,1.0*\height,3.0) -- (0.33*\width,1.0*\height,0.0);
    \draw[black, dashed, ultra thick] (0.66*\width,1.0*\height,3.0) -- (0.66*\width,0.0*\height,3.0);
    \draw[black, dashed, ultra thick] (0.66*\width,1.0*\height,3.0) -- (0.66*\width,1.0*\height,0.0);
    \draw[black, dashed, ultra thick] (1.0*\width,0.33*\height,3.0) -- (1.0*\width,0.33*\height,0.0);

    % Draw line that select one of the blocks
    \draw[selected-color, ultra thick] (0.66*\width,0.66*\height,3.0) -- (1.0*\width,0.66*\height,3.0);
    \draw[selected-color, ultra thick] (0.66*\width,0.66*\height,3.0) -- (0.66*\width,1.0*\height,3.0);
    \draw[selected-color, ultra thick] (1.0*\width,0.66*\height,3.0) -- (1.0*\width,1.0*\height,3.0);
    \draw[selected-color, ultra thick] (0.66*\width,1.0*\height,3.0) -- (1.0*\width,1.0*\height,3.0);
    \draw[selected-color, ultra thick] (0.66*\width,1.0*\height,3.0) -- (0.66*\width,1.0*\height,0.0);
    \draw[selected-color, ultra thick] (1.0*\width,0.66*\height,3.0) -- (1.0*\width,0.66*\height,0.0);
    \draw[selected-color, ultra thick] (1.0*\width,1.0*\height,3.0) -- (1.0*\width,1.0*\height,0.0);
    \draw[selected-color, ultra thick] (1.0*\width,0.66*\height,0.0) -- (1.0*\width,1.0*\height,0.0);
    \draw[selected-color, ultra thick] (0.66*\width,1.0*\height,0.0) -- (1.0*\width,1.0*\height,0.0);

    \coordinate (input-block-center) at ({0.83*\width}, {\height}, 1.5);
}

% --- Draw the output volume ---
\foreach \i in {0,...,6} {
    \pgfmathsetmacro{\z}{\i*\depth}
    \pgfmathparse{\outputcolorlist[\i]}
    \edef\layercolor{\pgfmathresult}

    % Draw a filled rectangle as a slice of the volume
    \filldraw[fill=\layercolor, opacity=0.9, draw=black, very thick]
    (\offset,0,\z) -- ++(\width,0,0) -- ++(0,\height,0) -- ++(-\width,0,0) -- cycle;

    % Define a coordinate at the top center of the layer
    \coordinate (output-layer\i) at ({\offset+0.5*\width}, {\height + 0.5}, \z);

    % Optional: draw top edge
    %\draw[black, very thick] (0,0,\z+\depth) -- ++(\width,0,0) -- ++(0,\height,0) -- ++(-\width,0,0) -- cycle;

    % Optional: connect corners to next layer
    \ifnum\i<6
    \foreach \dx/\dy in {0/0, \width/0, \width/\height, 0/\height} {
        \draw[black, very thick] (\offset+\dx,\dy,\z) -- (\offset+\dx,\dy,\z+\depth);
    }
    \fi

    % Draw splitting lines
    \draw[black, dashed, ultra thick] (\offset+0.0*\width,0.33*\height,3.0) -- (\offset+1.0*\width,0.33*\height,3.0);
    \draw[black, dashed, ultra thick] (\offset+0.0*\width,0.66*\height,3.0) -- (\offset+1.0*\width,0.66*\height,3.0);
    \draw[black, dashed, ultra thick] (\offset+0.33*\width,1.0*\height,3.0) -- (\offset+0.33*\width,0.0*\height,3.0);
    \draw[black, dashed, ultra thick] (\offset+0.33*\width,1.0*\height,3.0) -- (\offset+0.33*\width,1.0*\height,0.0);
    \draw[black, dashed, ultra thick] (\offset+0.66*\width,1.0*\height,3.0) -- (\offset+0.66*\width,0.0*\height,3.0);
    \draw[black, dashed, ultra thick] (\offset+0.66*\width,1.0*\height,3.0) -- (\offset+0.66*\width,1.0*\height,0.0);
    \draw[black, dashed, ultra thick] (\offset+1.0*\width,0.33*\height,3.0) -- (\offset+1.0*\width,0.33*\height,0.0);

    % Draw line that select one of the blocks
    \draw[selected-color, ultra thick] (\offset+0.66*\width,0.66*\height,3.0) -- (\offset+1.0*\width,0.66*\height,3.0);
    \draw[selected-color, ultra thick] (\offset+0.66*\width,0.66*\height,3.0) -- (\offset+0.66*\width,1.0*\height,3.0);
    \draw[selected-color, ultra thick] (\offset+1.0*\width,0.66*\height,3.0) -- (\offset+1.0*\width,1.0*\height,3.0);
    \draw[selected-color, ultra thick] (\offset+0.66*\width,1.0*\height,3.0) -- (\offset+1.0*\width,1.0*\height,3.0);
    \draw[selected-color, ultra thick] (\offset+0.66*\width,1.0*\height,3.0) -- (\offset+0.66*\width,1.0*\height,0.0);
    \draw[selected-color, ultra thick] (\offset+1.0*\width,0.66*\height,3.0) -- (\offset+1.0*\width,0.66*\height,0.0);
    \draw[selected-color, ultra thick] (\offset+1.0*\width,1.0*\height,3.0) -- (\offset+1.0*\width,1.0*\height,0.0);
    \draw[selected-color, ultra thick] (\offset+1.0*\width,0.66*\height,0.0) -- (\offset+1.0*\width,1.0*\height,0.0);
    \draw[selected-color, ultra thick] (\offset+0.66*\width,1.0*\height,0.0) -- (\offset+1.0*\width,1.0*\height,0.0);

    \coordinate (output-block-center) at ({\offset+0.83*\width}, {\height}, 1.5);
}

% --- Draw the encoder-decoder neurons ---
% --- Draw the encoder-decoder architecture ---
% Input layer
\foreach \i in {0,1,2,3,4,5,6}
    \pgfmathparse{\colorlist[\i]}
    \edef\layercolor{\pgfmathresult}
    \node[circle, draw=\layercolor, ultra thick, minimum size=1pt, name=einput\i, scale=0.75] at (3.75,1.5 - \i*0.4) {};

% Latent layer
\foreach \i in {1,2,3,4}
    \node[circle, draw=gray!50!black, ultra thick, minimum size=0.75pt, name=latent\i, scale=0.75] at (5.0,1.7 - \i*0.5) {};

% Output layer
\foreach \i in {0,1,2,3,4,5,6}
    \pgfmathparse{\outputcolorlist[\i]}
    \edef\layercolor{\pgfmathresult}
    \node[circle, draw=\layercolor, ultra thick, minimum size=0.75pt, name=eoutput\i, scale=0.75] at (6.25,1.5 - \i*0.4) {};

% Connections to latent
\foreach \i in {0,1,2,3,4,5,6}
    \foreach \j in {1,2,3,4}
        \draw[-, gray!50!black, very thick] (einput\i) -- (latent\j);

\foreach \i in {0,1,2,3,4,5,6}
    \foreach \j in {1,2,3,4}
        \draw[-, gray!50!black, very thick] (latent\j) -- (eoutput\i);

% Gated mechanism label below the last latent neuron
%\node[below=30pt of latent4, align=center] (gated-mechanism-label) {Encoder-Decoder}; 

\draw[-{Stealth[length=8pt,open]}, ultra thick, selected-color] 
  (input-block-center) 
  |- ($(einput0)+(0,0.75cm)$) 
  node[xshift=-0.9cm, below] {Reshape} 
  -| (einput0);

\draw[-{Stealth[length=8pt,open]}, ultra thick, selected-color] 
  (eoutput0) 
  |- ($(output-block-center)+(0,0.75cm)$) 
  node[xshift=-3.0cm, below] {Reshape} 
  -| (output-block-center);
\draw[-{Stealth[length=8pt,open]}, ultra thick, dashed, gray] 
    (einput6) |- ($(eoutput6)+(0,-0.75cm)$) node[xshift=-1.25cm, below=0.0cm, name=residual] {If $WHC=W'H'C'$} -| (eoutput6);

\node at (0.5*\width,-0.25*\height,3.0) {\large \textcolor{blue70}{Input $\boldsymbol{X} \in \mathbb{R}^{W \times H \times C}$}};
\node at (\offset+0.5*\width,-0.25*\height,3.0) {\large \textcolor{green!85!black}{Output $\boldsymbol{Y} \in \mathbb{R}^{W' \times H' \times C'}$}};

\end{tikzpicture}
\caption{The encoder-decoder convolutional layer. The input volume is split into
blocks of size $P \times P \times C$ and the encoder-decoder layer is applied to each
block. The output is obtained by combining the outputs of processed blocks with
the proper reshaping.}
\label{fig:encoder-decoder-conv}
\end{figure}

First, we split the input volume into blocks of size $P_W\times P_H$. For
simplicity, assume $P_W=P_H=P$ and naturally assume $P \mid W,H$. Assuming the
input volume is $\boldsymbol{X} \in \mathbb{R}^{W \times H \times C}$, we get
$P^2$ blocks $\boldsymbol{X}_{i,j} \in \mathbb{R}^{\frac{W}{P} \times
\frac{H}{P} \times C}$ for $i,j \in [P]$. Then, we flatten each block into the
vector of size $\frac{WHC}{P^2}$. We connect the flattened blocks to the
encoder-decoder layer, each with the $K$ hidden units. Finally, assume we want
to get $W'$, $H'$ and $C'$ as the output volume. This way, we connect the latent
layer (of size $K$) to the output layer of size $\frac{W'H'C'}{P^2}$. Finally,
we reshape the output volume into the volume of size $\frac{W'}{P} \times
\frac{H'}{P} \times C'$ and combine the blocks together to get the final output
volume $\boldsymbol{Y} \in \mathbb{R}^{W' \times H' \times C'}$.

Now, let us analyze the cost of such convolution.
\begin{proposition}
    Given that the non-linearity in the encoder-decoder layer costs $b$
    constraints, the total cost of the encoder-decoder convolutional layer is
    approximately $P^2Kb$.
\end{proposition}

\textbf{Reasoning.} The only non-linear operation in the encoder-decoder layer
is the activation function. Since each processing of encoder-decoder costs $Kb$
constraints and we need to process $P^2$ blocks, the total cost is $P^2Kb$.

Note that while the regular convolutional layer costs $W'H'C'b$ constraints, we
reduce it significantly to $P^2Kb$ constraints.

\section{Quantization Scheme Proof}\label{section:quant-proof-appendix}

\begin{theorem}
    Let $\widehat{x} := Q_{\rho}(x)$ and $\widehat{y} := Q_{\rho}(y)$ be the
    quantized values of $x$ and $y$, respectively. Then, if we define circuit
    $\mathtt{C}_f(\widehat{x},\widehat{y}) = \widehat{x} + \widehat{y}$ for
    the addition $f(x,y)=x + y$, the following relation for an error
    $\varepsilon_{\rho}$ holds:
    \begin{equation*}
        \varepsilon_{\rho} := \left| D_{\rho}(\mathtt{C}_f(\widehat{x},\widehat{y})) - f(x,y) \right| \leq 2\cdot 2^{-\rho}
    \end{equation*}

    Consequently, the error $\varepsilon_{\rho}$ is negligible in $\rho$.
\end{theorem}

\textbf{Proof.} For simplicity, assume $x,y>0$ (the negative case, with more
care, can be handled similarly). Since $\widehat{x} = Q_{\rho}(x)$ and
$\widehat{y} = Q_{\rho}(y)$, we have $\widehat{x} = 2^{\rho}x + \delta_x$ for
$|\delta_x|<1$. Same holds for $y$: we have $\widehat{y} = 2^{\rho}y+\delta_y$
with $|\delta_y|<1$. Then, it is easy to see that:
\begin{equation*}
    \mathtt{C}_f(\widehat{x},\widehat{y}) = \widehat{x} + \widehat{y} = (2^{\rho}x + \delta_x) + (2^{\rho}y + \delta_y) = 2^{\rho}(x+y) + (\delta_x + \delta_y)
\end{equation*}

Dequantization of such expression gives:
\begin{equation*}
    D_{\rho}(\mathtt{C}_f(\widehat{x},\widehat{y})) = x + y + 2^{-\rho}(\delta_x + \delta_y)
\end{equation*}

From which follows that $|D_{\rho}(\mathtt{C}_f(\widehat{x},\widehat{y})) -
f(x,y)| \leq 2^{-\rho} (|\delta_x| + |\delta_y|) \leq 2 \cdot 2^{-\rho}$.

\begin{theorem}
    Let $\widehat{x} := Q_{\rho}(x)$ and $\widehat{y} := Q_{\rho}(y)$ be the
    quantized values of $x$ and $y$, respectively. Then, if we define circuit
    $\mathtt{C}_f(\widehat{x},\widehat{y}) = \widehat{x}\cdot \widehat{y}$ for
    the multiplication $f(x,y)=x \cdot y$ and assume $\beta := 2\max\{|x|,|y|\}$
    to be relatively small, the following relation for an error
    $\varepsilon_{\rho}$ holds:
    \begin{equation*}
        \varepsilon_{\rho} := \left| D_{2\rho}(\mathtt{C}_f(\widehat{x},\widehat{y})) - f(x,y) \right| \leq 2^{-\rho} \beta + 2^{-2\rho}
    \end{equation*}

    Consequently, the error $\varepsilon_{\rho}$ is negligible in $\rho$.
\end{theorem}

\textbf{Proof.} For simplicity, assume $x,y>0$ (the negative case, with more
care, can be handled similarly). Since $\beta$ is assumed to be small, we assume
that $\widehat{x} = [2^{\rho}x]$ and $\widehat{y} = 2^{\rho}y$ as well as
$\widehat{x} \cdot \widehat{y}$ do not overflow the field. Then, it 
is easy to see that $\widehat{x} = 2^{\rho}x + \delta_x$ for $|\delta_x|<1$. 
Same holds for $y$: we have $\widehat{y} = 2^{\rho}y+\delta_y$ with $|\delta_y|<1$. Then,
\begin{equation*}
    \mathtt{C}_f(\widehat{x},\widehat{y}) = \widehat{x} \cdot \widehat{y} = (2^{\rho}x + \delta_x)(2^{\rho}y + \delta_y) = 2^{2\rho}xy + 2^{\rho}(\delta_x y + \delta_y x) + \delta_x \cdot \delta_y
\end{equation*}

Now, we can dequantize the result with precision $2\rho$:
\begin{equation*}
    D_{2\rho}(\mathtt{C}_f(\widehat{x},\widehat{y})) = xy + 2^{-\rho}(\delta_x y + \delta_y x) + 2^{-2\rho}\delta_x \delta_y
\end{equation*}

Now note that $|\delta_x y + \delta_y x| \leq |x| + |y| \leq 2\max\{|x|,|y|\} = \beta$. It is also 
easy to see that $|2^{-2\rho}\delta_x\delta_y| < 2^{-2\rho}$. Therefore,
\begin{align*}
    \varepsilon_{\rho} &= \left|D_{2\rho}(\mathtt{C}_f(\widehat{x},\widehat{y})) - f(x,y)\right| \\
                    &= \left|2^{-\rho}(\delta_x y + \delta_y x) + 2^{-2\rho}\delta_x \delta_y\right| \\
                    &\leq 2^{-\rho}|\delta_x y + \delta_y x| + 2^{-2\rho}|\delta_x \delta_y| \\
                    &\leq 2^{-\rho}\beta + 2^{-2\rho}. \hfill \blacksquare
\end{align*}
    
\end{appendices}

\end{document}